Automatic techniques for cochlear implant
CT image analysis

By

Yiyuan Zhao

Dissertation

Submitted to the Faculty of the

Graduate School of Vanderbilt University

in partial fulfillment of the requirements

for the degree of

DOCTOR OF PHILOSOPHY

in

Electrical Engineering

May 11th, 2018

Nashville, Tennessee

Approved:

Benoit M. Dawant, Ph.D.

Jack H. Noble, Ph.D.

Robert F. Labadie, M.D., Ph.D.

Bennett A. Landman, Ph.D.

Richard A. Peters, Ph.D.

To my beloved grandparents Jufa Zhu and Chunxian Yang. You raised me up and gave me unconditional love and supports to pursue my dream.

# ACKNOWLEDGMENTS

The path towards finishing this work and writing up this dissertation is long and arduous – and it is not possible for me to finish this dissertation without the guidance and support and advice from my committee members and the colleagues at Vanderbilt University Medical Image Processing (MIP) lab and the Vanderbilt Biomedical Image Analysis for Image Guided Interventions Laboratory (BAGL).

First and foremost, I express my heartiest gratitude to my respected adviser Dr. Benoit M. Dawant for introducing me to this amazing research field and giving me his trust for letting me join his research team in MIP lab. During my Ph.D. years, Dr. Dawant not only financially supported me but also taught me the advanced medical image processing knowledge. He had always been very patient when giving me constructive feedbacks on my research. Working under his guidance helped me grow to be a professional researcher. The highly motivated attitude, the honest and hard-working work ethics I have learned from Dr. Dawant will benefit me for life.

I also owe my deepest gratitude to my other respected adviser Dr. Jack H. Noble. Dr. Noble had directly mentored me through my early days at the MIP lab. I would not be able to finish my first project, my first research paper, and my first conference presentation without his patient guidance. I had been deeply impressed by his dedication to research, his detail-oriented working ethic, and his passion to explore the new areas of the research field. Those spirits will always be inspirations for me to achieve more progress in the future.

I would also like to express my appreciation to other members on my dissertation committee. Besides giving feedbacks on my dissertation, they have also helped me progress both professionally and personally during my Ph.D. years. Thanks to Dr. Bennett A. Landman for choosing me to be his teaching assistant for one year. This process had

# ACKNOWLEDGMENTS

strengthened my skillset for presenting knowledge. Thanks to Dr. Robert F. Labadie for teaching me the clinical background of cochlear implant. Through the meetings with him I learned the clinical impact of my research and the potential improvements that should be made. Thanks to Dr. Richard A. Peters. He had been my academic adviser for my first two years at Vanderbilt University. I appreciate his efforts in helping me with exploring my research interests and his encouragements for me to pursue my career goal.

I want to thank for the colleagues and some former colleagues from MIP lab, BAGL lab, and Vanderbilt University Medical Center for supporting my research. Thanks to Rui Li for providing software support for the research. Thanks to Dr. Raul Wirz and Ms. Priyanka Prasad for managing the cochlear implant patient database. Thanks to Au.D. Robert Dwyer for giving feedbacks on the performance of my algorithms. Thanks to Dongqing Zhang, Xiaochen Yang, Srijata Chakravorti, Ahmet Cakir, Dr. Yuan Liu, Dr. Fitsum A. Reda, Robert Shults and Bill Rodriguez in my lab for all the valuable discussions and help on my projects. The experience I spent with you was invaluable and unforgettable.

I would also like to thank for Dr. Julie A. Adams for recommending me to join MIP lab at the second year and giving me the confidence in continuing finishing my degree. Your words inspired me to stick to my goals whenever I met any obstacles.

Special thanks to financial support by NIH and NIDCD grants R01DC014037, R01DC008408, R21DC012620, R01DC014462, 5R01DC014462, and 5R01DC014037. Special thanks also to financial support by Vanderbilt Institute for Surgery Engineering.

Last, I want to thank my parents Mr. Ming Zhao and Ms. Min Zhu, for the unconditional love, support and understanding they have given me over the years. Thank you and I dedicate my work to you.

TABLE OF CONTENTS

# LIST OF TABLES

LIST OF TABLES

# LIST OF FIGURES

# LIST OF FIGURES

# LIST OF ABBREVIATIONS

AB ............................................................................................................... Advanced Bionics®

AB1 ....................................................................................................... Advanced Bionics 1J array

AB2 ................................................................................................ Advanced Bionics Mid-scala array

AB3 .................................................................................................... Advanced Bionics Helix array

AL .................................................................................... Automatic localization results for CI electrodes

AR ........................................................................................................... Active region (of modiolus)

ASM ................................................................................................................. Active shape model cGAN ..................................................................................... Conditional Generative Adversarial Network

CI ......................................................................................................................... Cochlear implant

CL ................................................................................................... Centerline-based localization method

CO ................................................................................................................................ Cochlear®

CO1 ......................................................................................... Cochlear contour advance (512) array

CO2 ............................................................................................... Cochlear CI422 (522) array

CO3 ................................................................................................... Cochlear CI24RE-ST array

COI .................................................................................... Candidate voxel of interest for CI electrodes

DtoBM .............................................................................................. Distance to basilar membrane

DtoM .................................................................................................. Distance to modiolar surface

DOI .................................................................................................... Angular depth of insertion

CT ..................................................................................................................... Computed tomography

DVF ........................................................................................................ Distance-vs.-Frequency

eCT ................................................................................................... Extended Hounsfield Unit CT

GL .................................................................................... Ground truth localization results for CI electrodes

GP ........................................................................................... Graph-based path finding algorithm

INTRODUCTION

1.1 Cochlear implant

The cochlea is the auditory portion of the inner ear. As shown in Figure 1.1a, it is a spiral-shaped cavity which makes 2.5 turns around its axis. In a natural hearing process, when the sound waves reach the inner ear, the malleus, the incus, and the stapes vibrate. These vibrations cause the oval window of cochlea to send pulsating fluid waves that stimulate the spiral ganglions (SG) in the cochlea [1]. The SG nerves are the nerve pathways that branch to the cochlea from the auditory nerves, which are tonotopically ordered by decreasing characteristic frequency along the length of the cochlea [2, 3] as shown in Figure 1.2a. A SG nerve is stimulated if the incoming sound contains the frequency associated with it. This stimulation generates hearing impulses and the hearing impulses are sent to the brain to induce a sense of hearing. Cochlear implants (CIs) are neural prosthetics that provide a sense of sound to people who experience severe to profound hearing loss [1]. As shown in Figure 1.1b, a CI consists of two components: an external component and an internal component. The external component contains a microphone, a processor, and a transmitter, which are used to process sounds and send signals to the implanted electrodes. The internal component contains a CI electrode array, which receives the signals sent by the external component and bypasses the damaged cochlea and directly stimulates the SG nerves. During a CI surgery, a CI electrode array is blindly threaded into the cochlea by a surgeon. After the surgery, audiologists need to program the CI device by defining a series of CI instructions we refer to as the "MAP". The tuning of the "MAP" involves a process that specifies stimulation levels for implanted CI electrodes based on the measurements of the recipient's perceived loudness, and a process that selects a frequency allocation table, which defines activation levels for individual electrodes when specific frequencies are detected in the sound. According to the frequency allocation table, the electrodes associated with the specific frequencies that are present in the incoming sound are activated in a CI-assisted hearing process. The electrode activation stimulates the SG nerves and provides a sense of hearing to the CI recipients [4]. CIs have achieved a significant successful rate in hearing restoration among users with an average postoperative sentence recognition rates over 70% correct for unilaterally implanted users and 80% correct for bilaterally implanted users [5, 6]. However, there are still a number of recipients suffering from a marginal experience in hearing restoration.

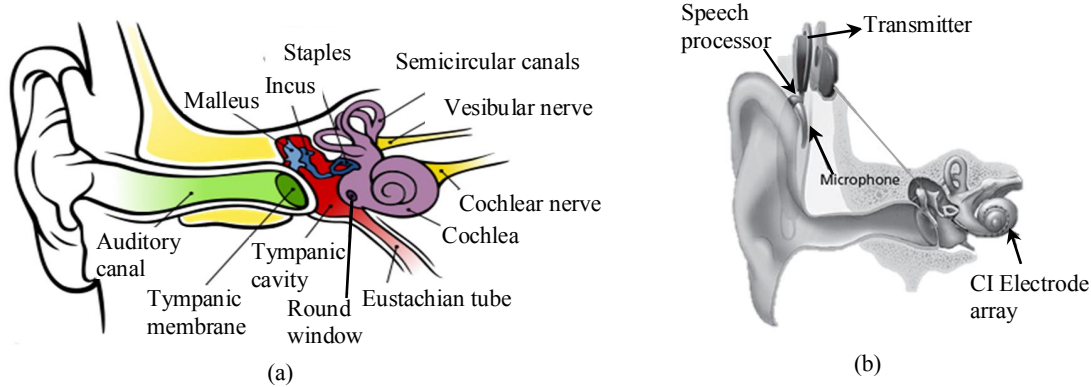

**Figure 1.1.** Panel (a) [35] shows the anatomical structure of an inner ear. Panel (b) [1] shows the components of a cochlear implant device.

Recent studies have demonstrated a correlation between hearing outcomes and the intra-cochlear locations of CI electrodes [7-12]. Competing stimulation, which is also known as "electrode interaction" at the neural level, is one major factor causing hearing outcomes to decline. Electrode interaction occurs when multiple CI electrodes stimulate the same auditory neural site (competing stimulation) [13, 14]. This can be avoided by having CI experts manually deactivate the CI electrodes causing the competing stimulations. To perform this deactivation process, the spatial relationship between the electrodes and the auditory neural sites needs to be determined before analyzing the possibility for individual electrodes to cause electrode interaction [15, 16], as shown in Figure 1.2b. However, determining the spatial relationship between the CI electrodes and the intra-cochlear anatomy is a difficult task because (1) electrode arrays are blindly threaded into cochlea by surgeons during the surgeries. There is no knowledge about the final locations of the electrodes after the surgery, and (2) it is hard to locate the intra-cochlear anatomy in the post-implantation CTs due to the image artifacts introduced by the metallic implants. Figure 1.3a shows an example of the cropped volume of interest (VOI) pre-implantation CT image with intra-cochlear anatomy structures segmented. Figure 1.3b shows the post-implantation CT image of the same case shown in Figure 1.3a. As can be seen in Figure 1.3b, the metallic electrodes lead to relatively high intensities around the electrode contacts, which makes it possible to manually pick them out. However, the metallic electrodes also distort the intensity around the electrode array due to the beam hardening artifacts, which makes it difficult to segment the intra-cochlear anatomy structures directly from the post-operative CT images. In Figure 1.2b we show the CI electrodes activation patterns. When the optimal electrode configuration is selected, some electrodes are deactivated in order to reduce electrode interactions. The traditional clinical workflow assumes all the electrodes are placed within the cochlea at predefined positions and the audiologists use a default frequency allocation table to program the CIs. This generates sub-optimal electrode configurations which negatively affects hearing outcomes.

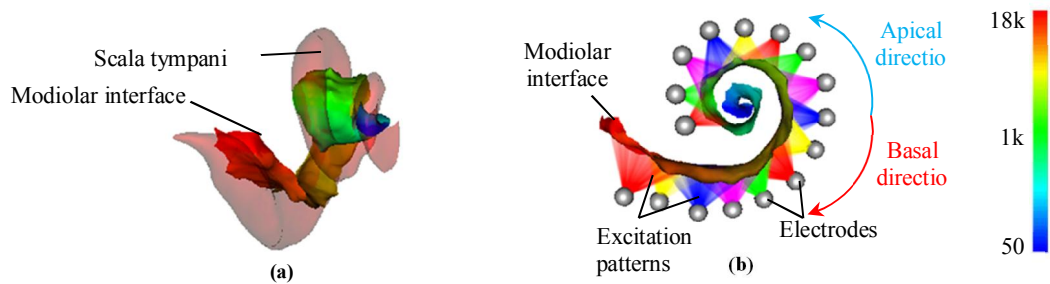

**Figure 1.2.** Visualization of CI electrode activation patterns. In (a), the scala tympani (an intracochlear cavity) is shown with the modiolar surface, which represents the interface between of the SG nerves and the intra-cochlear cavities and is color-coded with the tonotopic place frequencies of the SG in Hz. In (b), synthetic examples of stimulation patterns on the modiolar interface created by the implanted electrodes are shown in multiple colors to illustrate the concept of stimulation overlap.

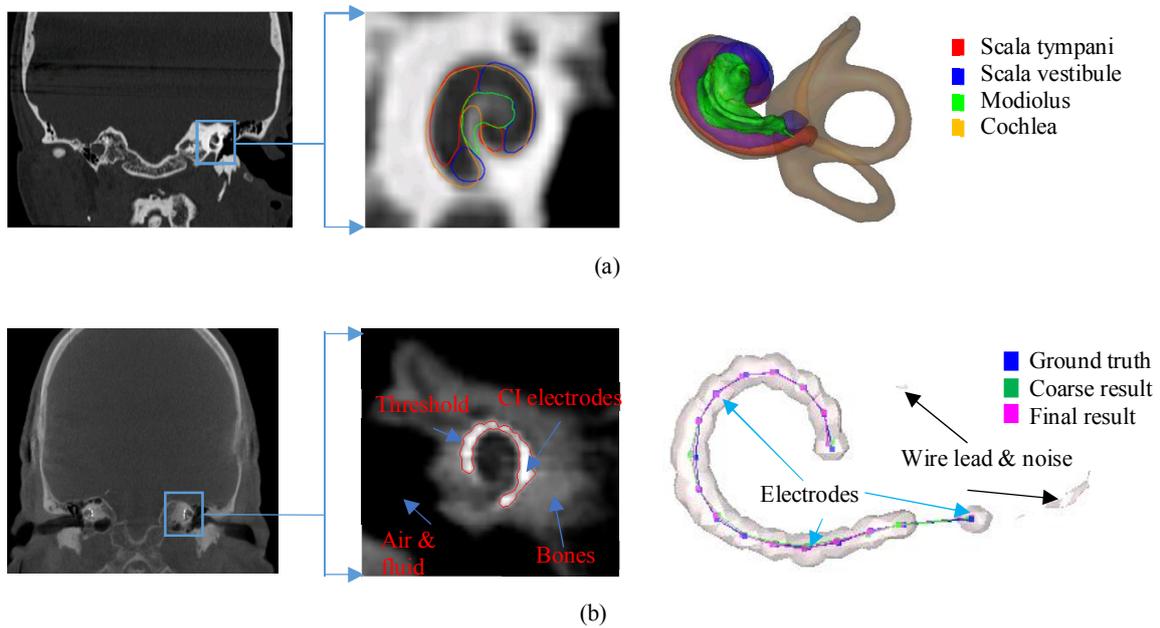

**Figure 1.3.** Examples of CT images in the coronal view. Panel (a) shows the cropped volume of interest (VOI) containing the cochlea and the segmented intra-cochlear anatomy. Panel (b) shows the VOI with the automatically localized and manually localized CI electrode array.

## 1.2 Image-guided cochlear implant programming

With the goal of providing patient-specific electrode configurations for CI recipients to improve their hearing outcomes, a process referred to as image-guided cochlear implant programming (IGCIP) [17] has been developed. Figure 1.4 visualizes the workflow for this IGCIP process. It relies on a series of image processing techniques and consists of two main stages: the pre-operative stage and the post-operative stage.

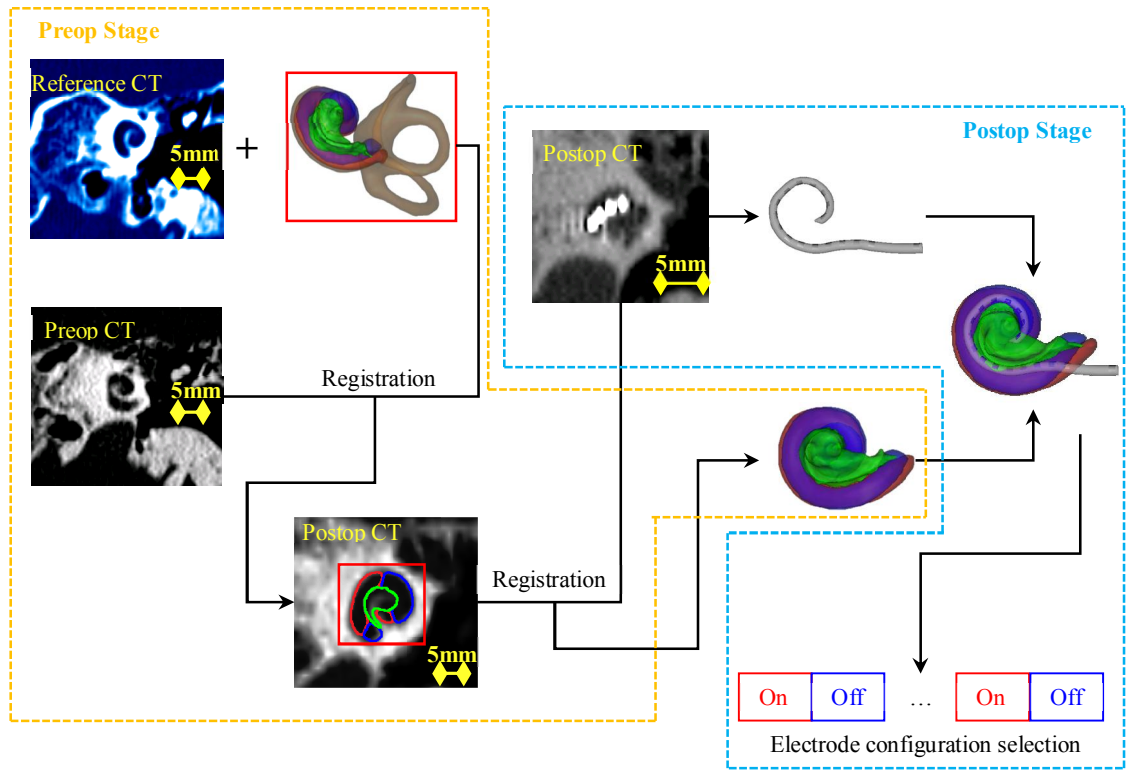

**Figure 1.4.** Workflow of IGCIP

In the pre-operative stage of IGCIP, the intra-cochlear anatomy, i.e., the modiolus (MOD), the scala tympani (ST) and scala vestibuli (SV) are segmented using pre-implantation CTs [18-22]. For patients who do not have pre-implantation CTs, the intra-cochlear anatomy is segmented using only the post-implantation CTs [21].

In the post-operative stage, an expert manually localizes the positions of the electrodes in the post-implantation CTs. Then the post-implantation CT images, where the locations of the CI electrodes are identified, are registered to the pre-implantation CTs, where the intra-cochlear anatomical structures are segmented, to find the electrode array position relative to the auditory nerves. This permits to analyze electrode interaction patterns. Lastly, an experienced CI programmer is asked to select an electrode deactivation plan based on the analysis result. Studies have shown that when the set of active electrodes is selected to reduce competing stimulations, hearing outcomes are improved and these improvements are statistically significant [23-25]. Although substantial progress has been made toward automating IGCIP [18-22, 26-30], several steps still require manual intervention, especially in the post-operative stage.

In the remainder of this chapter, we present brief reviews on the methods that are currently used for IGCIP, we identify their limitations, and we introduce the contributions of this dissertation to the full automation of the programming process. IGCIP involves three main phases that we will discussed: (1) intra-cochlear anatomy segmentation, (2) implanted CI electrodes localization, and (3) automatic electrode configuration selection.

1.2.1. Intra-cochlear anatomy segmentation in CT

Segmenting intra-cochlear anatomy in clinical pre-implantation CTs is difficult because the membrane that separates the two major cavities, i.e, the ST and SV, in the cochlea cannot be seen in conventional CTs. To solve this problem, an active shape model-based method has been developed [18]. In this method, models are created with µCT scans of the cochlea in which intra-cochlear structures are visible. The model is then fitted to the regions that are visible in the conventional CTs. It is subsequently used to estimate the position of the anatomical structures that are not visible in the CT scans. This method thus makes it possible to segment automatically the intra-cochlear anatomy in pre-implantation CTs, which is crucial for the following steps in IGCIP. Among all the intra-cochlear anatomy segmentation methods used for IGCIP, the method described in [18] is the only one that had been validated with µCTs prior to this work. It is also the most accurate intra-cochlear anatomy segmentation method developed at our institution. It has been used to evaluate three other methods detailed in [19], [21], and [22] that have been developed to segment the intra-cochlear anatomy when pre-operative images are not available.

For CI recipients who do not have a pre-implantation CT, the method [18] introduced above cannot be directly applied. This is because in post-implantation CTs, image artifacts introduced by the electrode array obscure the intra-cochlear anatomy. To solve this problem, techniques that permit segmenting intra-cochlear anatomy with only post-implantation CTs have neem developed. The method described in [19] is applied to post-implantation CTs of unilateral CI recipients. This method firstly segments the labyrinth of the normal contralateral ear. Then, it uses the position of the labyrinth and leverages the intra-subject inter-ear symmetry to segment the intra-cochlear anatomy of the implanted ear. However, for bilateral CI recipients or CI recipients who only have CTs of the implanted ear, i.e., the contralateral ear is not visible in the image, this method cannot be applied. The method described in [21] addresses this problem. It relies on the observation that parts of the inner ear that are not typically affected by the image artifacts can be used to infer the locations of the intra-cochlear anatomical structures that are affected. It firstly localizes the former parts. Then, it uses a library of segmented cochlear labyrinth shape to build an active shape model. With this active shape model, the labyrinth of the cochlea in the post-implantation CT is segmented. Then, another pre-defined active shape model of the ST, SV and MOD is used to segment those structures of interest (SOIs). Recently, we have also explored the possibility to use the method developed for pre-operative images directly on post-operative images processed to reduce the electrode-induced artifacts. This approach relies on deep learning techniques to synthesize from a post-operative image a corresponding image in which the artifact is eliminated.

1.2.2. Cochlear implant electrode array segmentation in CT

Localizing CI electrodes automatically in post-implantation CTs is also a challenging problem. The first challenge is that the image quality of the CTs that are acquired clinically is limited for our needs. First, the resolution of typical CT images is coarse (the voxel size in clinical scans is typically 0.2 x 0.2 x 0.3 mm$^3$) compared to the typical size of the CI electrodes which is on the order of 0.3 x 0.3 x 0.1 mm$^3$. Due to the partial volume effect, it is difficult to localize small-sized CI electrode array in clinical CTs. The images resolution is also coarse relative to the spacing between electrodes. This makes it difficult to separate the individual electrodes from the array, as shown in Figure 1.5. Second, because the electrodes are composed of radiodense platinum, beam hardening artifacts distort the intensities in the region around the electrode array, resulting in erroneous intensities assigned to voxels around the electrodes during reconstruction. This complicates the identification of individual electrodes in CTs. Third, even though the CI electrodes usually appear as high intensity voxel groups in CTs, it is difficult to select a threshold such that the thresholded image only contains voxels occupied by CI electrodes. This is because voxels occupied by wire lead, receiver coils, and cortical bones are usually assigned high intensity values too. A fourth challenge is the fact that the CT images are reconstructed with different algorithms. In an image reconstructed with an "extended" Hounsfield Unit (HU) range (eCT), the metallic structures are assigned higher intensity values than the cortical bones. In an image reconstructed with a "limited" HU range (lCT), the maximum intensity is limited to the intensity of cortical bones. Thus, in an eCT, the false positive voxels are usually occupied by the metallic wire lead as shown in Figure 1.5a. In a lCT, there are many more false positive voxels as shown in Figure 1.5c. The last challenge is the fact that several models of electrode arrays are manufactured and used. These have different specifications, e.g., number of contacts, size of contacts, or spacing between contacts. As a results they appearance in CT images can be substantially different. The most common electrode arrays are manufactured by the three leading manufacturers, i.e., Med-El® (MD) (Innsbruck, Austria), Advanced Bionics® (AB) (Valencia, California, USA), and Cochlear® (CO) (Sydney, New South Wales, Australia). Table 1.1 shows the specifications of the commonly used models of CI arrays. Figure 1.6 illustrates the geometric models of typical CI electrode arrays produced by the three manufacturers. Based on their inter-electrode spacing, we classify CI electrode arrays into two broad categories: closely-spaced and distantly-spaced arrays. Closely-spaced arrays are such that individual electrodes cannot be resolved in the images and the set of electrodes thus usually form a single connected region as shown in Figure 1.5b. When localizing a closely-spaced electrode array in a post-implantation CT, there is usually not enough intensity contrast to separate the individual electrodes. To estimate the locations of closely-spaced electrodes in CT images, human experts need first to manually delineate the centerline of the ROI that contains all the electrodes. Then, they use their experience and visual clues to determine the locations of the basal and apical electrodes on the centerline. Last, they fit a 3D model of the implanted array to the centerline to estimate the locations of individual electrodes.

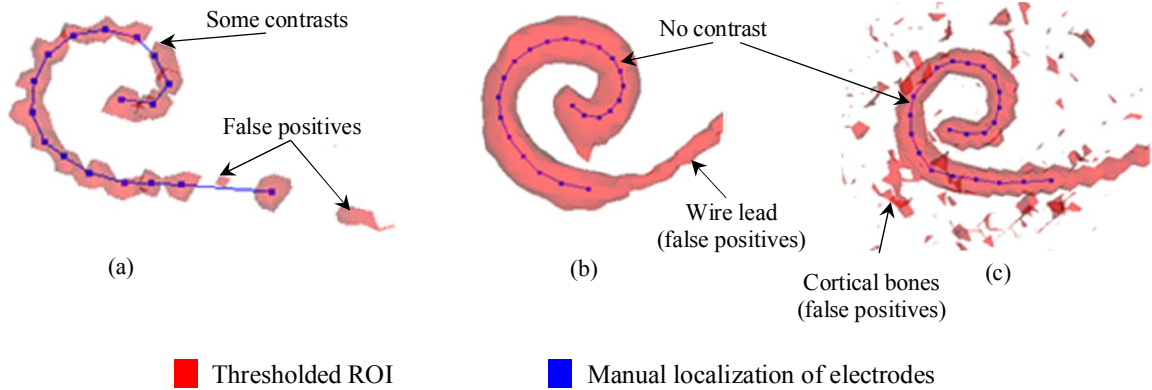

Figure 1.5. Panels (a) and (b) show examples of distantly and closely-spaced arrays in eCTs. Panel (c) shows an example of a closely-spaced array in a lCT.

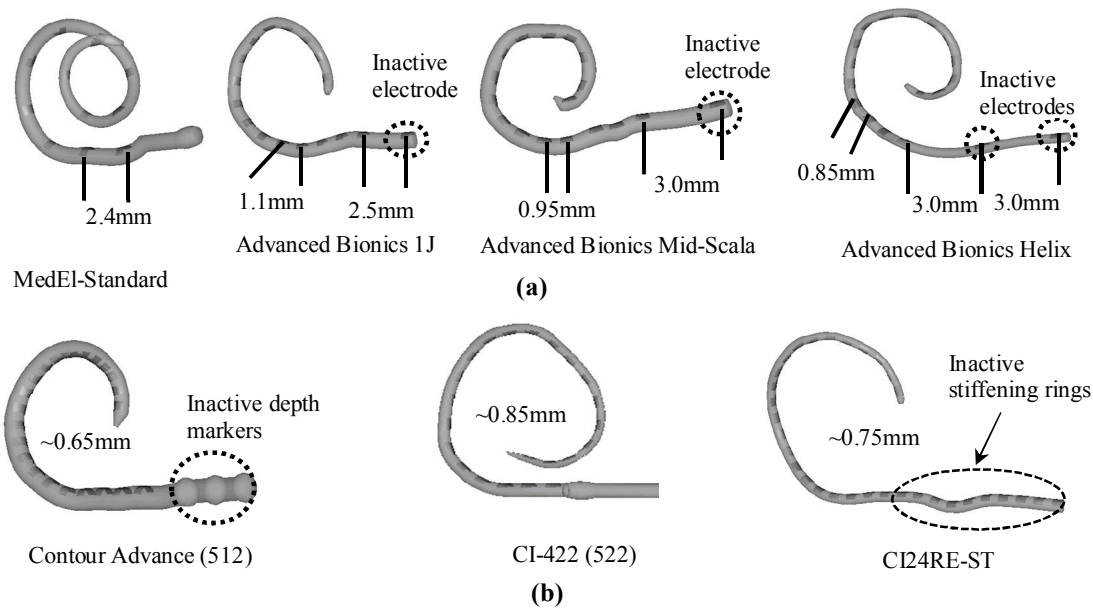

**Figure 1.6.** Seven major types of CI electrode arrays provided by the three major manufacturers. Panel (a) presents four typical examples of distantly-spaced electrode arrays and panel (b) presents three typical examples of closely-spaced electrode arrays.

**Table 1.1** Specifications of different FDA-approved CI electrode arrays

| Type | Electrode array brand | Total electrodes | Electrode spacing distance (mm) |
|---|---|---|---|
| Distantly-spaced | Med-El standard (MD1) | 12 | 2.4 |
| | Med-El Flex28 (MD2) | 12 | 2.1 |
| | Advanced Bionics 1J (AB1) | 17 (1 inactive electrode) | 1.1 and 2.5 |
| | Advanced Bionics Mid-Scala (AB2) | 17 (1 inactive electrode) | 0.95 and 3.0 |
| | Advanced Bionics Helix (AB3) | 18 (2 inactive electrodes) | 0.85 and 3.0 |
| Closely-spaced | Contour Advance (512) (CO1) | 22 | ~0.65 |
| | CI-422 (522) (CO2) | 22 | ~0.90 |
| | CI24RE-Straight (CO3) | 32 (10 stiffening rings) | ~0.75 |

This manual localization method requires time and experience and is prone to error. As can be seen from Figure 1.5b, the intensity contrast may not be obvious around the most basal electrode. This leads to a mis-localization of the basal electrode (e.g., a point on the wire lead) and this error is propagated to the whole array when fitting the 3D model. When localizing a distantly-spaced electrode array in a post-implantation CT, experts manually select a threshold to separate contacts from the rest of the images. Next they need to manually select the center of the each contact to which the 3D model is fitted. This is also a time-consuming process that requires expertise.

Two preliminary methods designed for localizing distantly- and closely-spaced electrode arrays in post-implantation CTs have been described in [28] and [26], respectively. The method described in [28] relies on two graph-based path finding algorithms. Given a post-implantation CT of a CI recipient implanted with a distantly-spaced array with $N$ electrodes, this method first generates the volume of interest (VOI) that contains the cochlea by using a reference image. Then, it thresholds the VOI to generate regions of interest (ROIs) that are regions that potentially contain individual contacts. By applying a voxel thinning method [32] to the ROIs, it generates a set of candidates of interest (COIs) that represent the possible locations of the electrodes. The COIs are treated as nodes in a graph for the following two path-finding algorithms. By using two path-finding algorithms, it finds a fixed-length path connecting $N$ COIs together as the localization result. But, when applying this method to a large-scale dataset of clinical CTs, we found it to lack robustness. As part of this dissertation, we have proposed several improvements that have substantially increased this earlier method.

The method described in [26] is a snake-based method driven by the Gradient Vector Flow (GVF) [33-34] designed to localize contacts in closely-spaced arrays. It is based on the assumption that the centerline of the electrode corresponds to the medial axis of the artifact region because, as can be seen from Figure 1.5b, the metallic artifact is much brighter than the surrounding anatomy. Given a post-implantation CT of a CI recipient implanted with a closely-spaced array, this method localizes the centerline of the electrode array and then fits a 3D model of the implanted array to the extracted centerline to localize individual contacts. To initialize the centerline initialization algorithm, a curve representing the typical locations for a cochlear implant is defined in a reference image. This curve is then projected from the reference image to the target post-implantation image using non-rigid registration. The initialized curve is updated with a snake-based method that uses GVF as the external force. Again, when applying this method to large data sets, we found the centerline initialization to be too coarse because the electrode array can be inserted much deeper or shallower than the manually defined curve in the reference image. This results in large errors in the initialization step that are propagated to the following steps. We also found that the GVF was not always capable of driving the initialized curve to the centerline of the implanted array. As part of this dissertation, we have developed and evaluated contact localization methods for closely-space electrode arrays that substantially outperform earlier ones.

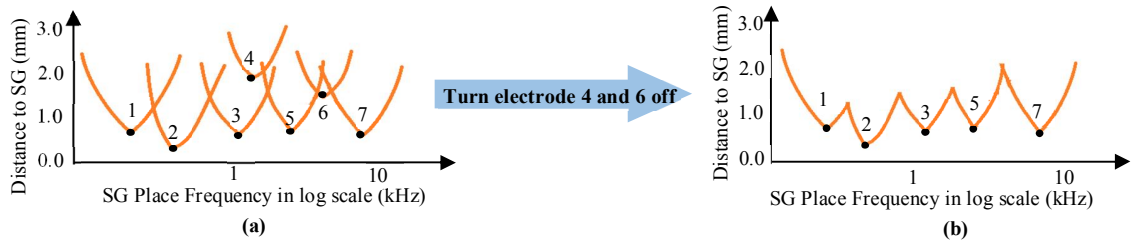

**Figure 1.7.** Visualization of DVF curves. (a) shows an example of a combination of the DVF curves formed by 7 electrodes. Each single curve represents the distance from the corresponding electrode to the frequency mapped sites along the length of the modiolus. (b) shows the DVF curves after electrode configuration adjustment.

1.2.3. Automatic electrode configuration selection for IGCIP

As mentioned in Section 1.2, knowledge of the spatial relationship between the electrodes and the SG nerves is crucial for the CI programmer to be able to select the subset of active electrodes, i.e., the electrode configuration. In order to permit the analysis this spatial relationship a visualization method called electrode distance-vs.-frequency (DVF) curves [17] has been developed. The DVF curves are 2D plots which capture the patient-specific 3D spatial relationship between the individual electrode and the SG nerve as is shown in Figure 1b. Figure 1.7 shows an example of DVF curves for 7 electrodes. The horizontal axis represents the positions along the length of the modiolus in terms of the characteristic frequencies of the SG nerves. A number is assigned to each DVF curve to represent the corresponding electrode. The height of each DVF curve on the vertical axis indicates the distance from the corresponding electrode to the frequency-mapped modiolar surface. Each DVF curve is constructed by finding the distance to the corresponding electrode from the frequency-mapped neural activation sites on the modiolar surface. As can be seen from Figure 1.7a, electrode 3 is approximately 1mm from the modiolar surface around the 1kHz characteristic frequency. The current assumption in IGCIP is that if one electrode's DVF curve is not the closest DVF curve in the region around its minimum, it is likely that it is interfering with another electrode. As shown in Figure 1.7a, the minimum of the DVF curve of electrode 4 falls above the DVF curve of electrode 3, which indicates that electrode 4 is likely to stimulate the same neural region as electrode 3. Furthermore, even if the minimum of the DVF curve of electrode 6 falls below the other DVF curves, its depth of concavity relative to the minimum envelope of the neighboring DVF curves is small. This also indicates a high possibility for electrode 6 to interfere with electrode 5 and 7. The strategy used by the expert when selecting an electrode configuration is to keep active as many electrodes as possible that are not likely to cause stimulation overlap. Thus, in this case, electrodes 4 and 6 would be deactivated, as shown in Figure 1.7b.

When manually selecting the electrode configuration, the expert makes his/her decision using a set of heuristics based on a series of DVF-based features. Automating the electrode configuration process is challenging because algorithms have to be developed to compute the DVF-based features and because the relative importance of these features in the expert's decision process need to be estimated. One contribution of this dissertation is an automatic method capable of producing deactivation plans.

## 1.3 Sensitivity of IGCIP

As discussed in Section 1.2, IGCIP relies on an intra-cochlear anatomy segmentation method and electrode localization techniques to analyze the patient-specific spatial relationship between the implanted CI electrodes and the auditory nerves. This permits to provide patient-customized electrode deactivation configurations. The accuracy of each method could affect the shape of the DVF curves and the generation of the deactivation configurations. Among the intra-cochlear anatomy segmentation methods, only the method in [18] has been validated with μCTs, from which an anatomical ground truth can be created. The methods in [19], [21], and [22] were validated by comparing them to the method in [18]. Electrode localization methods were validated only by comparing automatic and manual localization in clinical post-operative CT scans. Manual localization is an imperfect ground truth because: (1) as discussed above the clinical CTs have a coarse resolution (typical voxel size 0.2 x 0.2 x 0.3mm$^3$) compared to the electrode sizes (typical size 0.2 x 0.2 x 0.1mm$^3$). When localizing small-sized objects in clinical CTs, partial volume artifacts make it difficult to identify the center of the electrodes; (2) beam hardening artifacts in clinical CTs also make it difficult to localize the centers of the electrodes because the voxels around those positions are also assigned high intensity. Thus, to better characterize the performance of IGCIP and to fully understand the limitations of IGCIP, a thorough validation study needs to be completed with a better ground truth dataset. This has been accomplished as part of this dissertation.

## 1.4 Goals and Contributions of the Dissertation

The goals of this dissertation are to fully automate the image processing techniques needed in the post-operative stage of IGCIP and to perform a thorough analysis of (a) the robustness of the automatic image processing techniques used in IGCIP and (b) assess the sensitivity of the IGCIP process as a whole to individual components. The automatic methods that have been developed include the automatic localization of both closely- and distantly-spaced CI electrode arrays in post-implantation CTs and the automatic selection of electrode configurations based on the stimulation patterns. Together with the existing automatic techniques developed for IGCIP, the proposed automatic methods enable an end-to-end IGCIP process that takes pre- and post-implantation CT images as input and produces a patient-customized electrode configuration as output.

The specific contributions of this dissertation are summarized below:

**Chapter II** presents a snake-based automatic method which aims at extracting the centerline of the implanted array in CTs. It is an improvement on the method presented in [26] designed for localizing closely-spaced array in post-implantation CTs. This method is validated on 15 eCTs of CI recipients implanted with CO1 arrays.

**Chapter III** presents an automatic graph-based method for localizing distantly-spaced CI electrode arrays in clinical CT with sub-voxel accuracy. This method is an extension of the method described in [28] and is validated on a large scale dataset of clinical CTs of CI recipients implanted with various types of distantly-spaced arrays. Its robustness with respect to several acquisition parameters (the HU range, resolution, dose, and type of the implanted arrays) is further validated on a set of phantom CTs acquired with different acquisition parameters. The method is the state of art technique for localizing distantly-spaced electrode arrays in clinical CTs.

**Chapter IV** proposes an automatic centerline-based method for localizing closely-spaced electrode arrays in clinical CTs. This method is an extension of the method described in Chapter II and is a more generic method for closely-spaced array localization.

It is validated on a large scale dataset of clinical CTs of CI recipients implanted with CO1, CO2 and CO3 arrays. This method is the state of art technique for localizing closely-spaced electrode arrays in clinical CTs.

**Chapter V** presents an automatic electrode configuration selection method based on the spatial relationship generated by the anatomy segmentation and electrode localization procedures used in IGCIP. This method is trained on 36 subjects implanted with electrode arrays produced by the three major manufacturers. It is further validated on a dataset of 60 subjects and the validation study results are evaluated by two experts in electrode deactivation configuration. This is the first automatic method that is capable of emulating human experts for selecting electrode configurations.

**Chapter VI** creates a gold-standard ground truth dataset for electrode localization and intra-cochlear anatomy segmentation that relies on pre- and post-implantation μCTs of 35 temporal bone specimens. The gold-standard ground truth dataset is used to rigorously evaluate the accuracy of the intra-cochlear anatomy segmentation methods [18], [21], and [22], and the accuracy of the electrode localization method described in Chapter III. The method described in [19] is not evaluated with this gold standard because the specimens do not have a normal contralateral ear. The methods described in Chapter II and IV are not evaluated because the electrode arrays implanted in the specimens are only distantly-spaced arrays. We also use the gold-standard ground truth dataset to evaluate the sensitivity of the IGCIP process as a whole to each step. This is the first thorough sensitivity analysis of IGCIP with respect to errors introduced by automatic image processing techniques. The dataset and the framework used in in this study can be extended to other validation studies.

**Chapter VII** provides the summary of the work and discusses possible future work.

Chapter II

# AUTOMATIC LOCALIZATION OF COCHLEAR IMPLANT ELECTRODES IN CT

Yiyuan Zhao[1], Benoit M. Dawant[1], Robert F. Labadie[2], and Jack H. Noble[1]

[1]Department of Electrical Engineering and Computer Science, Vanderbilt University, Nashville, TN, 37232, USA

[2]Department of Otolaryngology – Head & Neck Surgery, Vanderbilt University, Nashville, TN, 37232, USA

Abstract

Cochlear Implants (CI) are surgically implanted neural prosthetic devices used to treat severe-to-profound hearing loss. Recent studies have suggested that hearing outcomes with CIs are correlated with the location where individual electrodes in the implanted electrode array are placed, but techniques proposed for determining electrode location have been too coarse and labor intensive to permit detailed analysis on large numbers of datasets. In this paper, we present a fully automatic snake-based method for accurately localizing CI electrodes in clinical post-implantation CTs. Our results show that average electrode localization errors with the method are 0.21 millimeters. These results indicate that our method could be used in future large scale studies to analyze the relationship between electrode position and hearing outcome, which potentially could lead to technological advances that improve hearing outcomes with CIs.

## 2.1. Introduction

Cochlear Implants (CI) are surgically implanted neural prosthetic devices used to treat severe-to-profound hearing loss. In CI surgery, an electrode array is threaded into the cochlea. After surgery, a processor worn behind the ear sends signals to the implanted electrodes, which activate auditory nerve pathways inducing the sensation of hearing. Although CIs have been remarkably successful, a significant number of CI recipients experience marginal hearing restoration. Recent research has suggested that hearing outcomes with CIs are correlated with the location where the electrodes are placed [1-5]. However, without post-implantation imaging, the position of the electrodes is generally unknown since the array is blindly threaded into a small opening of the cochlea during surgery, with its insertion path guided only by the walls of the spiral-shaped intra-cochlear cavities.

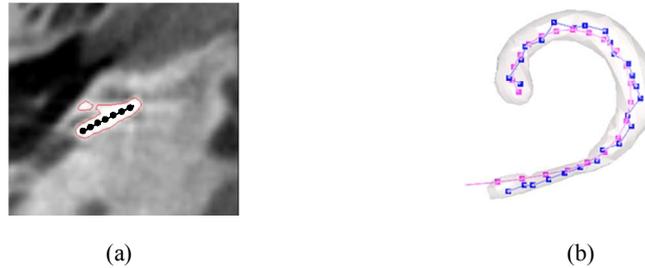

(a)                  (b)

**Figure 2.1.** Panel (a) shows a portion of an electrode array in an axial slice of a CT. Black dots indicate locations of individual electrodes. An isocontour around high intensity voxels is shown in red. Panel (b) shows a 3D isosurface of an electrode array with a manually determined centerline in purple. The blue curve is the coarse approximation to the centerline determined using our automatic initialization process discussed in Section 2.2.2.

In efforts to analyze the relationship between electrode location and outcome, several groups have proposed coarse electrode position measurements that can be visually assessed in CT images, e.g., whether all electrodes are within one of the two principal intra-cochlear cavities, depth of insertion of the first and last electrode, etc. [1-5]. Studies using these techniques have indicated that placement and outcome are indeed correlated, but it has not been possible to determine specific factors that affect outcome because dataset size was limited and because the electrode positions were never precisely quantified with these techniques. One factor that has limited the size of the datasets in the studies is the amount of manual effort that must be undertaken to analyze the images. Our group has shown that knowledge of electrode location can be used to select better CI processor settings to significantly improve hearing out-comes compared to standard clinical results [6]. In the current work, we propose a fully automatic approach for localizing CI electrodes in CT images. An electrode localization approach that is automatic and accurate would be significant as it could facilitate precise quantification of electrode position on large numbers of datasets to better analyze the relationship between electrode position and outcome, which may lead to advances in implant design or surgical techniques. It could also automate the electrode localization process in systems designed to determine patient-customized CI settings such as the one proposed in [6], reducing the technical expertise required to use such technologies and facilitating transition to large scale clinical use.

Figure 2.1 shows an example of an electrode array in a CT slice. Localizing the electrodes in CT images is difficult because (a), as seen in the figure, the beam hardening artifacts caused by the metallic electrodes distort intensities in the region around the electrode array, leading to incorrect assignment of very high intensities during image reconstruction to nearby voxels that are not occupied by metal, thus making it difficult to segment electrodes via thresholding; and (b) the individual electrodes are so close that there is no contrast between them in standard CT images, even when acquired at very fine slice thickness and resolution. Our solution is to identify the centerline of the voxels occupied by the CI electrodes using a snake-based localization approach [7] and then to fit a 3D model of the electrode array to the extracted centerline. This is a similar approach to that which we proposed in [8]. However, the technique we presented in that paper leads to inaccurate results around the first and last electrodes due to curve shrinkage. This shrinking phenomenon is caused by the use of an intensity-based attraction function since the image intensity decreases mildly at the array endpoints relative to the rest of the array. Further, we found that the "forward energy," an external energy term designed to counteract endpoint shrinking errors by expanding the curve, became unstable and led to failures when applying the technique on clinical image datasets. As will be described in the following section, in this work, we propose a new technique to counteract the shrinking effect by localizing and fixing the endpoints prior to snake optimization. Our results, presented and discussed in Sections 2.3 and 2.4, will show that this fully automatic

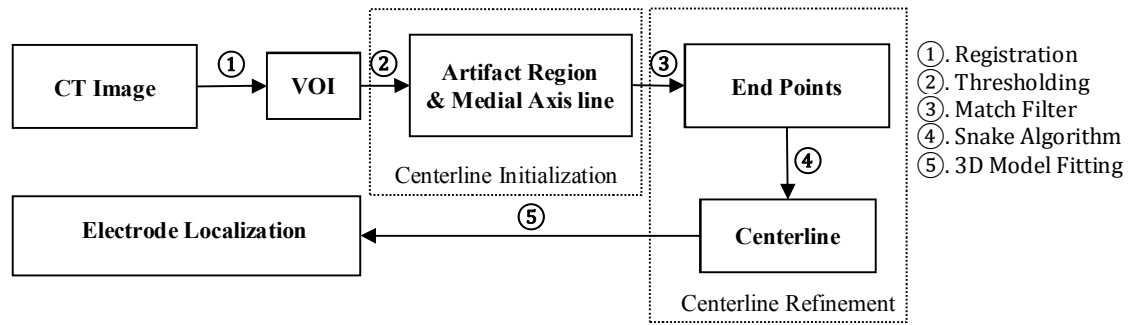

**Figure 2.2.** Flow chart of the electrode array centerline localization process

approach can reliably be applied to clinical images.

## 2.2. Methods

The automatic segmentation method we propose is outlined in Figure 2.2. As can be seen in the figure, the first step (1) involves coarsely estimating the location of the region of interest (ROI), which is a local region ~1 cm$^3$ around the cochlea. This is done through registration with a known volume. The subsequent processing steps are then performed solely within the ROI. The next step (2) is to initialize our electrode array centerline localization. This is done by segmenting via thresholding the region of the image that contains the metallic electrodes and then computing the initialized centerline as the medial axis of the result. The thresholding step will produce a segmentation that includes electrode voxels as well as those that appear bright due to partial volume or beam hardening artifacts, but the medial axis extraction step is able to reliably and coarsely approximate the centerline of the electrode array. After initialization, the next steps (3-4) are to refine the centerline using a snake-based optimization approach [7]. In the third step, the curve endpoints are first localized within the neighborhood of their initialized positions using an endpoint detection filter we have designed. In the fourth step, the endpoints are fixed and the points in the rest of the curve are optimized. This is done using a snake with its external energy defined using the output of a vesselness filter that is applied to the original image to enhance the centerline of the electrode array [9]. By detecting and fixing the endpoints prior to snake optimization, curve shrinking effects discussed in the previous section are eliminated. The final step (5) is a straightforward resampling of the extracted centerline to determine individual electrode locations using *a-priori* knowledge about the distance between neighboring electrodes. The following subsections detail this approach.

2.2.1. Data

The images in our dataset include images from 15 subjects acquired with a Xoran xCAT®. The images have voxel size 0.4 x 0.4 x 0.4 $mm^3$. As a pre-processing step, an VOI bounding the region around the electrode array in each target image is automatically localized by using a mutual information-based affine registration computed between the target image and a known reference image [10]. The ROI is then automatically cropped from the original target image and all subsequent steps are performed on the cropped image. Each cropped image includes approximately $30 \times 30 \times 30$ $mm^3$. Each subject in this study was implanted with a Cochlear™ Contour Advance®. Thus, the methods presented are focused on segmenting this type of electrode array but could prove in future studies to be applicable to other implant models.

2.2.2. Centerline Initialization

The centerline is initialized by thresholding the region of the image that includes the electrode array and computing the medial axis of the result. We determine the threshold dynamically using a maximum likelihood estimation-based (MLE) threshold selection approach [11] since the best threshold can vary across subjects due to the relatively low signal-to-noise ratio (SNR) achieved using the low-dose acquisition protocols on a flat panel scanner. We would also expect that a dynamic threshold would account for differences between scanners, but this was not tested in this study. The MLE approach we have designed is to fit a model, defined as the sum of two Gaussian distributions, to the VOI image histogram and compute a threshold based on this result. One distribution $G(\mu_1, \sigma_1)$ corresponds to soft tissue and another $G(\mu_2, \sigma_2)$ corresponds to bony tissue. While air and metal are present in the VOI image, their relatively small volumes contribute little to the shape of the histogram, and thus these intensity classes are ignored in the histogram fitting. Once the distributions are estimated, the threshold is selected based on the upper tail of the Gaussian that models the intensity distribution of bone to be $\mu_2 + 5\sigma_2$, which was empirically determined to lead to good results. We chose to use this MLE-based approach, rather than a simpler percentile-based approach, because this approach is not sensitive to differences in VOI volume or differences in volume of metal present in the VOI, which can vary across subjects. After a threshold is determined, the medial axis of the resulting thresholded volume is computed using the medial axis extraction techniques presented by *Bouix et al.* [12]. The resulting curve provides a close but coarse approximation to the centerline of the electrode array. An example result of this process is shown in blue in Figure 2.1b.

2.2.3. Centerline Refinement

After the curve is initialized, we refine its position using a snake-based algorithm. The traditional snake algorithm localizes a contour by minimizing the energy equation:

$$E = \int_0^1 \rho_1 \|x'(s)\|^2 + \rho_2 \|x''(s)\|^2 + E_{ext}(x(s)) \, ds, \qquad (2.1)$$

where $x(s)$ is the position of the parameterized curve at $s$, $\rho_1$ and $\rho_2$ are the tension and rigidity weighting terms, and $E_{ext}$ is the external energy term. In our experiments, we set

$\rho_1 = 0.03$ and $\rho_2 = 0.08$ as these values were empirically determined to lead to good results, and we define $E_{ext}$ to be the output of a vesselness response filter applied to the ROI image [9]. We apply the filter at scales $\sigma = \{0.08, 0.16, \ldots, 0.8\}$ mm and set the other internal parameters to be $\alpha = 0.5$, $\beta = 0.5$, and $\gamma = 500$. Vesselness response, rather than, for example, a direct function of image intensity is used as an external energy because the high intensity voxels in the region around the electrode array can be noisy, and voxels with intensity that is locally maximal often do not fall on the centerline of the homogeneous bright region in the image (see Figure 2.1). Since the electrode array has the appearance of a tubular structure, a vesselness response filter is a natural choice to enhance the centerline of the electrode array.

The robustness of the vesselness filter in detecting the centerline of the electrode array is high along the length of the array but diminishes at the endpoints. Thus, with no additional information, optimizing the snake would result in a shrinking of the curve at the endpoints. To address this, we determine the endpoint positions using an endpoint detection filter and fix them during the snake optimization. The endpoint detection filter we have constructed, $M_{\hat{v}}(\boldsymbol{\omega})$, is a match filter. For the sake of simplicity, we define $M_{\hat{v}}(\boldsymbol{\omega})$ such that $\boldsymbol{\omega} = \mathbf{0}$ lies at the center of the filter (see Figure 2.3a). We also orient the filter using $\hat{v}$, which represents the orientation of the centerline of the electrode array at the endpoint. To define $M_{\hat{v}}(\boldsymbol{\omega})$, we first define $M'_{\hat{v}}(\boldsymbol{\omega})$ as

$$M'_{\hat{v}}(\boldsymbol{\omega}) = \begin{cases} r^2 - \|\boldsymbol{\omega}\|^2 & \boldsymbol{\omega} \cdot \hat{v} \geq 0 \\ r^2 - \|\boldsymbol{\omega} - (\boldsymbol{\omega} \cdot \hat{v})\hat{v}\|^2 & \boldsymbol{\omega} \cdot \hat{v} < 0 \end{cases}, \quad (2.2)$$

This equation defines $M'_{\hat{v}}(\boldsymbol{\omega})$ such that when $\cdot \hat{v} \geq 0$, i.e., in the $\hat{v}$ direction from the origin as seen in Figure 3a, $M'_{\hat{v}}(\boldsymbol{\omega})$ matches a semispherical structure, whereas in the opposite direction where $\boldsymbol{\omega} \cdot \hat{v} < 0$, the filter matches a tubular structure. The radius, $r$, of the sphere and tube are set to be 0.3 mm, which is approximately the radius of the electrode arrays in our images. The final form of the filter is defined as

$$M_{\hat{v}}(\omega) = M'_{\hat{v}}(\omega)\left(\rho_3 H(M'_{\hat{v}}(\omega)) + (1-\rho_3)H(-M'_{\hat{v}}(\omega))\right),$$ where $H(\cdot)$ is the Heaviside function and $\rho_3 = 0.97$ is a parameter we chose empirically to optimize results and tunes the weighting between the fore- and background regions of the filter.

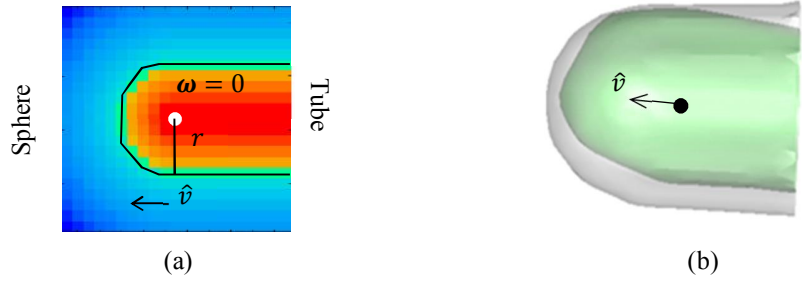

(a)                                              (b)

**Figure 2.3.** (a) shows a slice of $M'_{\hat{v}}(\omega)$ with $M'_{\hat{v}}(\omega) = 0$ isocontour in black and $\omega = 0$ shown as white dot. (b) shows the 3D isosurface of $M'_{\hat{v}}(\omega)$ (white) aligned with the tip of an electrode array (green).

To find each endpoint using this filter, we set $\hat{v}$ to be the orientation of the central axis of the electrode array as estimated by the vesselness response at $x_e^i$, the location that the endpoint was initialized using the methods described in Section 2.2, and then compute the endpoint location $x_e$ as:

$$x_e = \mathrm{argmax}_{x \in N(x_e^i)} \sum_{y \in L(x)} I(y)\, M_{\hat{v}}(y - x)\,, \tag{2.3}$$

$N(x_e^i)$ is a neighborhood function that we define as the set of 16 x 16 x 16 points uniformly sampled in a 1.2 x 1.2 x 1.2 mm³ box surrounding $x_e^i$, $I$ is the ROI image, and $L(x)$ is a neighborhood function defined as the set of 21 x 21 x 21 points uniformly sampled in a 1.2 x 1.2 x 1.2 mm³ box oriented in the $\hat{v}$ direction surrounding $x$. In summary, Eqn. (2.3) selects the endpoint as the point in a local region around the initial endpoint that maximizes the response of the endpoint enhancement filter, and the filter response should be maximized when it is aligned with and centered on the tip of the electrode array.

After the endpoints are determined, they are fixed and the positions of the remaining points in our curve are optimized by iterating the standard snake update equations [7] until convergence or until reaching 100 iterations. Once the final curve is localized it is straightforward to resample the curve to identify the location of individual electrodes based on *a priori* knowledge of the distance between electrodes in the array.

2.2.4. Validation

We quantified the accuracy of our automatic electrode array extraction technique in a dataset of fifteen head CT images by comparing centerlines computed automatically using the proposed technique (PT) to ground truth (GT) curves, which were created by averaging of three sets of curves independently defined by an expert. Metrics used to characterize distance between two curves include mean and max curve distance (mean and max of the distances computed from each point on curve 1 to the closest point on curve 2 and vice versa), mean and max electrode distance (distance between each electrode location in curve 1 to the corresponding electrode in curve 2 after determining electrode locations along the curves as described in Section 2.2.3), and distance between corresponding endpoints in curves 1 and 2. To show the benefit our matched filter provides, we also report quantitative errors that result from computing the curve when (a) endpoints are fixed at their initialization position without the matched filter update (NM) and (b) when the endpoints are not fixed but optimized with the snake method similarly to the rest of curve (NF).

To assess whether the PT produces acceptable results, we conducted a second study in which an expert was asked to select between the GT and PT endpoints, blind to their identity. We focused on the endpoints because, as our results will show, this is the area in which there are the largest discrepancies between GT and PT curves.

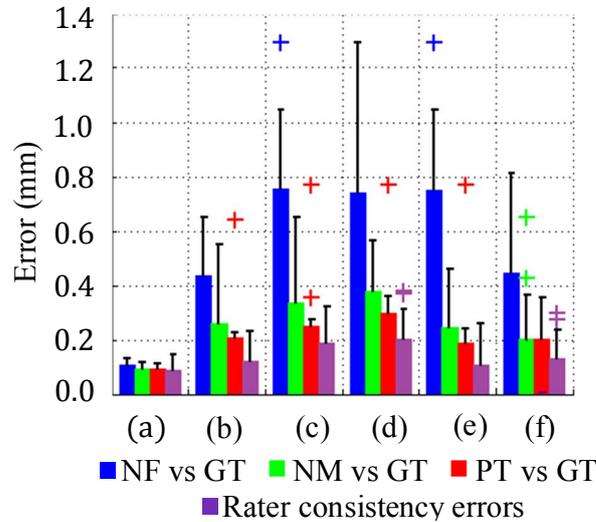

**Figure 2.4.** Barplots of mean (a) and max (c) curve distances; mean (b) and max (d) electrode distances; and tip (e) and base (f) endpoint distances.

## 2.3. Results

The quantitative comparisons between the GT and PT centerlines for all the datasets are shown in Figure 4 in red, and Figure 5 shows visualizations of two cases. In Figure 4, for each barplot, the height of the bars, crosses, and black whiskers denote the mean, outlier data, and maximum non-outlier value. Data are considered outliers if they fall above $q_3 + 1.5(q_3 - q_1)$, where $q_3$ and $q_1$ are the 25$^{th}$ and 75$^{th}$ percentiles of the dataset. As can be seen in the figure, our proposed method results in mean curve errors of 0.09 mm (0.13 of a voxel diagonal) and average maximum curve errors of 0.25 mm (0.36 of a voxel diagonal) with an overall maximum of 0.80 mm. Our method extracts a much more accurate centerline compared to prior work in which we achieved mean curve errors of 0.2 millimeters [8]. Further, the mean electrode localization error with our currently proposed method is only 0.21 mm. The utility of fixing the endpoints and optimizing them with our matched filter is also apparent in Figure 2.4 as NF and NM lead to much larger electrode and endpoint localization errors. This difference is not as pronounced in mean curve errors since curve distances along the length of the curve are not sensitive to errors at the endpoints. The mean tip and base endpoint errors with PT are 0.19 mm and 0.2 mm. These quantities are slightly higher for NM and substantially higher for NF. The outlier values for PT that fall above 0.6 mm all correspond to the case shown in Figure 2.5b, where the tip of the array was localized incorrectly due to lower than normal SNR in the image. We also show in purple in Figure 4 rater consistency errors computed among the three sets of curves manually delineated by an expert. We find mean and overall maximum consistency curve errors of 0.09 and 0.35 mm, suggesting that except for the outlier case, errors in our PT are close to the level of rater repeatability.

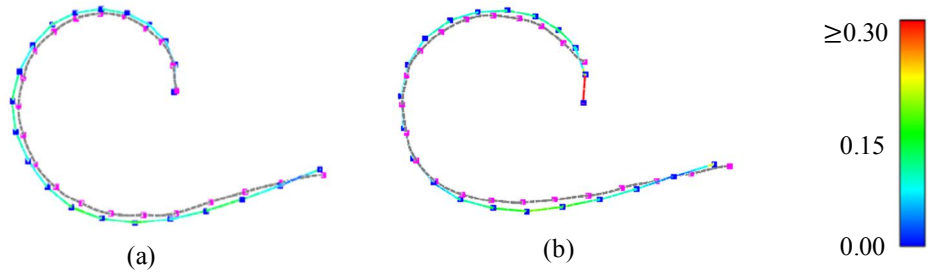

**Figure. 2.5.** 3D renderings of GT (color-coded with curve distance in mm) and PT (shown in transparent black) curves for our best (a) and worst (b) case errors. Points indicate electrode locations along curves determined by distance priors.

In the expert endpoint selection test, among the 30 endpoints in the 15 cases, 8 PT endpoints were judged to be equally accurate to GT, and 29 of 30 PT endpoints were judged to be acceptable. The lone exception was the tip endpoint shown in Figure 2.5b.

## 2.4. Conclusion

In this work, we have designed an automatic cochlear implant electrode array centerline extraction method. Our experiments show that our method is highly accurate, even when applied to clinical images. Compared to our prior method reported in [8], the method we propose here achieves results with errors that are half as large on average. This improvement is due in large part to the use of our matched filter, which leads to better endpoint localization. Our approach requires approximately 3 minutes of computation time on a standard PC.

Our method did result in unacceptably large errors for one of fifteen images. Future studies will involve developing techniques to detect and handle such errors. Additionally, we plan to test our method with images acquired with different scanners and of subjects with different implant models. We also plan to apply our method to large numbers of datasets to facilitate studying how the location of individual electrodes correlates with outcomes with the goal of developing technologies that can improve hearing outcomes with CIs.

Chapter III

# AUTOMATIC GRAPH-BASED METHOD FOR LOCALIZATION OF COCHLEAR IMPLANT ELECTRODE ARRAYS IN CLINICAL CT WITH SUB-VOXEL ACCURACY

Yiyuan Zhao, Srijata Chakravorti, Benoit M. Dawant, and Jack H. Noble

Department of Electrical Engineering and Computer Science, Vanderbilt

University, Nashville, TN, 37232, USA

Abstract

Cochlear implants (CIs) are neural prosthetics that provide a sense of sound to people who experience severe to profound hearing loss. Recent studies have demonstrated a correlation between hearing outcomes and intra-cochlear locations of CI electrodes. Our group has been conducting investigations on this correlation and has been developing an image-guided cochlear implant programming (IGCIP) system to program CI devices to improve hearing outcomes. One crucial step that has not been automated in IGCIP is the localization of CI electrodes in clinical CTs. Existing methods for CI electrode localization do not generalize well on large-scale dataset of clinical CTs implanted with different brands of CI arrays. In this paper, we propose a novel method for localizing different brands of CI electrodes in clinical CTs. Our method firstly generates the candidate electrode positions at sub-voxel resolution in a whole head CT. Then, we use a graph-based path-finding algorithm to find a fixed-length path that consists of a subset of the candidates as the localization result. Validation on a large-scale dataset of clinical CTs shows that our proposed method outperforms the state-of-art CI electrode localization methods and achieves a mean error of 0.12mm. This represents a crucial step in translating IGCIP from the laboratory to large-scale clinical use.

## 3.1 Introduction

Cochlear implants (CIs) are surgically implanted devices for treating severe-to-profound hearing loss [11]. A CI device consists of an external and an internal component. The external component contains a microphone, a processor, and a transmitter. The transmitter is used to send signals to a receiver coil that is under the skin and connects via a wire lead to an electrode array implanted within the cochlea. The implanted CI electrodes then stimulate the spiral ganglion (SG) nerves to induce a sense of hearing. The SG nerves are tonotopically ordered by decreasing characteristic frequency along the length of the cochlea [10, 22]) as shown in Figure 3.1. A SG nerve is stimulated when the frequency associated with it exists in the incoming sound [26]. After the CI surgery, an audiologist needs to program the CI. This includes the selection of the stimulation level of each individual electrode based on perceived loudness from the patient and the selection of a frequency allocation table, which determines which individual electrodes are activated when the incoming sound contains specific frequencies. CIs lead to remarkable success in hearing restoration for the vast majority of recipients with average post-implantation sentence recognition rates over 70% correct for unilaterally implanted users and 80% correct for bilaterally implanted users, respectively [8-9]. However, there are a significant number of users experiencing only marginal benefits. Recent studies have demonstrated that there exists a correlation between hearing outcomes and the intra-cochlear locations of CI electrodes [1, 20, 21, 23, 24, 25]. One factor that negatively affects hearing outcomes is electrode interaction (or channel interaction). Electrode interaction leads to nerve groups being activated in response to multiple frequency bands [2, 7]. Electrode interaction can be alleviated by deactivating the electrodes that cause electrode interaction [13]. In Figure 3.1 we show the CI electrodes and their activation patterns for a subject. As can be seen, by deactivating some electrodes (labelled with red crosses), electrode interaction can be reduced.

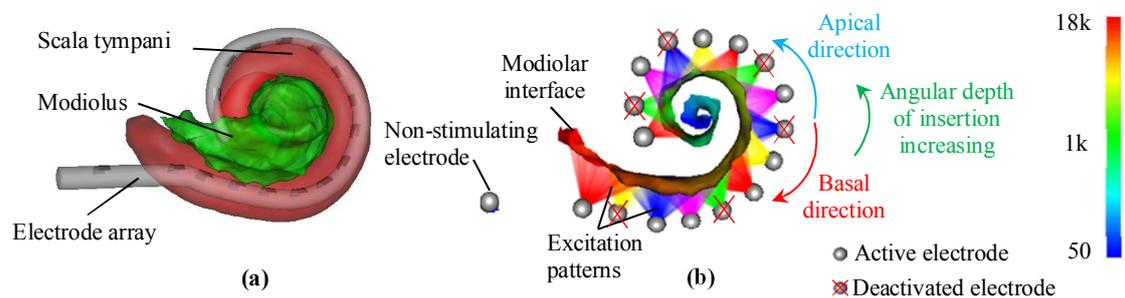

**Figure 3.1.** Visualization of the intra-cochlear anatomy and CI electrode array. Panel (a) shows the scala tympani in red and the modiolus in green. Modiolus is the interface between the auditory nerves of the SG and the intra-cochlear cavities. Panel (b) illustrates the stimulation patterns produced by electrodes on one array. The modiolar surface is color-coded with the tonotopic place frequencies of the SG in Hz.

Our group has developed methods for image-guided cochlear implant programming (IGCIP) [14] to assist audiologists with CI programming. IGCIP uses image processing techniques we have developed to analyze the spatial relationship between the CI electrodes and auditory neural sites for each individual recipients in order to estimate the occurrence of electrode interaction and select electrodes to deactivate to alleviate interactions. The major steps consist of (1) the segmentation of the intra-cochlear anatomy, [15, 17, 18, 19], (2) the localization of the implanted CI electrodes [28, 12, 29], (3) the analysis of the spatial relationship between the CI electrodes and the neural interface [14], and (4) the automatic electrode configuration selection [30-31]. Clinical studies have shown that hearing outcomes are significantly improved when the CI electrode deactivation plans generated by IGCIP are adopted [13, 32]. However, because the electrode localization procedure in IGCIP is still not fully automated, it is difficult to translate IGCIP from the laboratory to large scale clinical use.

Automating the electrode localization procedure is challenging. The first challenge is that the image quality of the clinical CTs is limited due to the current CT scanners. For instance, the resolution of clinical CT images is usually coarse (resolution obtained nowadays is typically 0.2 x 0.2 x 0.3 $mm^3$) compared to the typical size of the CI electrodes which is on the order of 0.3 x 0.3 x 0.1 $mm^3$. Due to the partial volume effects, it

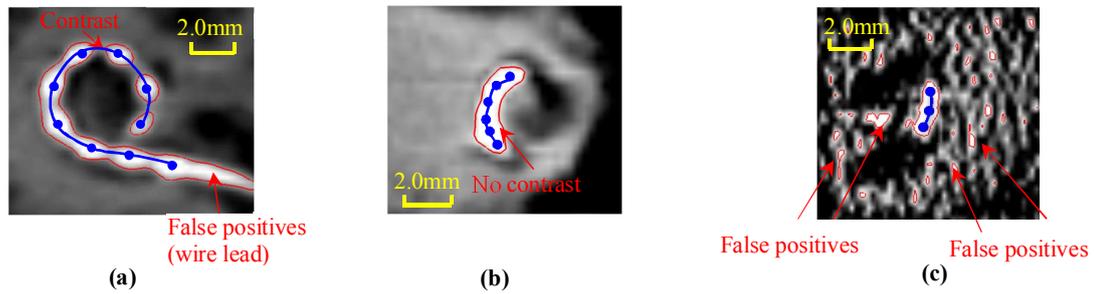

**Figure 3.2.** Panels (a) and (b) show examples of distantly and closely-spaced arrays in eCTs. Panel (c) shows an example of a distantly-spaced array in a lCT.

is difficult to localize small-sized CI electrode array in clinical CTs. The images resolution is also coarse relative to the spacing between electrodes. This makes it difficult to separate the individual electrodes from the array, as shown in Figure 3.2. Further, because the electrodes are composed of radiodense platinum, beam hardening artifacts distort the intensities in the region around the electrode array, resulting in erroneous intensities assigned to voxels around the electrodes during reconstruction. This complicates the identification of individual electrodes in CTs. The second challenge is that even though the CI electrodes usually appear as high intensity voxel groups in CTs, it is difficult to select a threshold such that the thresholded image only contains voxels occupied by CI electrodes because voxels occupied by wire lead, receiver coils, and cortical bones are usually assigned high intensity values too. In this article, the non-electrode voxels with intensity values higher than a selected threshold are defined as "false positive" voxels. CT images are also reconstructed with different algorithms. In an image reconstructed with an "extended" Hounsfield Unit (HU) range (eCT), the metallic structures are assigned higher intensity values than the cortical bones. In an image reconstructed with a "limited" HU range (lCT), the maximum intensity is limited to the intensity of cortical bones. Thus, in an eCT, the false positive voxels are usually occupied by the metallic wire lead as shown in

Figure 3.2a. In a lCT, there are many more false positive voxels as shown in Figure 3.2c. The third challenge is that there exist several models of electrode arrays, which lead to various intensity-based features in clinical CTs. The widely used models of electrode arrays made by the three leading manufacturers are: Med-El® (MD) (Innsbruck, Austria), Advanced Bionics® (AB) (Valencia, California, USA), and Cochlear® (CO) (Sydney, New South Wales, Australia). Arrays differ by the number of electrodes, the size of electrodes, and the spacing between electrodes. Based on inter-electrode spacing, we classify CI electrode arrays into two broad categories: Closely-spaced and Distantly-spaced arrays. Closely-spaced arrays are such that individual electrodes cannot be resolved in the images and the set of electrodes thus form a single connected region as shown in Figure 3.2b. We have proposed a centerline-based snake-based localization method [28] to localize individual electrodes in this type of array. This method fails for distantly-spaced arrays because electrodes do not form a single connected region as shown in Figure 3.2a. Thus, to fully automate IGCIP, we need an automatic method to localize distantly-spaced electrode arrays in clinical CTs.

Other groups have investigated methods for localizing CI electrodes in CTs [33-34]. In [33], *Bennink et al.* proposed a method for localizing closely-spaced arrays by using the *a-priori* knowledge of the CI array geometry. This method requires a manual initialization on the whole head CT by defining a bounding box that includes all the electrode contacts for the subsequent CI array centerline localization algorithm. Then, it uses a curve tracking method and an intensity profile matching algorithm to localize individual electrodes on the array. However, the manual definition of the bounding box requires expertise in recognizing the intensity-based features of the endpoints of the implanted CI array and can also be complicated due to the existence of the false positive voxels on the wire lead. Due to the requirement for manual input, this method could not be directly applied for fully automatic IGCIP. Further, the curve tracking and intensity profile matching algorithms in this method would also need to be modified to be used for localizing distantly-spaced arrays. The curve tracking algorithm only aims to find the voxels with maximum intensity in a small local search range. When localizing distantly-spaced arrays, the local search range would need to be set larger, however this could lead to erroneous results. Consider the Med-El Standard array case shown in Figure 3a. The Euclidean distance between electrodes 5-6 and electrodes 5-11 are close. Thus, both electrode 6 and 11 could be present in the search range of electrode 5. The curve tracking process could wrongly select electrode 11 as the next electrode after electrode 5. Further, the existence of false positive voxels in CTs could make the process even more difficult. In [34], *Braithwaite et al*. proposed a method for localizing distantly-spaced arrays in CTs by using spherical measures. This method uses a thresholding step and a specialized filter chain to segment the electrodes and then uses a linear model to determine the order to connect the segmented electrodes. This method is also not fully automated as it requires a manual definition of a bounding box including all the intra-cochlear electrodes so that the order of the electrodes can be defined. Moreover, the method had only been validated on a small dataset of Cone Beam CTs of specimens implanted with CI arrays produced by one manufacturer, where all the CTs being used have the same intensity range. Thus, a pre-defined threshold for the thresholding step can generate a response image in which the $N$ greatest local maxima correspond to the $N$ electrodes. When applying this method to CTs acquired with different scanners, the pre-defined threshold will not work. From our experiments on a large-scale dataset of CTs acquired by using different scanners, even a threshold determined by using an automatic method [28] could generate many false positive voxels in the thresholded image. The graph-based path finding method we present in this article is designed to localize individual electrodes in distantly-spaced arrays. For simplicity, in the remainder of this article, we refer to our proposed method as GP. It builds upon and substantially improves a limited graph-based method (lGP) [12] proposed by our group. In Section 3.2, we describe this method in detail. In Section 3.3, we evaluate GP and we compare it to lGP and to an early implementation of GP (pGP) [29] that does not provide sub-voxel accuracy. This is done on a large-scale dataset of clinically acquired CT images of subjects implanted with 4 different types of CI arrays. In Section 3.4, we summarize our work and discuss possible directions for extending it.

**Table 3.1.** Specifications of different FDA-approved distantly-spaced CI electrode arrays in our dataset

| Manufacturer | Brand | Total electrodes | Electrode spacing distance (mm) |
|---|---|---|---|
| Med-El | Standard (MD1) | 12 | 2.4 |
|  | Flex28 (MD2) | 12 | 2.1 |
| Advanced Bionics | 1J (AB1) | 17 (1 inactive electrode) | 1.1 and 2.5 |
|  | Mid-Scala (AB2) | 17 (1 inactive electrode) | 0.95 and 3.0 |
|  | Helix (AB3) | 18 (2 inactive electrodes) | 0.85 and 3.0 |

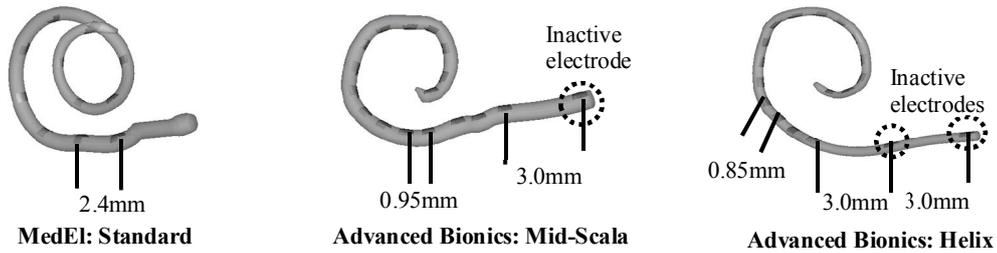

**Figure 3.3.** Three types of distantly-spaced CI electrode arrays provided by the two major manufacturers.

### 3.2   Methods

3.2.1. Dataset

Figure 3.3 shows geometric models for three representative types of distantly-spaced electrode arrays. In Table 3.1, the specifications of the distantly-spaced CI electrode arrays produced by the major manufacturers are summarized. Table 3.2 lists the datasets we use in this study. Dataset 1 consists of whole head CTs of 177 patients. Among these, 151 are eCTs and the remaining 26 are lCTs. 144 of the 151 eCTs are acquired with a Xoran xCAT® flat panel scanner at the Vanderbilt University Medical Center. The remaining 7 are acquired with various scanners at various institutions. The two typical voxel sizes for our eCTs are $0.4 \times 0.4 \times 0.4 mm^3$ and $0.3 \times 0.3 \times 0.3 mm^3$. The 26 lCTs are acquired with various conventional scanners at various institutions (Siemens Somatom Definition AS, Siemens Somatom Force, Siemens Sensation 64, Siemens Somatom Emotion 16, Philips iCT 128, Philip Brilliance 64, Philips Mx8000 IDT16, Philips Comer-256, GE LightSpeed VCT, and GE Medical System BrightSpeed). The typical voxel size for lCTs is $0.23 \times 0.23 \times 0.34 mm^3$. The coarsest voxel size for lCT in our dataset is $0.46 \times 0.46 \times 0.50 mm^3$. Since our method includes several parameters, we randomly select 52 CTs from Dataset 1 that contain different types of electrode arrays for a parameter tuning process. The rest of the 124 CTs from Dataset 1 are used to validate the localization accuracy of our proposed method. An experienced CI electrode localization expert manually generated three sets of localization results on all the post-implantation CTs in Dataset 1. Among the three sets of manual localization results, we randomly select two and average them to serve as the ground truth localization results. The third manual localization result is used to estimate the rater's consistency error (RCE) defined as the distance between the ground truth and the third localization.

**Table 3.2.** Datasets used in Chapter 3

| Dataset # | Purpose | Type of array | Number of eCTs | Number of lCTs | Total number of CTs |
|---|---|---|---|---|---|
| Dataset 1 (177 CTs) | Training (52 CTs) | AB1 | 15 | 0 | 15 |
| | | AB2 | 9 | 1 | 10 |
| | | AB3 | 3 | 0 | 3 |
| | | MD1 | 11 | 0 | 11 |
| | | MD2 | 12 | 1 | 13 |
| | Validation (125 CTs) | AB1 | 19 | 6 | 25 |
| | | AB2 | 25 | 7 | 32 |
| | | AB3 | 4 | 0 | 4 |
| | | MD1 | 17 | 0 | 17 |
| | | MD2 | 36 | 11 | 47 |
| Dataset 2 (28 CTs) | Robustness test | AB1 | 9 | 5 | 14 |
| | | AB2 | 9 | 5 | 14 |

Dataset 2 consists of 28 CTs of a cochlear implant imaging phantom. We use Dataset 2 to evaluate the robustness of GP to various acquisition parameters [4]. The phantom was created using a cadaveric skull implanted with CIs in both left (AB1) and right (AB2) ears. For each side, we have acquired 14 CT scans with a range of acquisition parameters (the HU range, resolution, dose, and type of the implanted arrays) and with different scanners. In this data set, the ground truth localization results are determined by averaging 10 sets of expert localization results.

3.2.2. Method overview

The workflow of GP is outlined in Figure 3.4. (1) We locate the volume of interest (VOI) that contains the cochlea region by registering the whole head CT to a reference image. (2) Next, we up-sample the VOI and the subsequent procedures are performed on the VOI. (3) Then, we determine the value of a set of parameters that will be used in the following steps using *a-priori* knowledge of the geometry of the array model. We call these parameters electrode spacing distance (ESD)-based parameters. As has been shown in Table 3.1, the distances between individual electrodes are known for each model. For a specific electrode array, we denote the distance between the centers of the $i^{th}$ and the $(i+1)^{th}$ electrodes as $D_i$ and we define $\{D_i\}$ as the set of inter-electrode distances. We then define the set of ESD-based parameters associated with this array as $\{d_{m, m=1,...,M}\}$, with $M$ the number of unique

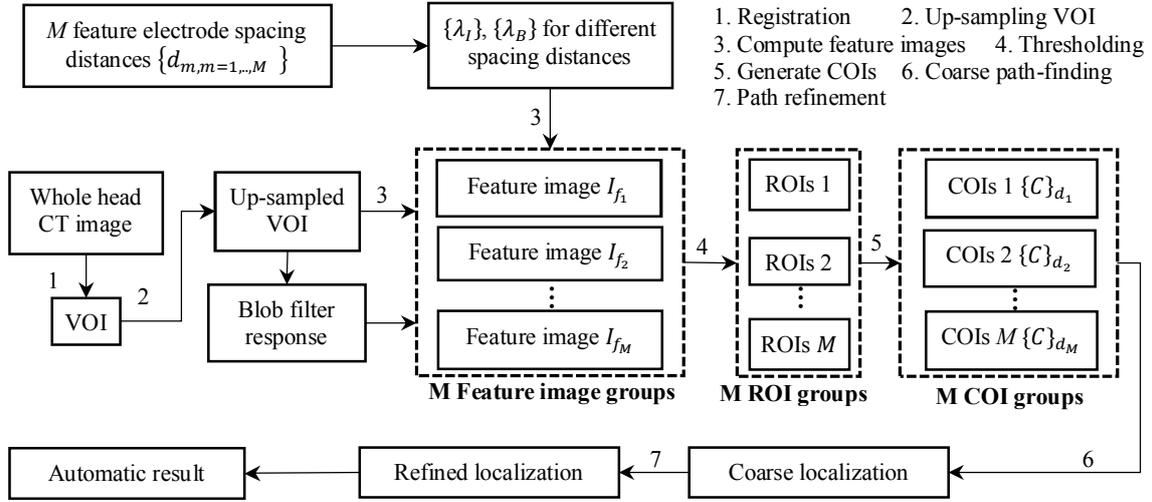

**Figure 3.4.** Workflow of GP.

values in $\{D_i\}$. For example, an AB1 array has a distance of 2.5 mm from the inactive electrode to the most proximal electrode ($D_1 = 2.5$ mm) and a distance of 1.1 mm between each other individual active electrode on the array ($D_2 = D_3 = \cdots = D_{16} = 1.1$ mm). Thus, for an AB1 array, there are $M = 2$ different ESDs, $d_1 = 2.5$ mm and $d_2 = 1.1$ mm. In the same way, we determine the ESD-based parameters for the other types of electrode arrays. In the dataset we use for this study, $M = 1$ for arrays manufactured by Med-El and $M = 2$ for arrays manufactured by Advanced Bionics. However, our design permits defining an arbitrary number of ESD-based parameters. Parameter values are used to tune filters or detection thresholds and produce $M$ feature images, each optimized to detect electrodes separated by the corresponding $d_i$ distance. (4) Next, we identify the regions-of-interest (ROIs) that contain voxels occupied by the CI electrodes by using the $M$ feature images. (5) Then, we perform a voxel thinning method on each of the ROIs to extract the medial axis points as candidates of interest (COIs). At this stage, COIs consists of voxels occupied by electrodes and false positive voxels. (6) Once the COIs are extracted, we perform a coarse path-finding algorithm to find a fixed-length candidate path linking $N$

COIs that minimizes a cost function to coarsely localize the electrodes. (7) Finally, we use a second path-finding algorithm to locally refine the location of each individual coarsely localized electrode. Each of these steps are detailed in the following subsections. In the remainder of this article the value of all the parameters denoted with Greek letters is determined through a parameter tuning process described in subsection 3.2.6.

3.2.3. COI generation

The first step in our method is to identify the VOI that contains the cochlea region (a local region ~30cm³ around the cochlea). We achieve this by registering a reference image where the VOI bounding box is defined [27] to the target CT. After determining the VOI, we up-sample it to a voxel size of $0.1 \times 0.1 \times 0.1 \text{mm}^3$ and then compute a feature image $I_f$ based on it. The feature image $I_f$ is used for generating the ROIs and is computed as:

$$I_f(v) = \lambda_B(d_m)\frac{I_B(v) - T_B(\alpha_B\%)}{T_B(\alpha_B\%)} + \lambda_I(d_m)\frac{I(v) - T_I(\alpha_I\%)}{T_I(\alpha_I\%)} \tag{3.1}$$

where $I$ is the intensity image of the VOI, $I_B$ is the response to a blob filter applied to the VOI that is inspired by Frangi's vesselness filter [6]. As does Frangi, we use the value of the three eigenvalues ($L_1$, $L_2$ and $L_3$) of the $3 \times 3$ Hessian matrix at a voxel $v$ to define the filter:

$$I_B(v) = \begin{cases} B_1(v) \cdot B_2(v) \cdot B_3(v), & L_1, L_2, L_3 < 0 \\ 0, & \text{otherwise}' \end{cases} \tag{3.2}$$

The three terms in Eqn. (3.2) are defined as $B_1 = 1 - \exp\left(-\frac{\sum_{i=1}^{3} L_i^2}{S_1^2}\right)$, $B_2 = \exp\left(-\frac{r_{12} + r_{23} + r_{13}}{S_2}\right)$, and $B_3 = 1 - \exp\left(-\frac{L_{\min}}{S_3}\right)$, where $r_{ij} = |L_i - L_j|$, $L_{\min} = \min(-L_1, -L_2, -L_3)$, $S_1 = T_I(\alpha_I)$, $S_2 = 5000$, $S_3 = 40000$. In Eqn. (3.2), $T_I(\alpha_I\%)$ is a function which takes a percentage value $\alpha_I\%$ as input argument and generates an intensity threshold applied to $I$ that corresponds to the top $\alpha_I\% = 0.048\%$ of the cumulative histogram. $S_2$ and $S_3$ were empirically selected. The term $B_1$ enhances the voxels with high intensity. The terms $B_2$ and $B_3$ enhance the voxels that have spherical structures. The scales for our blob filter are selected as [0.2, 0.4] mm with a step of 0.04mm, which is the typical range for the CI electrode radius. In Eqn. (3.1), as is $T_I(\alpha_I\%)$, $T_B(\alpha_B\%)$ is a function that generates a threshold applied to $I_B$ that corresponds to the top $\alpha_B\% = 0.028\%$ of the cumulative histogram of $I_B$. $\lambda_I(d_m)$ and $\lambda_B(d_m)$ are functions of the ESD-based parameters $d_m$ that return two weighting scalars. Because the weighting scalars returned by $\lambda_I(d_m)$ and $\lambda_B(d_m)$ are related to $d_m$, our method allows different weighting scalars to be assigned to the intensity and the blob filter response of the VOI depending on the spacing between electrodes. This is important because, for closer electrodes, heavier reliance on the blob filter image is necessary to differentiate electrodes. For more distant electrodes, the more reliable intensity image can be emphasized in the cost function and the blob filter image is less important. Thus, $\lambda_I$ and $\lambda_B$ are defined as:

$$\lambda_I(d_m) = (-\kappa_I d_m + \beta_I)H(-\kappa_I d_m + \beta_I), \qquad (3.3)$$

$$\lambda_B(d_m) = (\kappa_B d_m - \beta_B)H(\kappa_B d_m - \beta_B), \tag{3.4}$$

where $\beta_I = 2.72, \kappa_I = 1.82, \beta_B = 1.14, \kappa_B = 1.21$ are positive weighting scalars. $H(\cdot)$ is the Heaviside function.

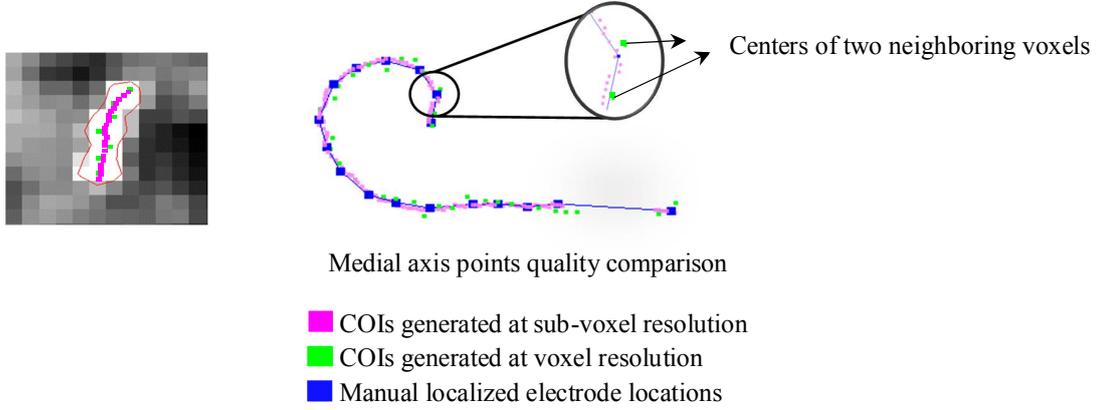

**Figure 3.5.** Quality comparison between the COIs generated by our method on the feature image at sub-voxel resolution and at voxel resolution on the VOI.

Each feature image $I_f$ created with the corresponding $d_m$ is then thresholded at 0. The thresholded regions are the ROIs for electrodes with a ESD value $d_m$. Next, we apply a voxel thinning method [3] to the ROIs to generate the COIs associated with $d_m$. For each ROI, the voxel thinning method generates a series of points that are ordered sequentially as a medial axis lines. Since we have up-sampled the VOI before generating feature images, ROIs and COIs, the COIs we generate also have higher resolution than the COIs that would be generated by using the ROIs produced by the original VOI. Figure 3.5 shows the difference between medial axis points generated by the voxel thinning method simply on the thresholded VOI without up-sampling and the medial axis points generated by our voxel thinning method on the up-sampled VOI. As can be seen, by up-sampling the VOI, our method permits to generate COIs with sub-voxel resolution. Among the COIs generated by using the up-sampled VOI (magenta), there exist candidate points that are closer to the actual locations of implanted electrodes (blue) than the COIs generated at voxel resolution (green). By up-sampling the VOI to a resolution higher than $0.1 \times 0.1 \times 0.1 \text{mm}^3$, we can generate COIs with a higher resolution. However, we found that the selected resolution leads to an adequate resolution for the COIs with an acceptable computational efficiency.

For a CT implanted with an array with $M$ ESD values $\{d_{m,m=1,..,M}\}$, GP generates $M$ sets of ROI groups, one for each ESD value. For each ROI, one set of COIs is generated. The complete set of COIs for the $M$ ESD values are denoted as $\{C\}_{d_1}, \{C\}_{d_2}, \ldots, \{C\}_{d_M}$. We denote a COI that is the $k^{\text{th}}$ medial axis point on the medial axis line of the $j^{\text{th}}$ ROI in the $m^{\text{th}}$ ROI group as $c_m^{j,k}$. These COIs serve as the candidate nodes in a graph search problem used to coarsely localize the individual electrodes. In the following descriptions, we note $p$ as a candidate path, $p_i$ as the $i^{\text{th}}$ COI on the path $p$, and $\{C\}_{d_m}^j$ as the set of COIs that are on the medial axis line of the $j^{\text{th}}$ ROI in the $m^{\text{th}}$ ROI group associated with $d_m$.

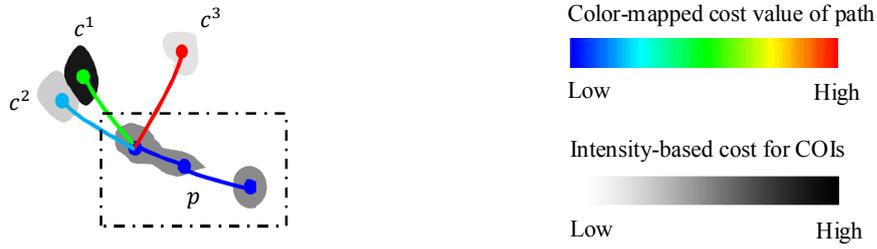

**Figure 3.6.** A simplified example of the coarse path-finding algorithm in GP. At the $i^{\text{th}}$ iteration, the existing path $p$ consisting of $i$-1 nodes has 3 reachable COIs $c^1$, $c^2$, and $c^3$. The path-finding algorithm computes the shape-based cost and intensity-based cost for the three COIs and adding $c_i^2$ to the existing path will result in lowest cost. Compared to $c^2$, $c^1$ has acceptable shape-based features but its intensity-based cost is high. Although $c^3$ has the lowest intensity-based cost, the sharp turn formed by $c^3$ and $p$ makes its shape-based cost high.

3.2.4. Coarse path-finding algorithm

The coarse path-finding algorithm aims to find a fixed-length path of $N$ COIs representing the electrodes on the array, where $N$ is the number of the electrodes on the array. While a standard technique such as Dijkstra's algorithm [5] is typically used for path-finding problems because it guarantees finding a globally optimum solution, we instead use a custom path-finding algorithm that provides no such guarantee because it permits using non-local geometric-based constraints during the search. At each iteration of our proposed path-finding algorithm, a grow stage and a prune stage are included. At the first iteration, the algorithm uses every node in $\{C\}_{D_1}$ as a seed COI representing a candidate path that are each of length 1 in a candidate path group $\{p\}$. The candidate path group $\{p\}$ is used to store the candidate paths during the path-finding algorithm. At the $i^{\text{th}}$ iteration, in the grow stage, each candidate path in $\{p\}$ is grown into a new set of candidate paths by connecting each of the reachable COIs in $\{C\}_{D_i}$ to it. The new set of candidate paths are added into $\{p\}$ to replace the candidate paths before the prune stage. Reachability is defined in the next paragraph. After the grow stage, the candidate path group $\{p\}$ contains a large number of candidate paths. Because the number of candidate paths would grow exponentially at each iteration and the problem would become computationally intractable if left unchecked, we use a prune stage to reduce the set of candidate paths after the grow stage. This is done by computing at each iteration the value of a candidate path cost function and keeping the $\eta_{\max} = 1200$ best candidate paths in $\{p\}$ in the prune stage. After $N$-1 iterations, $\{p\}$ consists of candidate paths of length $N$ and each node in these paths corresponds to a candidate electrode position. Node positions in the path with the lowest cost are used as coarse electrode positions. The cost function consists of a shape-based cost term and an intensity-based cost term, which capture the geometric and intensity features of the electrode arrays in clinical CTs. Figure 3.6 shows a grow stage step for one candidate path with 3 reachable COIs. Among the three reachable COIs for path $p$, the path formed by adding $c^2$ leads to the lowest overall cost.

At the $i^{\text{th}}$ iteration, a candidate path $p$ consists of $i - 1$ COIs. In the grow stage, a COI $c_m^{j,k}$ is considered reachable for a candidate path $p$ if it obeys the following 5 hard constraints. First, it should be such that $\gamma_1 D_{i-1} < \text{dist}(p_{i-1}, c_m^{j,k}) < \gamma_2 D_{i-1}$. In this equation, $\text{dist}(p_{i-1}, c)$ is defined as the Euclidean distance between a COI $c_m^{j,k}$ and the endpoint $p_{i-1}$ of the candidate path $p$. This constrains the distance between the current endnode of the path and the candidate node to be close to the expected *a-priori* distance $D_{i-1}$. The second constraint imposes that $c_m^{j,k}$ is only reachable for $p$ if $D_{i-1} = d_m$. This constrains the candidate node to belong to the corresponding ESD value. The third constraint imposes that $c_m^{j,k} \notin p$, which forbids to add a COI to a path if the COI is already in the path, keeping the path from looping back upon itself. The fourth constraint imposes that if $p_{i-1} \notin \{C\}_{D_{i-1}}^{j}$ (the ROI for $c_i^{j,k}$), then it is only permitted to add $c_i^{j,k}$ to $p$ if $p_z \notin \{C\}_{D_{i-1}}^{j}, \forall z \in [1, i-2]$. This constraint does not allow the path to return to the ROIs that the candidate path $p$ has already visited. The last constraint imposes that if $p_{i-1}, p_{i-2} \in \{C\}_{D_{i-1}}^{j}$ and $d_m = D_{i-1}$, $c_m^{j,k}$ is only reachable for $p$ if $p_{i-2}$, $p_{i-1}$, and $c_m^{j,k}$ are monotonically ordered in the medial axis line $\{C\}_{D_{i-1}}^{j}$. This constraint prevents the path from looping back within an ROI, since COIs belonging to a ROI should be ordered identically to the ROI's medial axis.

We use a cost function to evaluate the quality of each candidate path $p$ after adding a COI $c$. At the $i^{\text{th}}$ iteration, $p$ has $i-1$ COIs. The cost for adding a new COI $c$ into path $p$ is:

$$C_{O1}(c,p) = \rho C_{I1}(c) + C_{S1}(c,p) \tag{3.5}$$

where $\rho$ is a weighting scalar to specify how much we rely on the intensity-based term $C_{I1}(c)$ relative to the shape-based cost term $C_{S1}(c,p)$. $N$ is the total number of electrodes in the array. The intensity based cost term $C_{I1}(c)$ is defined as:

$$C_{I1}(c) = \omega \cdot \left( \mu_I \frac{I_{\max} - I(c)}{I_{\max}} + \mu_B \frac{I_{B\max} - I_B(c)}{I_{B\max}} + \mu_V \frac{I_{V\max} - I_V(c)}{I_{V\max}} \right), \tag{3.6}$$

where $I_{\max}, I_{B\max}, I_{V\max}$ are the maximum values of the image intensity, blob filter response, and vesselness filter response for all the COIs, respectively. $I(c), I_B(c)$, and

$I_V(c)$ are the same at the location of the COI $c$. The blob filter is as described in Eqn. (3.2). The vesselness filter is Frangi's vesselness filter [6] with a scale of 0.25mm. $\mu_I = 1$, $\mu_B = \lambda_B$, and $\mu_V = \lambda_I$ are weighting scalars. We include the image intensity and set $\mu_I = 1$ because voxels occupied by metallic electrodes are usually assigned high intensity. The blob filter response is included because the electrodes often have a blob-like appearance. When $d_m$ increases, $I_B$ becomes more reliable and $\mu_B$ increases. We also include the vesselness filter response because the electrodes sometimes have a tubular appearance if there is not much contrast between them in CT images. When $d_m$ decreases, $I_V$ becomes more reliable and $\mu_V$ increases. $\omega$ is a multiplier we use to punish solutions for which the first electrode is selected as a COI with low blob filter response. We do so to capture the fact that the first electrode usually has a high blob filter response because it only has a neighbor in one direction. At the $i^{\text{th}}$ iteration, $\omega$ is defined as:

$$\omega = \begin{cases} 100, & i = 1 \text{ and } I_B(c) < T_B(\alpha'_B\%) \\ 1, & \text{otherwise} \end{cases}, \quad (3.7)$$

where $T_B(\alpha'_B\%)$ is a function that gives a threshold value applied to $I_B$ that corresponds to the top $\alpha'_B\% = 0.007\%$ of the cumulative histogram of the blob filter response. Next, the shape-based cost term $C_{S1}(c, p)$ evaluates the geometric features of a candidate path $p$ when a COI $c$ is added. It is defined as:

$$C_{S1}(c, p) = \mu_d C_d(c, p_{i-1}) + \mu_s (C_a(c, p_{i-1}, p_{i-2}) + C_{\text{ins}}(c, p_{i-1})) \quad (3.8)$$

where $C_d(\cdot)$, $C_a(\cdot)$, and $C_{\text{ins}}(\cdot)$ are the distance-based, smoothness-based, and the angular depth of insertion (DOI) based cost terms, respectively. The first term $C_d(c, p_{i-1})$ is defined as:

$$C_d(c, p_{i-1}) = |\text{dist}(c, p_{i-1}) - D_{i-1}|, \quad (3.9)$$

$$\mu_d = \begin{cases} \mu_{d1} = 10, & \text{if } \text{dist}(c, p_{i-1}) < D_{i-1} \\ \mu_{d2} = 6, & \text{if } \text{dist}(c, p_{i-1}) \geq D_{i-1} \end{cases} \quad (3.10)$$

where $\text{dist}(c, p_{i-1})$ is the Euclidean distance between a COI $c$ to the endpoint of a candidate path $p$. Eqn. (3.9-3.10) punish the candidate path from growing an edge that is shorter or longer than the expected distance. $C_a(c, p_{i-1}, p_{i-2})$ is determined as:

$$C_a(c, p_{i-1}, p_{i-2}) = (\angle(c, p_{i-1}, p_{i-2}) - \tilde{Z}_{i-1}) H(\angle(c, p_{i-1}, p_{i-2}) - \tilde{Z}_{i-1}), \quad (3.11)$$

where $H(\cdot)$ is the Heaviside function, $\angle(c, p_{i-1}, p_{i-2})$ is the bending angle formed by adding c to the last two endpoints $p_{i-1}, p_{i-2}$ of an existing candidate path $p$ and is defined as:

$$\angle(c, p_{i-1}, p_{i-2}) = 1 - \frac{(c - p_{i-1}) \cdot (p_{i-1} - p_{i-2})}{\text{dist}(c, p_{i-1}) \cdot \text{dist}(p_{i-1}, p_{i-2})} \quad (3.12)$$

and $\tilde{Z}_{i-1}$ is a heuristically selected threshold bending angle value. Eqn. (3.11) punishes paths with bending angles that are sharper than the threshold value. From the ground truth localization results in our training dataset, we observed that (1) the electrodes inserted deeper in the cochlea have a sharper bending angle than the electrodes that are inserted shallower because the curvature of the cochlea increases with increasing the DOI, and (2) arrays from the MD family have sharper bending angles than arrays from the AB family due to a larger spacing distance between electrodes for MD arrays. Thus, we determine $\tilde{Z}$ values for arrays from AB ($\tilde{Z}_{AB}(\cdot)$) and MD ($\tilde{Z}_{MD}(\cdot)$) families separately. $\tilde{Z}_{AB}(\cdot)$ and $\tilde{Z}_{MD}(\cdot)$ are set as:

$$\tilde{Z}_{AB}(i) = \begin{cases} 0.30, & i \leq E_{\text{Half}} \\ 0.59, & i > E_{\text{Half}} \end{cases}, \quad (3.13)$$

$$\tilde{Z}_{MD}(i) = \begin{cases} 0.56 & i \leq E_{\text{Half}} \\ 1.27, & i > E_{\text{Half}} \end{cases}, \quad (3.14)$$

where $E_{\text{Half}} = \frac{N}{2}$ is used to empirically distinguish the electrodes that are inserted deeply versus shallowly in the cochlea. The values in Eqn. (3.13) and Eqn. (3.14) were selected as 130% of the maximum bending angles observed among training AB and MD arrays when $i \leq E_{\text{Half}}$ and $i > E_{\text{Half}}$. The DOI cost $C_{\text{ins}}(c, p_{i-1})$ is defined as:

$$C_{\text{ins}}(c, p_{i-1}) = \left(H\big(\text{DOI}(p_{i-1}) - \text{DOI}(c)\big) + H\big(|\text{DOI}(c) - \text{DOI}(p_{i-1})| - 180°\big)\right) \quad (3.15)$$

where $\text{DOI}(c)$ is the angular depth of insertion value for COI $c$. As the cochlea has a spiral shape with 2.5 turns, the depth of any position within the cochlea is quantified in terms of an angle from 0 to 900 degrees. To obtain the DOI(·) values, we register a pre-implantation CT, in which the intra-cochlear anatomy is segmented, to our post-implantation target CT. For recipients that do not have pre-implantation CTs, our group also has developed a method to segment the intra-cochlear anatomy from post-implantation CTs directly [19]. These two methods generate a DOI map for each individual voxels in the post-implantation CT. The first term in Eqn. (3.15) punishes paths in which a newly added COI $c$ has a $\text{DOI}(c)$ value that is smaller than the endpoint $p_{i-1}$ on the path $p$. The second term in Eqn. (3.15) punishes adding a COI $c$ into an existing path $p$ when the COI $c$ is more than a half turn (180°) ahead or behind the endpoint of $p$. The DOI term constrains the candidate path to grow in the correct direction and to not cross two turns of the cochlea. However, we have observed that in some training cases, some electrodes fall on the boundary of two turns of the cochlea. During the COIs generation step for those cases, due to (1) the registration errors between pre- and post-implantation CTs, (2) the localization errors for intra-cochlear anatomy segmentation, and (3) the limited accuracy of voxel thinning method for generating COIs from ROIs, the DOI values of the COIs close to the boundary of two turns of cochlea could be wrongly estimated. Figure 3.7 shows one example that is implanted with a MD2 array. In Figure 3.7, we label the DOI(·) values for each individual ground truth location of the electrodes. As we can see, the 11$^{th}$ electrode is on the boundary between the second turn and the first turn of cochlea. The DOI of the COIs for the 11$^{th}$ electrode would be estimated in the wrong turn if selected by using the DOI value of the nearest voxel of those COIs. In the path-finding algorithm, this will lead to an inaccurate large cost value when growing a path from the 10$^{th}$ electrode to the COIs for the 11$^{th}$ electrode. To solve this issue, for each COI, we find the maximum and minimum ($DOI_{Max}$ and $DOI_{Min}$) among the DOI values for each voxel in a $3 \times 3 \times 3$ neighborhood around its nearest neighbor voxel. If $\text{DOI}_{\text{Max}} - \text{DOI}_{\text{Min}} \geq 180°$, the COI is near a border and so we create an additional "phantom" COI for the original COI at the same location in the image. The DOI values of the phantom COI and the COI are assigned $\text{DOI}_{\text{Max}}$ and $\text{DOI}_{\text{Min}}$, respectively. Aside from DOI values, the phantom COI has the same information as the original COI. Thus, the path-finding algorithm has equal chance to visit the phantom COI and the original COI and evaluate the cost value for the candidate path with two estimates of the DOI values.

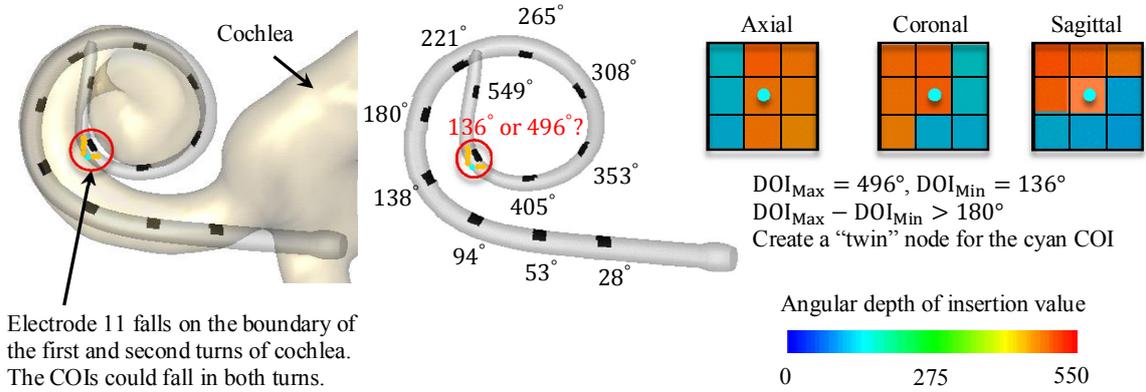

**Figure 3.7.** One example of the problem in the traditional computation method for DOI(·). The 11$^{th}$ electrode falls on the boundary of the cochlea, which is close to the boundary of the part of the cochlea that is one turn before ($-360°$) the actual turn of the electrode. On the right side, the color-coded of the angular DOI map around the electrode is shown. The DOI map is generated by resampling a $3 \times 3 \times 3$ voxels rectangular grid around the closest voxel to the 11$^{th}$ electrode with 27 points on the grid.

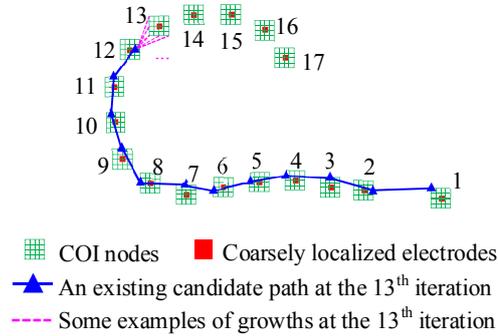

**Figure 3.8.** Visualization of the path-refinement process at iteration 13 for an existing candidate path. This path grows by adding all the COI nodes (the re-sampled rectangular grids) around the 13$^{th}$ electrode to it. The prune step keeps only $\eta_{max2} = 500$ candidate paths with lowest costs for the next iteration.

With the cost function defined above, GP runs the first path-finding algorithm to coarsely localize the location of the electrodes. After the completion of the first path-finding algorithm, the candidate path with the lowest overall cost is selected as the coarsely localized electrode array.

### 3.2.5. Path refinement

The process described in sub-section 2.4 coarsely localizes the electrodes. The second path finding procedure is to refine the coarse result in a local region around each coarsely localized electrode. In this step, the method defines a set of COIs $\{c\}^i$ around each coarsely localized electrode $p_i$ by sampling a fine rectangular grid of points (Shown in Figure 3.8). The set of candidate COIs around the $i^{\text{th}}$ coarsely localized electrode is defined as:

$$\{c\}^i = \{p_i + \varphi_q[x, y, z]\}_{x,y,z \in [-\varphi_r, \varphi_r]} \quad (3.16)$$

In the path refinement algorithm, our method aims at localizing $N$ electrodes after $N$ iterations. We use a candidate path group $\{p\}$ which is similar to the one being described in sub-section 2.4 to store the candidate paths during the path finding algorithm. At the first iteration, all the nodes in $\{c\}^1$ are treated as seed nodes which represent candidate paths with length 1. At the $i^{\text{th}}$ iteration, the method grows the candidate paths by adding the candidate nodes $\{c\}^i$ to the existing candidate paths in the candidate path group (Shown in Figure 3.8). Then the method prunes the candidate path group by keeping only $\eta_{\max 2} = 500$ paths with the lowest cost in the group after each iteration. The cost function to evaluate the quality of a new candidate path constructed by adding a COI $c$ to an existing candidate path $p$ consists of an intensity-based cost term and a shape-based cost:

$$\text{Cost}_2(c,p) = C_{I2}(c) + C_{S2}(c,p) \tag{3.17}$$

The intensity-based cost term $C_{I2}(c)$ is defined as:

$$C_{I2}(c) = -\left(\varphi_I G_\sigma(I(c)) + \varphi_B I_B(c)\right) \tag{3.18}$$

where $G_\sigma(I(c))$, and $I_B(c)$ are the Gaussian filter response, and the blob filter response at $c$, respectively. $\sigma$ is the scale of the Gaussian filter, which is selected as 0.275mm. The shape-based cost is defined as:

$$C_{S2}(c,p) = |\text{dist}(c,p) - D_{i-1}| \cdot \begin{cases} \varphi_{d1}, & \text{dist}(c,p) < D_{i-1} \\ \varphi_{d2}, & \text{dist}(c,p) \geq D_{i-1} \end{cases} \tag{3.19}$$

where $\text{dist}(c,p)$ is the Euclidean distance between node $c$ to the endpoint electrode on path $p$. After $N$ iterations, the path with the lowest overall cost is selected as the final localization result generated by GP.

### 3.2.6. Parameter tuning for GP

The parameter tuning process is performed by using the CTs in our training dataset. The initial values of the parameters are heuristically determined. Then, parameters were optimized sequentially and iteratively until a local optimum was reached for each parameter with respect to the mean localization errors in the training dataset. The parameters used in the coarse localization step were optimized first and then the parameters used in the refinement step were optimized. After determining the optimized values of all the parameters, we fixed those parameter values and performed validation study of the GP on the testing dataset.

**Table 3.3** The selected values for parameters in GP

| Coarse path-finding algorithm | | Path refinement algorithm | |
|---|---|---|---|
| $\eta_{\max}$ | 1200 | $\eta_{\max 2}$ | 500 |
| $\alpha_I$ (%) | 0.048 (%) | $\varphi_q$ | 0.03 |
| $\alpha_B$ (%) | 0.028 (%) | $\varphi_r$ | 3 |
| $\beta_I$ | 2.72 | $\sigma$ | 0.275 |
| $\kappa_I$ | 1.82 | $\varphi_I$ | 32 |
| $\beta_B$ | 1.14 | $\varphi_B$ | 16 |
| $\kappa_B$ | 1.21 | $\varphi_{d1}$ | 0.6 |
| $\gamma_1$ | 0.6 | $\varphi_{d2}$ | 2.5 |
| $\gamma_2$ | 1.2 | | |
| $\rho$ | 2.0 | | |
| $\alpha'_B$ (%) | 0.007 (%) | | |
| $\mu_{d1}$ | 10 | | |
| $\mu_{d2}$ | 6 | | |
| $\mu_S$ | 450 | | |

### 3.3. Results of validation studies

#### 3.3.1. Parameter tuning

Table 3.3 lists the parameter values after the tuning process. To show the effectiveness of the parameters we select, we visualize the parameter sweeping procedure in Figure 3.9a with respect to the mean localization errors in log-scale. Each parameter was swept from 0 to the double of its final selected value with uniform step size. Two exceptions are $\eta_{\max}$ and $\eta_{\max 2}$. For $\eta_{\max}$ and $\eta_{\max 2}$, we start by setting them as 1 because the two path-finding algorithms need to store at least one candidate path.

From Figure 3.9a, we observe that every parameter contributes to the coarse localization step and setting any of them to 0 increases the mean localization error. This indicates that the cost terms we have designed are useful, and the parameters we selected are effective in achieving low localization errors. Among the parameters in the coarse path-finding algorithm, $\alpha_I$, $\alpha_B$, $\beta_I$, $\beta_B$, $\kappa_I$, $\kappa_B$, and $\mu_{d2}$ are sensitive because adjusting them from their selected values results in much larger errors. Aside from $\mu_{d2}$, the other sensitive parameters are all related to feature image construction and COIs generation, which shows that the COI generation step plays a crucial role for the following path-finding algorithms to localize the array. $\eta_{max}$, $\rho$, $\alpha'_B$, $\mu_{d1}$, $\gamma_1$, $\gamma_2$, and $\mu_S$ are not sensitive around the selected values. However, using the selected values for those parameters lead to a lowest mean localization error in our parameter tuning process.

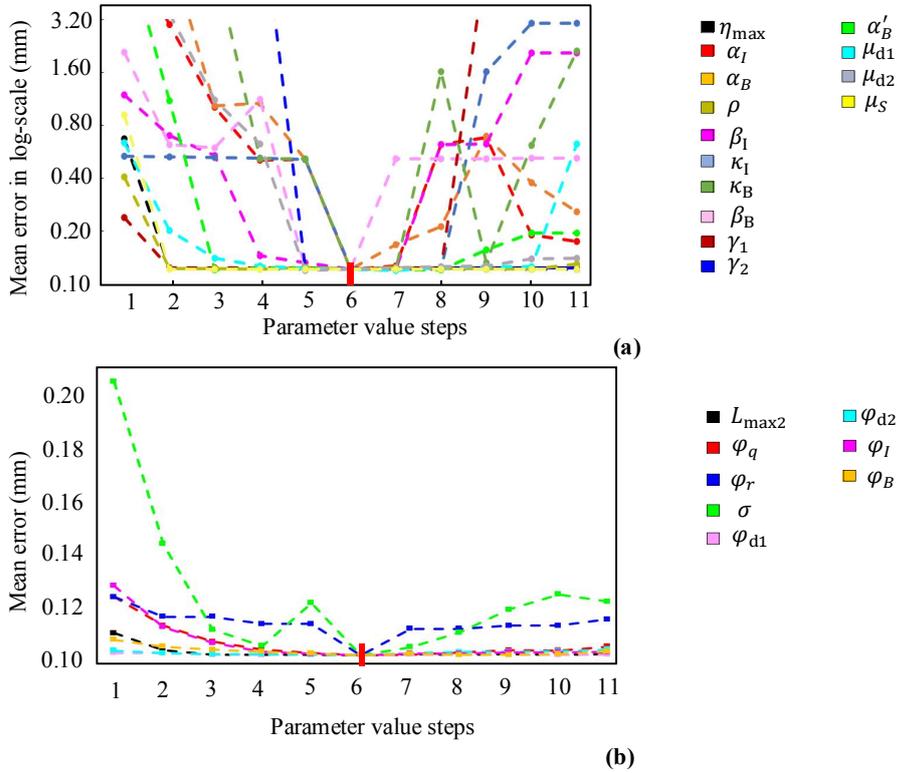

**Figure 3.9.** Visualization of errors when testing each parameter used in the coarse path-finding algorithm and the path refinement in GP. Each parameter is tested over a range from 0 to the double of the optimal values.

From Figure 3.9b, we can observe that the refined localization errors are relatively flat around the selected values of each individual parameters. The most sensitive parameter is the scale $\sigma$ for the Gaussian blur filter. The other parameters are not sensitive around their selected values. However, setting any parameter as 0 increases the mean localization error on the training dataset. After the parameters were selected through the training process, they were fixed to validate the performance of our electrode localization methods on our testing dataset.

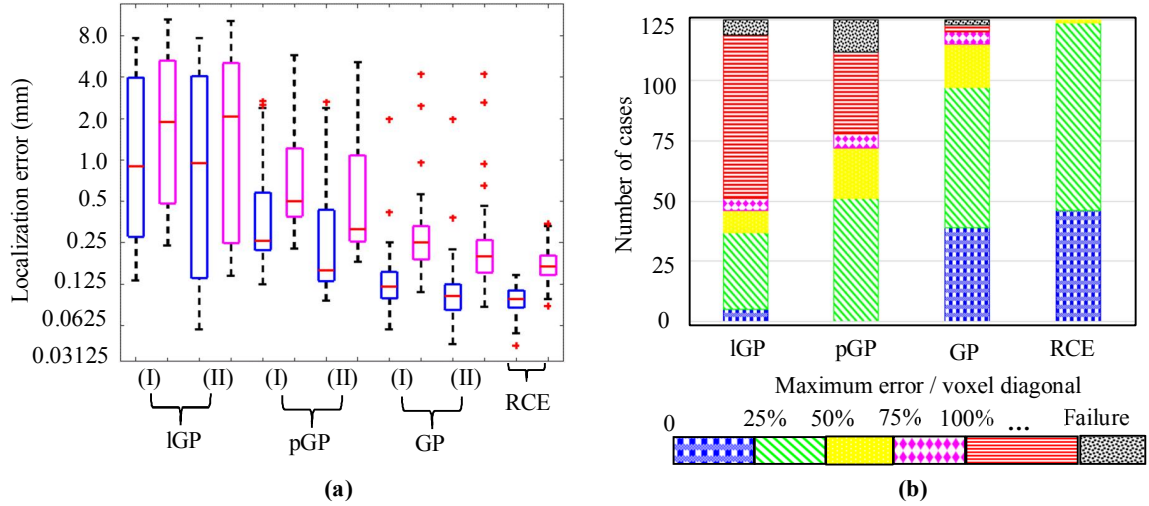

**Figure 3.10.** Panel (a) shows the boxplot (in log-scale) of mean (blue) and maximum (magenta) coarse (I) and refined (II) localization errors between the automatic generated results by lGP, pGP, GP and the rater's consistency errors (RCEs) on CTs in testing dataset. Panel (b) shows the bar plot of the number of cases on which lGP, pGP, GP, and RCE achieves maximum final localization errors lower than 25% (blue), lower than 50% (green and blue), lower than 75% (magenta, green, and blue), lower than 100% (yellow, magenta, green, and blue), over 100% (red) voxel diagonal of the CTs, and the failure subjects (black).

3.3.2. Electrode localization accuracy study on clinical CTs in Dataset 1

In our validation study, we compare the performance of the proposed method GP with the baseline method lGP [12] and a preliminary implementation of GP (pGP) [29] on our testing dataset in Dataset 1 with 125 clinical CTs implanted with different types of distantly-spaced electrode arrays. The baseline method lGP relies solely on image intensity of VOI to generate ROI and COIs. Because of this limitation, it generates less accurate results for most eCTs and unacceptable results for most lCTs because the false positive COIs in lCTs are assigned the same maximum intensity as the true positive COIs. pGP is a preliminary implementation of GP. It uses a set of two fixed weighting scalars ($\lambda_B$ and $\lambda_I$ in Eqn. (3.1)) to generate feature images for ROIs and COIs generation. For lCTs, to reduce false positives among COIs pGP performs image opening on the ROIs with an empirically selected kernel size, which may accidentally remove true positive COIs. With GP, a cost function term is used as soft-constraint so that true positive COIs are not eliminated. In contrast to lGP and pGP, GP generates COIs with sub-voxel resolution, permitting more accurate results with the subsequent path-finding algorithms. The average running time for GP from CT registration to electrode localization is ~40 seconds on a standard Windows Server PC [Intel (R), Xeon (R) CPU X5570, 2.93 GHz, 48GB Ram], which is longer than pGP (~8 seconds) and lGP methods (~5 seconds). GP has a longer run time because it up-samples VOIs to generate COIs with sub-voxel resolution. This COI generation process takes ~32 seconds. The two path-finding algorithms in GP takes ~8 seconds.

We define a "failure" a case for which a method fails to find a fixed-length path from the COIs it generates or for which the method generates a solution that has a maximum error that is larger than one voxel diagonal. Among 125 clinical CTs in our testing dataset, lGP, pGP, and GP fails to find a fixed-length path for 6, 13, and 2 subjects, respectively. One major reason for the methods to fail is that COIs cannot be produced for one or more electrodes, and thus the subsequent coarse path-finding algorithm is not able to find a fixed-length path with $N$ COIs representing the electrodes on the array that obeys the hard constraints. Figure 3.10 shows the quantitative analysis of the localization results generated by lGP, pGP, GP, and the rater's consistency errors (RCEs) in boxplots. Besides the failure cases, our GP generate coarse localization results with a mean error of 0.15mm and final localization results with a mean error of 0.12mm. The mean error of GP's final localization error is close to the mean RCE error, which is 0.1mm. Figure 3.10b shows the distribution of the number of cases that have localization errors that fall into the intervals [0, 25%), [25%, 50%), [50%, 75%), [75%, 100%), and larger than or equal 100% of the voxel diagonal as well as the failure cases. As can be seen from Figure 3.10b, GP generates 120 out of 125 (96%) localization results that have maximum errors within one voxel diagonal, which is close to the RCE (100%) and outperforms pGP (58%) and lGP (41%). We perform a paired t-test between the mean localization errors among lGP, pGP, GP and RCE. The $p$-values are: $8.36 \times 10^{-12}$ for lGP-pGP, $1.07 \times 10^{-15}$ for lGP-GP, $3.21 \times 10^{-16}$ for lGP-RCE, $2.20 \times 10^{-8}$ for pGP-GP, $8.16 \times 10^{-9}$ for pGP-RCE, and $1.24 \times 10^{-1}$ for GP-RCE. According to the $p$ values, the results generated by GP are significantly different from lGP, pGP, but are not significantly different from RCE.

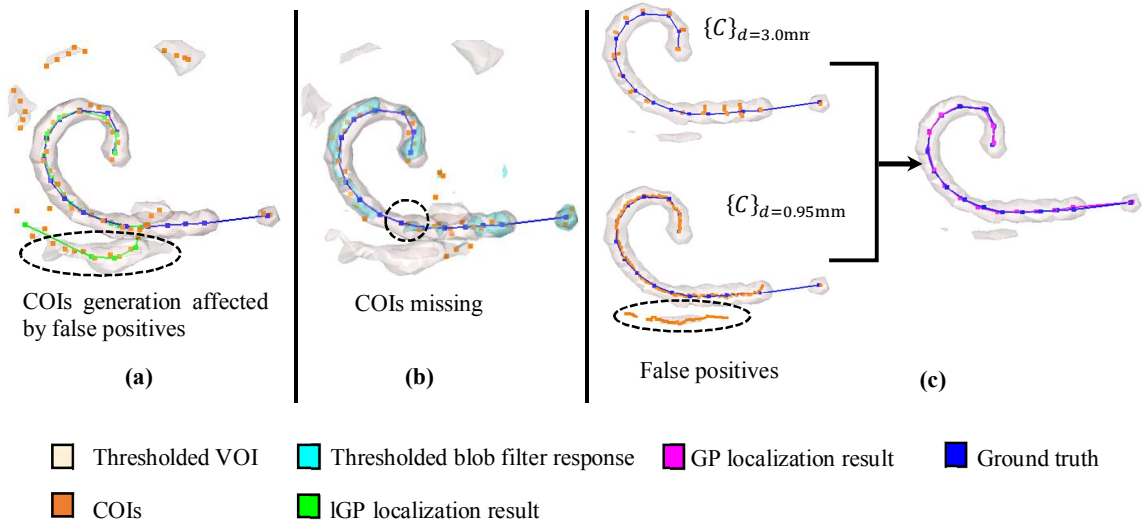

| | Thresholded VOI | | Thresholded blob filter response | | GP localization result | | Ground truth |
| | COIs | | lGP localization result | | | | |

**Figure 3.11.** Visualization of localization results generated by (a) lGP and by (c) GP. In (b), pGP fails to generate a fixed-length path as final localization result because the COIs are missing around two electrodes.

Figure 3.11 shows the localization results generated by GP (panel c) and lGP (panel a) for one example case. In this case, the lGP method generates an inaccurate result in an eCT implanted with an AB2 array. This is because the threshold selected for generating the ROIs and COIs is not high enough to eliminate the false positive voxels in the VOI, and lGP only relies on the image intensity for COIs generation. Thus, in Figure 3.11a, we see many false positives that do not represent the CI electrodes. The final localization result is affected by them. We also perform pGP on the same case but it fails to generate a result (Figure 3.11b). pGP uses the blob filter response to enhance the high intensity blob-like structures in the VOI, but because pGP uses a single set of fixed weighting scalars for image intensity and blob filter response to construct a feature image rather than ESD-based parameters, the method removes some ROIs that contain relatively closely-spaced electrodes on the array. This is so because those ROIs have less blob-liked features. Consequently, the method fails to find a fixed length path with 17 COIs representing all the electrodes on the array. GP generates two sets of COIs for two ESD values. As can be seen, for $d = 0.95$mm, the COIs generation relies more on the image intensity, which results in more false positives but is less likely to miss electrodes. For $d = 3.0$ mm, the COIs generation step relies on the blob filter response, which enhances the distantly-spaced electrodes that have a more obvious blob-like shape in the CT image. GP also up-samples the VOI, which permits the generation of more accurate COIs.

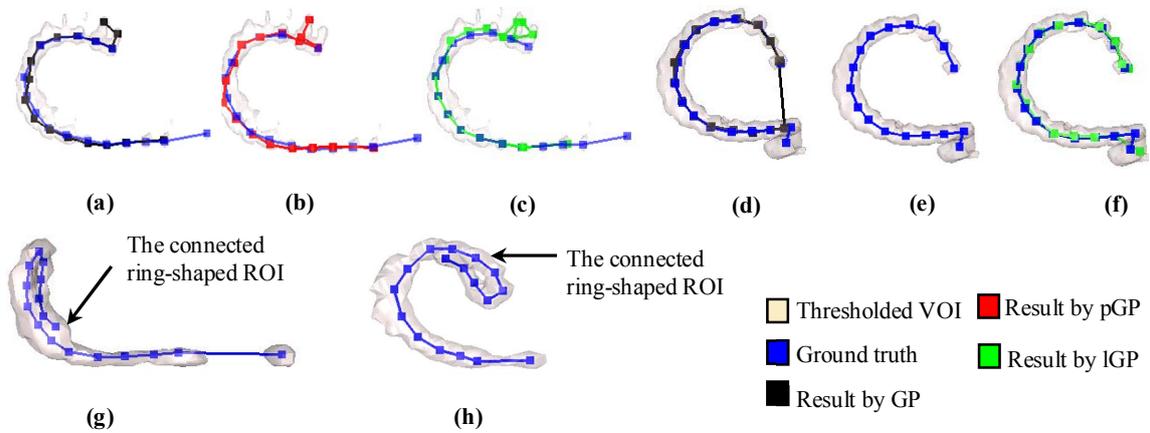

**Figure 3.12.** Panel (a)-(c) and (d)-(f) show localization results generated by GP, lGP, and pGP for two cases, respectively. Panel (g) and (h) show two failure cases for GP.

In Figure 3.12, we show 4 complicated cases for which GP fails to generate localization results with maximum errors within one voxel diagonal. Panels (a)-(c) and panels (d)-(f) show three sets of localization results generated by GP, pGP, and lGP for two cases implanted with AB1 arrays. In Case 1 shown in Figure 3.12a-c, the CT has abnormal intensity features due to beam hardening artifact. Around the most apical electrodes, several false positive voxels are assigned similar high intensity values as the voxels occupied by the actual electrodes. Meanwhile, the inactive electrode has low intensity and blob filter response. This causes all three localization methods to miss the inactive electrode and wrongly select one of the false positives as the most apical electrode. In Case 2 shown in Figure 3.12d-f, the inactive electrode lies much closer than usual to the first active electrode because the array is kinked between the electrodes. This leads to poor localization results generated by all the three methods. Figure 3.12g-h shows 2 cases on which GP fails to generate a fixed-length path as localization results. This is because the electrode array in these two cases are folded. The ROIs generated by GP could not distinguish the electrodes that are pushed together. These 4 cases indicate our method is not robust to extreme cases where the array is kinked or folded or with severe image artifacts. Since such cases are uncommon, we treat them as outliers in our validation study results.

Table 3.4. Mean localization errors for each image group in mm.

|  | HU range | | Resolution | | Dose | | Array | |
| --- | --- | --- | --- | --- | --- | --- | --- | --- |
|  | lCTs | eCTs | Low | Mid | Mid | High | AB1 | AB2 |
| lGP | 1.59±1.97 | 0.14±0.10 | 1.59±1.97 | 0.15±0.10 | 1.22±1.83 | 0.18±0.11 | 0.50±1.17 | 0.56±1.22 |
| pGP | 0.40±0.42 | 0.19±0.06 | 0.20±0.07 | 0.33±0.36 | 0.38±0.38 | 0.17±0.06 | 0.26±0.27 | 0.45±0.43 |
| **GP** | **0.13±0.06** | **0.08±0.05** | **0.13±0.06** | **0.08±0.05** | **0.12±0.06** | **0.10±0.06** | **0.10±0.06** | **0.11±0.07** |
| IL | 0.11±0.05 | 0.07±0.04 | 0.10±0.06 | 0.07±0.04 | 0.10±0.05 | 0.08±0.04 | 0.08±0.05 | 0.14±0.15 |

3.3.3. Robustness test on Dataset 2 with cochlear phantom CTs

Dataset 2 (Shown in Table 3.2) consists of 14 CTs of a cochlear phantom implanted with AB1 and AB2 arrays, acquired with different scanners by varying three parameters – HU range of reconstruction, image resolution and CT dose [4]. The localization sensitivity of any method over a variety of image acquisition parameters can be tested on this dataset. The CI arrays are automatically localized by all three methods under discussion – lGP, pGP and GP. Five results from lGP and three results from pGP are not included in robustness testing because they were too inaccurate and might lead to spurious inferences. The component of localization error expected just as a result of the imaging technique, i.e., the image-based localization error, was calculated separately. Table 3.4 lists the mean localization errors of the automatic methods along with the image-based localization error.

Using Bonferroni corrected unequal variances t-test, we determine that both lGP and pGP add significant algorithmic errors beyond the image-based localization errors for both AB1 and AB2, which is not unexpected. However, the automatic localization error for GP is not significantly different from the image-based localization errors (IL) within a corrected $p$-value of 0.05. This shows that we have achieved the best levels of localization accuracy that can be reasonably expected from these images given the imaging technique employed. The accuracy also isn't unduly affected by the variations of the parameters when compared to the image-based localization errors. Unlike lGP and pGP, GP does not produce poor localizations in case of low resolution or low dose images. The proposed method is thus highly accurate and robust to changes of the four CT acquisition parameters.

### 3.4. Conclusion

In this paper, we propose an automatic graph-based method for localizing distantly-spaced

CI electrode arrays in clinical CTs with sub-voxel accuracy. We use a method to generate candidate voxels of interests that are around electrodes at a sub-voxel resolution and use two path-finding algorithms to find a fixed-length path whose nodes represent electrodes on the array. We perform a parameter tuning process for our proposed method on a training dataset with clinical CTs implanted with different types of distantly-spaced arrays. The results of the validation studies on a large-scale testing dataset including 125 clinical CTs, and 28 phantom CTs show the accuracy and robustness of our proposed method. Comparing with the other two previously developed methods, our proposed GP achieves the lowest mean localization error of 0.12mm and fails to generate localization results with maximum errors within one voxel for only 4 cases. Our proposed automatic method generates localization results that are not significantly different from the localization results generated by an expert. The validation study on 28 CTs acquired from a cochlear implant imaging phantom indicate that our proposed method is robust to several CT acquisition parameters. The overall localization errors of GP are significantly different from the errors of the previously developed methods and are close to the rater consistency errors. One limitation of our proposed method is that it is not robust to electrode arrays that are kinked or folded. Future work will be aimed at addressing this limitation. Another limitation of this study is that the accuracy of the ground truth is limited by the resolution of clinical CTs we have in our dataset. In the future, we plan to construct a dataset with paired $\mu$CTs and clinical CTs. $\mu$CTs have higher resolution and can be used to manually generate ground truth localization results with high accuracy. We plan to apply our GP on clinical CTs and manually segment the electrodes on the paired $\mu$CTs. Then we will register the paired CTs and $\mu$CTs together to evaluate the accuracy of GP. The success of the GP represents a crucial step for fully automating our IGCIP techniques and translating

IGCIP into clinical use. It also enables us to conduct comprehensive large scale studies on the correlation between hearing outcomes and the intra-cochlear locations of CI electrodes.

Chapter IV

AUTOMATIC LOCALIZATION OF CLOSELY-SPACED COCHLEAR IMPLANT
ELECTRODE ARRAYS IN CLINICAL CTS

Yiyuan Zhao, Benoit M. Dawant, and Jack H. Noble

Department of Electrical Engineering and Computer Science, Vanderbilt

University, Nashville, TN, 37232, USA

Abstract

**Purpose:**

Cochlear Implants (CIs) are neural prosthetic devices that provide a sense of sound to people who experience profound hearing loss. Recent research has indicated that there is a significant correlation between hearing outcomes and the intra-cochlear locations of the electrodes. We have developed an image-guided cochlear implant programming (IGCIP) system based on this correlation to assist audiologists with programming CI devices. One crucial step in our IGCIP system is the localization of CI electrodes in post-implantation CTs. Existing methods for this step are either not fully automated or not robust. When the CI electrodes are closely-spaced, it is more difficult to identify individual electrodes because there is no intensity contrast between them in a clinical CT. The goal of this work is to automatically segment the closely-spaced CI electrode arrays in post-implantation clinical CTs.

**Methods:**

The proposed method involves firstly identifying a bounding box that contains the cochlea by using a reference CT. Then, the intensity image and the vesselness response of the VOI are used to segment the regions of interest (ROIs) that contain the electrode arrays. For each ROI, we apply a voxel thinning method to generate the medial axis line. We exhaustively search through all the possible connections of medial axis lines. On each possible connection, we define CI array centerline candidates by selecting two points on the connected medial axis lines as the array endpoints. For each CI array centerline candidate, we use a cost function to evaluate its quality, and the one with the lowest cost is selected as the array centerline. Then, we fit an *a-priori* geometric model of the array to the centerline to localize the individual electrodes. The method was trained on 29 clinical CTs of CI recipients implanted with 3 models of the closely-spaced CI arrays. The localization results are compared with the ground truth localization results manually generated by an expert.

**Results:**

A validation study was conducted on 129 clinical CTs of CI recipients implanted with 3 models of closely-spaced arrays. 98% of the localization results generated by the proposed method had maximum localization errors lower than one voxel diagonal of the CTs. The mean localization error was 0.13mm, which was close to the rater's consistency error (0.11mm). The method also outperformed the existing automatic electrode localization methods in our validation study.

**Conclusion:**

Our validation study shows that our method can localize closely-spaced CI arrays with an accuracy close to what is achievable by an expert on clinical CTs. This represents a crucial step towards automating IGCIP and translating it from the laboratory into the clinical workflow.

## 4.1   Introduction

Cochlear implants (CIs) are neural prosthetic devices used for treating severe-to-profound hearing loss [1]. A CI device has a microphone, a processor, and a transmitter in the external component. The external component receives and processes sound signals and sends them to the internal component, which consists of an internal receiver coil and an electrode array implanted within the cochlea. The implanted CI electrodes receive the electrical signals delivered by the receiver coil, then stimulate the spiral ganglion (SG) nerves to induce a sense of hearing. The SG nerves are the nerve pathways that branch to the cochlea from the auditory nerves, which are tonotopically ordered by decreasing characteristic frequency along the length of the cochlea [2-3] (Shown in Figure 4.1). A SG nerve is stimulated if the frequency associated with it is present in the incoming sound [4]. During a CI surgery, a CI electrode array is blindly inserted into the cochlea by a surgeon. After the CI surgery, for each CI recipient, based on the hearing response, the audiologist adjusts stimulation levels for each individual electrode and selects a frequency allocation table to determine which electrodes should be activated when specific sound frequencies are detected. CIs lead to remarkable success in hearing restoration among the vast majority of recipients [5-6]. However, there are a significant number of users experiencing only marginal benefits.

Recent studies have demonstrated that there exists a correlation between hearing outcomes and the intra-cochlear locations of CI electrodes [7-12]. When multiple CI electrodes stimulate the same nerve pathways, those nerve pathways are activated in response to multiple frequency bands [13-14]. This is known as electrode interaction (or "competing stimulation"). Clinical studies conducted by our group have shown that hearing outcomes of CIs can be significantly improved by using an image-guided cochlear implant programming technique we have designed [15]. In Figure 4.1 we show the CI electrodes activation patterns. With IGCIP techniques, we select an active electrode set in which the electrodes causing competing stimulations are identified and then deactivated [16-18]. To program the CI with IGCIP, we need to know the locations of the CI electrodes with respect to the intra-cochlear anatomy. However, CI placement is unique to each patient. Thus, identifying the intra-cochlear locations of CI electrodes for each individual CI recipient is a critical procedure in the IGCIP system.

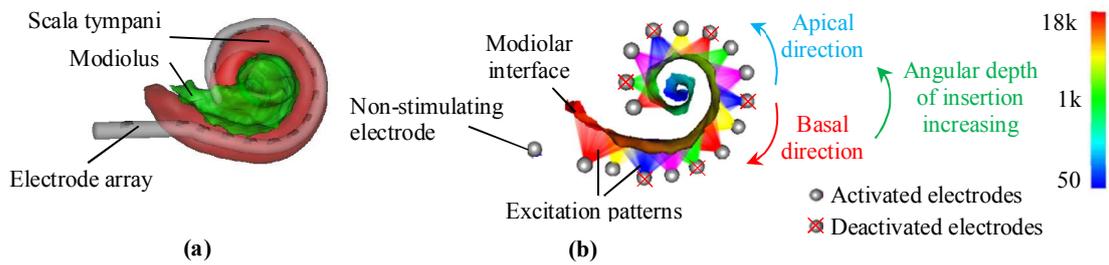

**Figure 4.1.** Visualization of a CI electrode array and intra-cochlear anatomy after CI implantation surgery. In (a), the scala tympani (an intracochlear cavity) is shown with the modiolus, which represents the interface between the auditory nerves of the SG and the intra-cochlear cavities. In (b), a subject implanted with an Advanced Bionics 1J electrode array and stimulation patterns of the electrodes are shown. The modiolar surface is color-coded with tonotopic place frequencies of the SG in Hz.

Identifying the locations of the CI electrodes relative to intra-cochlear structure is difficult. First, segmenting the intra-cochlear structures is difficult because they are not visible in CT images. To solve this problem, we have proposed several methods that use a statistical shape model to estimate the location of the invisible intra-cochlear anatomy by using the visible part of the external walls of the cochlea as landmarks [19-21]. Second, localizing CI electrodes in post-implantation CTs requires expertise. One challenge for localizing CI electrodes in clinical CTs is the limitation of the resolution of clinical CTs. The typical resolution of a clinical CT nowadays is on the order of $0.2 \times 0.2 \times 0.3 mm^3$. Typical CI electrode size is around $0.3 \times 0.3 \times 0.1 mm^3$, which is smaller than the size of a typical voxel in clinical CT. Thus, partial volume effects make it difficult to accurately localize CI electrodes in a clinical CT, even with expertise. In a clinical CT, the voxels occupied by the metallic CI electrodes are assigned high intensity. For electrode arrays with electrodes pitched further than 1mm, the individual electrodes are separable thanks to the obvious intensity contrast between them. Thus, for localizing distantly-spaced CI electrode arrays, our group uses a graph-based method that relies on the intensity contrast between electrodes [22]. However, for arrays with electrodes spaced closer than 1mm, the method for localizing distantly-spaced electrode arrays does not generalize well because, as shown in Figure 4.2, there is typically no intensity contrast between them due to the lack of resolution and beam hardening artifacts. The second challenge is that there exist many FDA-approved closely-spaced CI electrode array models. The spacing of the electrodes on the array differs between models, which leads to different intensity features in the post-implantation CTs. Table 4.1 shows the three major types of closely-spaced electrode array models produced by Cochlear® (Sydney, New South Wales, Australia). Among the three types, CO1 and CO3 have electrode spacing distances that are lower than 0.8mm. In a clinical CT implanted with these two types of arrays, the voxels occupied by the electrodes are usually connected in a high intensity region, as shown in Figure 2c-d. CO2 has a relatively larger electrode spacing distance compared to CO1 and CO3. In a clinical CT implanted with CO2, voxels occupied by those electrodes can be grouped into several regions, as shown in Figure 4.2a-b. The third challenge is the existence of false positive voxels. The wire lead and receiver coils are two sources of false positive voxels since they are also composed of metallic materials and have an appearance similar to the array in CTs. Another source of false positive voxels is the high density structures such as cortical bones. This is more common in a CT acquired with limited range of Hounsfield Unit (lCT). In a CT acquired with extended Hounsfield Unit (eCT), the intensity of the metallic material is much higher than the intensity of cortical bones, which makes the electrode array more separable from the cortical bones. In a lCT, the maximum intensity is limited to the intensity of cortical bones. Thus, the electrodes and the cortical bones are assigned the same maximum intensity, as shown in Figure 4.2c. All these three challenges complicate the automatic localization of closely-spaced electrode array in clinical CTs.

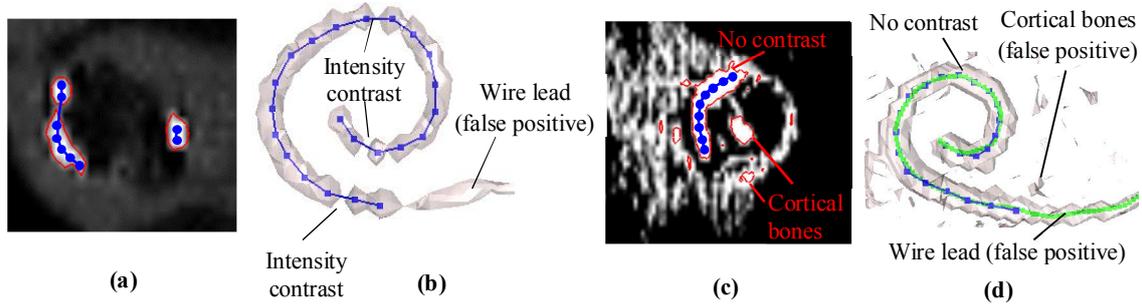

**Figure 4.2.** Panel (a) and (c) show examples of two slices of CT in coronal view of recipients implanted with closely-spaced arrays. Blue points indicate the locations of individual electrodes. An iso-contour around high intensity voxels is shown in red. Panels (b) and (d) show 3D iso-surfaces of the electrode arrays with the manually determined electrode locations generated by an expert. In panel (d), we also show the medial axis line (in green) of the largest ROI extracted by our proposed method. As can be seen, the endpoints of the medial axis line do not always correspond to the electrodes on the two ends of the array.

**Table 4.1.** Specifications of different FDA-approved closely-spaced electrode arrays in our dataset

| Electrode array brand | Total electrodes | Electrode spacing distance (mm) |
|---|---|---|
| Contour Advance (512) (CO1) | 22 | ~0.65 |
| CI-422 (522) (CO2) | 22 | ~0.90 |
| CI24RE-Straight (CO3) | 32 (10 stiffening rings) | ~0.75 |

Other groups have been exploring possible methods [23-24] for electrode localization in clinical CTs. *Braithwaite et al.* proposed a method using a simple thresholding step and then a specialized filter chain for distantly-spaced CI electrode arrays localization in CTs. This method relies on the intensity contrast between individual distantly-spaced electrodes. Thus, it cannot be directly applied to localizing closely-spaced electrode arrays due to the limited to no intensity contrasts between individual contacts. This method is also not fully automated as human intervention is required for the initialization of the step for connecting the segmented electrodes in the right order. *Bennink et al.* proposed a method for localizing closely-spaced electrode arrays in CTs. However, it also requires a manual procedure to define a bounding box that includes all the electrodes as a start point of its algorithm. Meanwhile, this method uses an intensity profile matching algorithm to localize the individual electrodes on the initialized centerline extracted by a curve tracking algorithm. This intensity profile matching algorithm was designed and validated on a small set of lCTs, in which the voxels occupied by electrodes are always have the same intensity value as 3071. In eCTs acquired by different scanners, the intensity values for individual electrodes are not homogeneous. The intensity profile matching algorithm needs to be modified to be applicable to eCTs. The existing electrode localization methods cannot be directly adopted for fully automating IGCIP. Thus, we still need a reliable method for automatic localization of closely-spaced arrays in CTs.

In this article, we present an automatic centerline-based method (CL) for localizing closely-spaced CI electrode arrays in clinical CTs. The method is detailed in Section 4.2. We present the validation study of CL on a large-scale dataset of clinical CTs implanted with the three major types of closely-spaced CI arrays shown in Table 4.1. In our validation study, we compare CL with three existing methods developed by our group: (1) The Graph-based path-finding (GP) algorithm for localizing distantly-spaced array, (2) Snake-based localization (SL) method for localizing CO1 [24], and a preliminary implementation of CL (pCL) [25]. Quantitative comparison of the results generated by the three methods is discussed in Section 4.3 and 4.4.

## 4.2   Methods

### 4.2.1. Dataset

Table 4.2 lists the dataset we use in this study. It consists of post-implantation clinical whole head CTs from 157 subjects acquired with different CT scanners.  Among the 157 clinical CTs, 129 are eCTs and 28 are lCTs. Most of the eCTs in our dataset are acquired with a Xoran xCAT® from Vanderbilt University Medical Center. These eCTs have an isotropic voxel size $0.4 \times 0.4 \times 0.4 mm^3$ or $0.3 \times 0.3 \times 0.3 mm^3$. The 28 lCTs are acquired with CT scanners from other institutions. For these lCTs, the voxel sizes vary and are usually anisotropic. Among the 157 CTs, the coarsest resolution is $0.37 \times 0.37 \times 0.63 mm^3$. We randomly select 28 CTs as training dataset for parameter tuning for our proposed CL method. The remaining 129 CTs are used for validation. For each CT in our dataset, an expert manually generated 3 sets of electrode localization results. We average two sets of manual localization results to generate the ground truth localization results. The remaining set is used to compute the rater's consistency error (RCE), which is defined as the distance between the ground truth and the remaining localizations.

**Table 4.2.** Datasets used in this Chapter 4

| Purpose | Type of array | Number of eCTs | Number of lCTs | Total number of CTs |
|---|---|---|---|---|
| Training (28 CTs) | CO1 | 8 | 7 | 15 |
|  | CO2 | 8 | 2 | 10 |
|  | CO3 | 3 | 0 | 3 |
| Validation (129 CTs) | CO1 | 78 | 10 | 88 |
|  | CO2 | 27 | 6 | 33 |
|  | CO3 | 5 | 3 | 8 |

4.2.2. Method overview

The workflow of our proposed CL method is shown in Figure 4.3. The first step is to extract the volume of interest (VOI) that contains the cochlea. Then, we compute a feature image which is the weighted sum of the intensity and the Frangi vesselness filter response [27] of the up-sampled VOI. We threshold the feature image to generate the regions of interest (ROIs) which contain electrodes and false positives. For each generated ROI, we perform a voxel thinning method [28] to generate its medial axis line. As is shown in Figure 4.3, the points on the actual centerline of the electrode array (shown in blue in Figure 4.3) are distributed across disconnected true positive ROIs. Meanwhile, there also

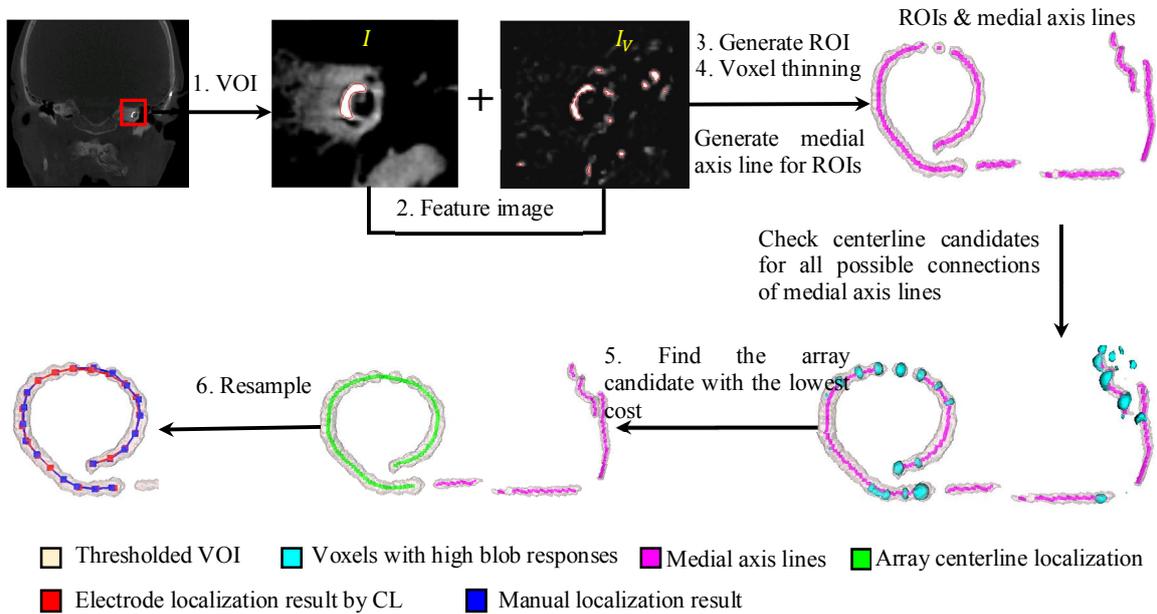

**Figure 4.3.** Workflow of our proposed centerline-based method.

exist several false positive ROIs that do not contain electrodes. When there are multiple ROIs, there exist many possible ways for connecting any number of their medial axes together. We refer to a given connection of medial axes as a "centerline candidate". Note that any centerline candidate constructed in this way cannot be treated as the array centerline directly because the endpoints of the centerline candidates do not always correspond to the two endpoints (the most basal and apical, as shown in Figure 4.1) electrodes on the array, for example, when the electrodes ROI also contains the lead as is shown in Figure 4.2d. Thus, we propose an approach to find an "array candidate" by exhaustively searching all the centerline candidates for the positions of the most basal and apical electrodes, such that the path formed by connecting the basal and apical electrodes along the centerline candidate optimizes a cost function we have designed. The pseudo-code for this algorithm is shown in Algorithm 4.1. The array candidate with minimum cost among all the centerline candidates found in the set of all possible combinations of

**Algorithm 4.1.** Finding the array centerline from centerline candidates

**Input:** $\{c\}_N$ = all centerline candidates constructed by all possible connection of medial axis lines ($N$ is the total number of centerline candidates), $c^k$ = the $k^{th}$ centerline candidate in $\{c\}_N$, $L(c^k)$ = the total number of points on $c^k$, $c_{ij}^k$ = an array candidate constructed by selecting the $i^{th}$ and $j^{th}$ points on a centerline candidate $c^k$ as the most basal and apical electrodes, and then connecting the two points and the points between them along $c^k$.

**Initialize:** Minimum cost value $\text{Cost}_{final} = -1$. Final array centerline $\{i_m = -1, j_m = -1, k_m = -1\}$.

for $k = 1$ to $N$:
    Select centerline candidate $c^k$ from $\{c\}_N$,
    for $i = 1$ to $L(c^k)$
        for $j = 1$ to $L(c^k)$
            Construct an array centerline $c_{ij}^k$ and compute the cost $\text{Cost}(c_{ij}^k)$ for it
            if ($\text{Cost}_{final} == -1$) or $\text{Cost}_{final} > \text{Cost}(c_{ij}^k) \geq 0$
                $\text{Cost}_{final} = \text{Cost}(c_{ij}^k)$;
                $i_m = i; j_m = j; k_m = k$;
            end
        end
    end
end

**Output:** Array centerline $c_{i_m j_m}^{k_m}$.

connections of the medial axis lines is selected as the centerline of the implanted array. Last, we resample the centerline of the implanted array by using the known electrode spacing distance of the array. The points on the resampled curve correspond to the centers of the electrodes. The following subsections present CL in details. All the parameters denoted with Greek letters are selected through a parameter tuning process described in Section 4.3.1.

### 4.2.3. Medial axes generation

To extract the VOI from a whole head CT image, we register it to a reference CT where the VOI bounding box is known [29]. All the subsequent procedures are performed on the VOI. We up-sample the VOI to a voxel size $0.1 \times 0.1 \times 0.1 \text{mm}^3$ so that the following voxel thinning method permits generating a finer resolution medial axis with the subsequent voxel thinning step described below. The up-sampling process will generate an up-sampled VOI with around $270 \times 270 \times 270$ voxels. Next, we compute a feature image constructed as the weighted sum of the normalized intensity image $I$ and the normalized Frangi vesselness filter response $I_V$ of the up-sample VOI. The range of scales for the Frangi vesselness filter are selected as [0.5, 0.6]mm with a step of 0.05mm. The feature image is computed as:

$$I_f = (1-\rho)\frac{I - T_I(\alpha_I\%)}{T_I(\alpha_I\%)} + \rho\frac{I_V - T_V(\alpha_V\%)}{T_V(\alpha_V\%)} \tag{4.1}$$

where $T_I(\alpha_I\%)$, $T_V(\alpha_V\%)$ are functions which take percentage values $\alpha_I\%$ and $\alpha_V\%$ as inputs, and generate thresholds applied to $I$ and $I_V$ that correspond to the top $\alpha_I\% = 0.06\%$ and $\alpha_V\% = 0.06\%$ of the cumulative histogram of $I$ and $I_V$, respectively. We include the vesselness filter response $I_V$ in addition to the intensity $I$ in $I_f$ because it proved to effectively enhance the centerline of the electrode array in the previously developed snake-based localization method. $\rho = 0.29$ is a weighting scalar tuned for balancing the significance of $I$ and $I_V$ in Eqn. (4.1). After computing the feature image, we threshold it at 0 to generate ROIs. After generating the ROI, we perform a voxel thinning method [28] on the ROI to generate a medial axis line of the structure. The medial axis line consists of a set of ordered medial axis points. Those medial axis points are defined as the locus of locations which maximizes the Euclidean distance from the ROI's boundary.

### 4.2.4. Centerline localization and electrode localization

As mentioned in Section 4.2.2, array candidates are formed by evaluating all possible selections of basal and apical electrode position across the centerline candidates formed by all combinations of connections between ROIs. Since a centerline candidate is a set of medial axis points ordered on a curve, by selecting two different points and labeling them as apical and basal endpoints, we construct an array candidate by connecting the points on the curve between the two selected apical and basal endpoints. In a centerline candidate with $n$ points, we can construct $n(n-1)$ different array candidates. An exhaustive search among all the possible array candidates is quick because (1) the maximum number of ROIs generated by our proposed method is usually less than 4, and (2) a centerline candidate typically has $n < 200$ points. We exhaustively search all the array candidates and evaluate their quality by using a cost function defined as:

$$\text{Cost}(p) = \text{Cost}_I(p) + \text{Cost}_S(p) \tag{4.2}$$

where $\text{Cost}_I(p)$ is the intensity-based cost term for $p$, and $\text{Cost}_S(p)$ is the shape-based cost term for $p$. The cost function is designed to capture intensity and shape-based heuristics for a closely-spaced electrode array so that it returns a low cost for the actual centerline of the implanted electrode array and higher cost values for the other array candidates. First, the intensity-based cost term $\text{Cost}_I(p)$ evaluates a blob filter response at the selected apical ($a$) and basal ($b$) endpoints:

$$\text{Cost}_I(p) = \frac{I_{B\max} - I_B(a)}{I_{B\max}} + \mu_1 \frac{I_{B\max} - I_B(b)}{I_{B\max}} \tag{4.3}$$

where $I_B(a)$ and $I_B(b)$ are blob filter responses for the selected apical and basal endpoints in an array candidate, respectively. $I_{B\max}$ is the maximum blob filter response among all the medial axis points. The blob filter response at voxel $v$ is computed in a way that is similar to the Frangi vesselness filter [4] by using the three eigen-values $L_1, L_2, L_3$ of the Heissian matrix computed at $v$:

$$I_b(v) = \begin{cases} B_1(v) \cdot B_2(v) \cdot B_3(v), & L_1, L_2, L_3 < 0 \\ 0, & \text{otherwise} \end{cases} \tag{4.4}$$

The blob filter response is non-zero only when the three eigen-values of the Hessian matrix at $v$ are all negative. This is because the blob structures we detect are bright structures on a dark background. The three terms in Eqn. (4.4) are $B_1 = 1 - \exp\left(-\frac{\sum_{i=1}^{3} L_i^2}{S_1^2}\right)$, $B_2 = \exp\left(-\frac{r_{12}+r_{23}+r_{13}}{S_2}\right)$, and $B_3 = 1 - \exp\left(-\frac{L_{\min}}{S_3}\right)$, where $r_{ij} = |L_i - L_j|$, $L_{\min} = \min(-L_1, -L_2, -L_3)$, $S_1 = T_I(\alpha_I)$, $S_2 = 5000$, $S_3 = 40000$. Eqn. (4.3) captures the heuristic that we expect the voxels occupied by the endpoints to have a large blob filter response. Due to the fact that the electrodes are closely-spaced and the CT resolution is limited, the blob filter responses for the non-endpoint electrodes are much smaller than for the two endpoint electrodes, as shown in Figure 4.3. Thus, we use the high blob filter response as an indicator to find the most apical and basal electrodes. We select the scales for the blob filter as the radius of the basal and apical electrodes in the brand of the implanted array. In Eqn. (4.3), we use $\mu_1 = 1.47$ as a weighting scalar to place extra emphasis on the blob response feature of the basal electrode compared to the apical electrode.

The shape-based cost function $\text{Cost}_S(p)$ captures geometric heuristics for the centerline of the implanted array. First, we define one hard constraint for constructing an array candidate with selected apical electrode (*a*) and basal electrode (*b*) (See Figure 4.2) as:

$$\text{DOI}(a) > \text{DOI}(b) \quad (4.5)$$

In Eqn. (4.5), $\text{DOI}(\cdot)$ is the angular depth of insertion value. As the cochlea has a snail shape with 2.5 turns, the depth into the cochlea of any point can be quantified in terms of an angle from 0 to 900 degrees. To determine $\text{DOI}(\cdot)$, we register a pre-implantation CT in which the cochlea anatomy is segmented, to our target post-implantation CT. In general, the apical electrode is inserted deeper into cochlea than the basal electrode. Thus, we only permit constructing an array candidate when the selected apical electrode has a larger depth of insertion value than the basal electrode. For array candidates satisfying Eqn. (4.5), we define the shape-based cost $\text{Cost}_S(p)$ as:

$$\text{Cost}_S(p) = \mu_2 \frac{\text{DOI}(a)}{\text{DOI}_{\text{max}}} + |\|p\| - D_e| \begin{cases} \mu_3, & \mu_4 \leq \frac{\|p\|}{D_e} \leq 1 \\ \mu_3 + \mu_5 \left(\mu_4 - \frac{\|p\|}{D_e}\right), & \frac{\|p\|}{D_e} < \mu_4 \\ \mu_3 + \mu_5 \left(\frac{\|p\|}{D_e} - 1\right), & \frac{\|p\|}{D_e} > 1 \end{cases} \quad (4.6)$$

where $\text{DOI}_{\text{max}}$ is the maximum angular depth of insertion value among all the points on the initialized centerline. $\mu_2 = 8.89$, $\mu_3 = 0.27$, $\mu_4 = 0.9$, and $\mu_5 = 1.78$ are four tuned parameters. $\|p\|$ is the length of the array candidate $p$. $D_e$ is the *a-priori* expected length of the array when it is straight, given by a 3D model of the implanted array. In the first term of Eqn. (4.6), we expect the apical electrode to have a deep depth of insertion value. In the second term of Eqn. (4.6), we expect the length of the best array candidate to be close to the *a-priori* expected length. Since the array is not elastic, we should not expect the centerline of the implanted array to be longer than $D_e$. Curvature of the array can result in a small reduction of the centerline length. Thus, we design separate cost terms for centerlines with length falling within and outside a pre-defined length range. This pre-defined length range is empirically selected as [90%, 100%] of $D_e$. For array candidates with lengths out of the normal range, an extra cost weighted by $\mu_5$ is added.

The centerline of the implanted array is determined as the array candidate that results in the lowest cost among all the centerline candidates. The resulting centerline is then resampled using the known *a-priori* electrode spacing distance of the array so that the points that form the resulting curve correspond to the centers of the electrodes to generate the final electrode array localization result.

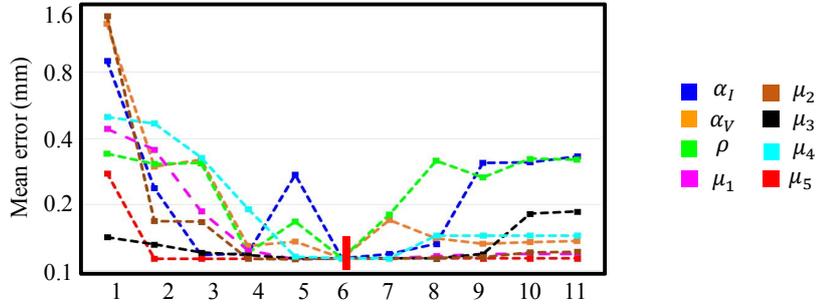

**Figure 4.4**. The parameter tuning process for all the parameters in CL. The red hash mark indicates the finally selected parameter value.

4.2.5. Parameter selection process

The parameter selection process is performed with the 28 CTs in our training dataset. The initial values of those parameters are heuristically determined. Then, we sequentially and iteratively optimize each parameter until a local optimum is reached with respect to the mean localization errors. After determining the parameter values, we use them to perform the validation study on the testing dataset.

## 4.3   Results

4.3.1 Parameter tuning

Table 4.3 lists the parameter values selected after the parameter tuning process. To show the effectiveness of the parameters selected, we visualize the parameter sweeping procedure in Figure 4.4 with respect to the mean localization error in log-scale. Each parameter was swept from 0 to the double of its selected value with 11 uniform step sizes. The mean localization error for the training cases with the selected parameters is 0.11mm. As can be seen from Figure 4.4, all the parameters reached a local minimum in mean localization error at their selected values. Setting any parameter as 0 would lead to an increase of mean localization errors in the training dataset. This shows that all the terms in our design contribute to the accuracy of the localization results.

**Table 4.3.** The selected values for parameters in our proposed method

| Parameters | Selected value | Parameters | Selected value |
|---|---|---|---|
| $\alpha_I(\%)$ | 0.06 (%) | $\mu_2$ | 8.89 |
| $\alpha_V(\%)$ | 0.06 (%) | $\mu_3$ | 0.27 |
| $\rho$ | 0.29 | $\mu_4$ | 0.90 |
| $\mu_1$ | 1.47 | $\mu_5$ | 1.78 |

4.3.2 Validation study

We apply GP, SL, pCL and our proposed CL to our testing dataset of 129 clinical CTs implanted with CO1, CO2, and CO3 arrays. GP and SL are two previously developed methods [10, 24] for localizing distantly-spaced electrode arrays and CO1 arrays, respectively. pCL [25] is a preliminary implementation of CL. GP, pCL and CL are implemented in C++ on a standard Windows Server PC [Intel (R), Xeon (R) CPU X5570, 2.93GHz, 48GB Ram]. SL was implemented with Matlab on the same platform. The average running time for GP, SL, pCL, and CL from extracted VOI to electrode localization are ~8s, ~55s, ~40s, and ~42 seconds.

Among the 129 testing cases, GP cannot generate localization results for 52 cases. This is because GP was designed for the localization of distantly-spaced arrays in CTs. One step in GP uses a voxel thinning method [2] to generate candidate nodes for the path-finding algorithms to find a fixed-length path with $N$ candidate nodes ($N$ is the number of the electrodes on the array). In the 52 cases, the candidate nodes cannot form a path that has the length of the implanted array. SL, pCL and CL can generate results for all the testing cases. The comparison of the mean/maximum electrode localization errors among GP (excluding the 52 cases for which GP cannot generate results), SL, pCL, CL, and RCE are shown as boxplots in Figure 4.5a. As can be seen from Figure 4.5a, CL generates localization results with a mean localization error of 0.13mm, which is close to the mean RCE (0.11mm). The three methods GP, SL and pCL have mean localization errors of 2.09mm, 2.06mm, and 0.94mm. In Figure 4.5b, we can see that CL generates 127 localization results among 129 subjects (98%) that have maximum errors within one voxel diagonal, which is close to the RCE (100%) and outperforms the preliminary version pCL (80%) and the previous developed GP (5%) and SL (51%) methods.

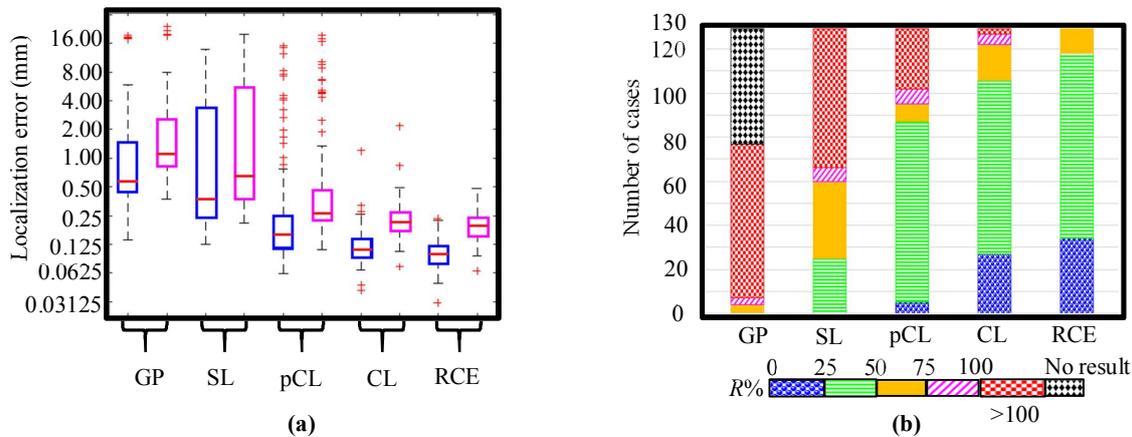

**Figure 4.5.** Panel (a) shows the boxplots of mean (blue) and maximum (magenta) electrode localization errors in log-scale among the different localization methods. Panel (b) shows the distribution of the ratio of the maximum localization errors with respect to the image voxel diagonal ($R\%$) for different localization methods.

## 4.4 Discussion

Figure 4.6 shows two localization results generated by GP, SL, pCL and CL in comparison with the ground truth localization results for two cases. In Figure 4.6a, we show an eCT implanted with a $CO_2$ array. In this case, we cannot find a threshold that makes all the electrodes appear in a connected region with high intensity voxels. The threshold we select also includes the wire lead in the ROIs. GP generates inaccurate localization result by selecting two points on the wire lead as the two most basal electrodes. This is because the path-finding algorithm in GP cannot distinguish the voxels on the wire lead from the voxels occupied by electrodes. SL and pCL both ignore the electrodes that are not in the largest ROI. This is because both of the two methods assume all the electrodes are within one connected group of voxels with high intensity. CL successfully localizes all the electrodes by evaluating the array centerline candidates produced by all the possible connections of ROIs. Figure 4.6b shows results of a lCT implanted with a CO1 array. In this case, the voxels occupied by the wire lead and the electrodes have the same maximum intensity value. Thus, both GP and SL fail to distinguish the wire lead from the electrodes. To avoid localizing the voxels on the wire lead as the basal electrode, pCL added one process before endpoints selection. After generating the medial axis line of the largest connected region after thresholding the VOI, pCL performs an image erosion operation on the thresholded VOI with an empirically selected kernel size to eliminate the false positives on the wire lead before blob filter response computation for endpoints localization. Then, pCL constrains the search of the basal and apical electrodes within the remaining voxel groups (labeled with grey color). As can be seen in Figure 4.6b, the image erosion operation eliminates the voxel groups around the most basal electrode in this specific case. CL localizes the electrodes with a maximum error within one voxel diagonal in this case. Without using image erosion for eliminating the false positives on the wire lead, CL uses the first cost term in Eqn. (4.6) to ensure that the points that are in the deeper region of cochlea are more likely to be selected as the apical point. Then, with an optimal apical electrode selected, CL uses the second term in Eqn. (4.6) so that the selection of the basal electrode and the apical electrode forms an array candidate that has a length close to the implant model.

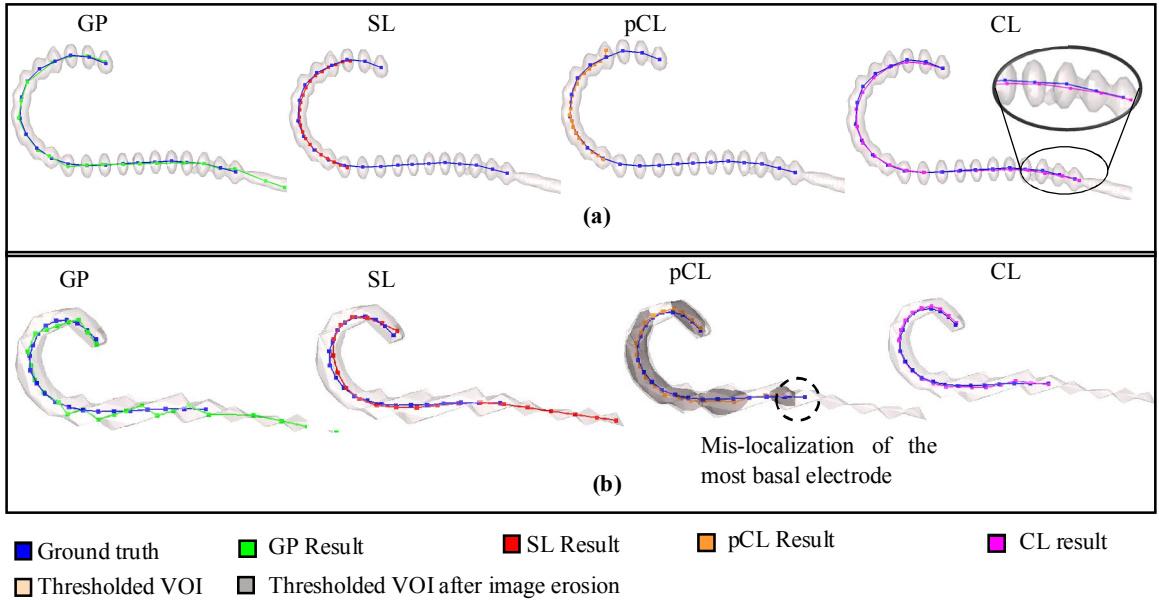

**Figure 4.6**. Visualization of localization results generated by GP, SL, pCL, and CL in comparison with the manual ground truth localization results.

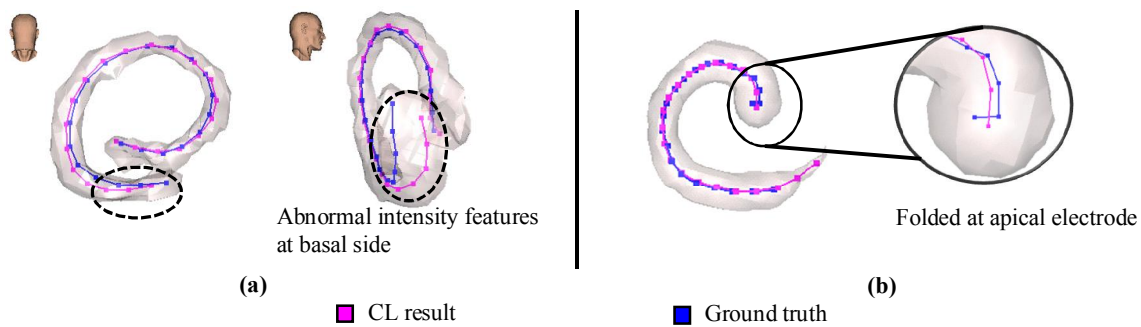

Figure 4.7. Visualization of two eCT cases on which CL generates localization results with maximum errors larger than one voxel diagonal.

Figure 4.7 shows two eCTs on which our proposed CL generates localization results with maximum localization errors larger than one voxel diagonal. In Figure 4.7a, the voxels between the most basal electrode and the third apical electrodes are abnormally assigned high intensity voxels. This causes CL to incorrectly localize the medial axis line, which further affects the centerline localization process. The automatically localized centerline deviates from the ground truth locations to the voxels that are closer to the apical end. In Figure 4.7b, the apical electrode is folded, which has been verified by an electrode localization expert (JN). However, the intensity feature does not show the folded electrode due to the limited resolution of clinical eCTs. Thus, CL mis-localizes the apical electrode and selects a false positive voxel on the wire lead as the basal electrode. The other existing methods all generate inaccurate localization results on these two cases. Failure cases as those shown in Figure 4.7 are rare, and our method generates accurate localization results on the remaining testing cases.

**Table 4.4.** p-value of t-test results among mean localization errors generated by GP, SL, pCL, CL and RCE

|      | GP | SL | pCL | CL | RCE |
|------|----|----|-----|----|-----|
| GP   | /  | $2.15 \times 10^{-1}$ | $2.16 \times 10^{-1}$ | $7.58 \times 10^{-5}$ | $6.81 \times 10^{-5}$ |
| SL   |    | /  | $1.03 \times 10^{-4}$ | $2.74 \times 10^{-12}$ | $1.42 \times 10^{-12}$ |
| pCL  |    |    | /   | $3.74 \times 10^{-4}$ | $2.44 \times 10^{-4}$ |
| CL   |    |    |     | /  | $6.30 \times 10^{-3}$ |
| RCE  |    |    |     |    | /   |

We perform paired t-tests with Bonferroni correction on the mean localization errors generated by GP, SL, pCL, CL, and RCE. The results are shown in Table 4.4. Our proposed method CL generates localization results that are significantly different than all the other methods and the RCE. However, the mean and maximum localization errors indicate CL can generate localization results with an accuracy that is close to the manual localization results generated by the CI electrode localization expert in our group.

Even though CL generates localization results close to the ground truth, its accuracy can still be improved. An example of errors that can be improved can be seen in Figure 4.6a. In this example, the five most basal electrodes have obvious deviations from the center of the high intensity blob in the CT images. This is because CL uses basal and apical electrodes as landmarks and the accuracy of the localization of electrodes in between them is not influenced by their local intensity-based features. However, when the CT has high resolution and the electrodes have larger spacing distance between each other, some contrast between electrodes could be used in the electrode localization process to further improve the electrode localization accuracy.

## 4.5  Conclusions

Localization of CI electrode arrays is a crucial step to analyze the electrode stimulation patterns with respect to the auditory nerves in our IGCIP system. In clinical CTs implanted with closely-spaced electrode arrays, the identification of each individual electrode is difficult because the intensity contrast between electrodes is small. In this paper, we have proposed an automatic centerline-based method for the localization of closely-spaced CI electrode arrays in clinical CTs. The validation study shows that our method outperforms the existing methods for localizing CI electrodes. Our proposed method generates localization results with mean localization error of 0.13mm. 98% of our localization results have maximum localization errors lower than one voxel diagonal. These results show that our proposed method can generate localization results with errors that are close to the rater's consistency errors and are smaller than the existing methods. This method represents a crucial step in fully automating IGCIP and translating it from the laboratory to clinical use. It also enables us to conduct large-scale studies on the electrode location and its effects on hearing outcomes. One limitation is that our proposed method uses intensity-based features only to localize the basal and apical electrodes. The other electrodes between them are localized by resampling a centerline defined by these two electrodes and the medial axis points between them. In future work, we will explore modifications to our approach to permit leveraging intensity contrast between electrodes when it is available. Another limitation of this study is the accuracy of the ground truth. Our ground truth is based on clinical CTs with limited resolutions that has inherent errors due to partial volume artifacts. In the future, we plan to use $\mu$CT-CT pairs of cochlear specimen for better characterizing the accuracy of our proposed electrode localization method. The $\mu$CTs will be used to generate ground truth localization results since they have a higher resolution. The paired clinical CTs will be used to generate automatic localization results. We will register the $\mu$CTs and CTs together so that we can analyze the electrode localization errors generated by our proposed method with more a reliable ground truth.

Chapter V

AUTOMATIC SELECTION OF THE ACTIVE ELECTRODE SET FOR IMAGE-GUIDED COCHLEAR IMPLANT PROGRAMMING

Yiyuan Zhao, Benoit M. Dawant, and Jack H. Noble

Department of Electrical Engineering and Computer Science, Vanderbilt

University, Nashville, TN, 37232, USA

Abstract

Cochlear implants (CIs) are neural prostheses that restore hearing by stimulating auditory nerve pathways within the cochlea using an implanted electrode array. Research has shown when multiple electrodes stimulate the same nerve pathways, competing stimulation occurs and hearing outcomes decline. Recent clinical studies have indicated that hearing outcomes can be significantly improved by using an image-guided active electrode set selection technique we have designed, in which electrodes that cause competing stimulation are identified and deactivated. In tests done to date, an expert is needed to perform the electrode selection step with the assistance of a method to visualize the spatial relationship between electrodes and neural sites determined using image analysis techniques. In this work, we propose to automate the electrode selection step by optimizing a cost function that captures the heuristics used by the expert. Further, we propose an approach to estimate the values of parameters used in the cost function using an existing database of expert electrode selections. We test this method with different electrode array models from three manufacturers. Our automatic approach generates acceptable active electrode sets in 98.3% of the subjects tested. This approach represents a crucial step towards clinical translation of our image-guided CI programming system.

## 5.1  Introduction

Over the last 20 years, cochlear implants (CIs) have become the most successful neural prosthesis and are used to treat severe-to-profound hearing loss [1]. In CI surgery, an array of electrodes is blindly threaded into the cochlea. After the surgery, the processor worn behind the ear sends signals to the implanted electrodes, which stimulate the auditory nerve pathways within the cochlea. After implantation, the CI is programmed by an audiologist. CI programming begins with the selection of a general signal processing strategy, e.g., continuous interleaved sampling [2]. Then the audiologist defines the "MAP", i.e., the CI processor instructions that determine what signals are sent to the implanted electrodes in response to incoming sounds. The MAP is determined by selecting the electrode configuration, i.e., the active electrode set, by specifying stimulation levels for each active electrode based on measures of the user's perceived loudness, and by selecting a frequency allocation table that specifies which electrodes will be activated when specific sound frequencies are detected. Electrode activation stimulates the spiral ganglion (SG) nerves, the nerve pathways that branch to the cochlea from the auditory nerve. In natural hearing, an SG nerve is activated when the characteristic frequency associated with that pathway is present in the incoming sound. The SG nerves, which are located within the modiolus of the cochlea, are tonotopically ordered by decreasing characteristic frequency along the length of the cochlea, and this precisely tuned spatial organization is well known [3-4] (see Figure 5.1a). The modiolar surface shown in Figure 5.1a represents the interface between the intra-cochlear cavities where the electrodes are placed and the modiolus where the SG nerves that are stimulated by the electrodes are located. Recent research has suggested that hearing outcomes with CIs are correlated with the location at which the electrodes are placed in the cochlea [5-10]. In surgery, the array is blindly threaded into the cochlea with its insertion path guided only by the walls of the spiral-shaped intra-cochlear cavities. The final position of the electrodes is not generally known in the traditional clinical workflow. However, we have developed techniques that enable accurately locating the electrodes using CT images [11-13].

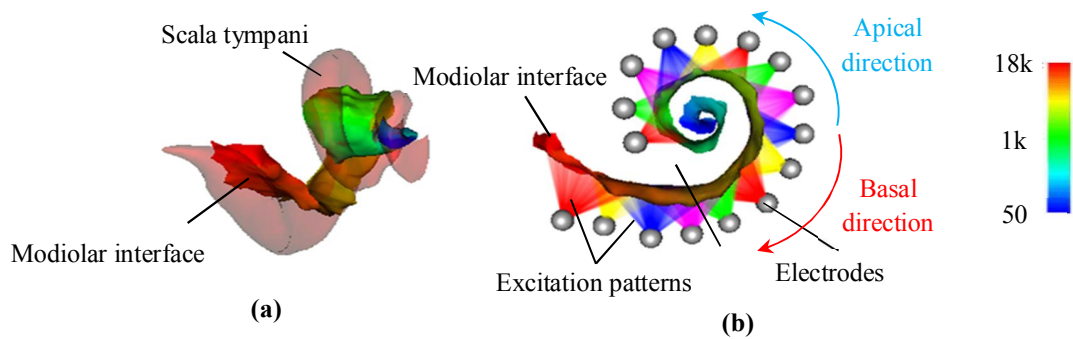

**Figure 5.1.** Visualization of CI electrode activation patterns. In (a), the scala tympani (an intracochlear cavity) is shown with the modiolar surface, which represents the interface between the nerves of the SG and the intra-cochlear cavities and is color-coded with the tonotopic place frequencies of the SG in Hz. In (b), synthetic examples of stimulation patterns on the modiolar interface created by the implanted electrodes are shown in multiple colors to illustrate the concept of stimulation overlap.

Recent research by our group [11, 14] has shown that the spatial relationship between the neural pathways and the electrodes can be used to estimate electrode interactions at the neural level, i.e., cross-electrode neural stimulation overlap (see Figure 5.1b), which is a phenomenon known to negatively affect hearing outcomes [15-16]. We have shown in a large clinical study that when stimulation overlap is detected and the configuration of active electrodes is adjusted to reduce that overlap, hearing outcomes are improved, and those improvements are statistically significant [17]. Our goal now is to fully automate our system so that clinical translation of this technology is feasible.

One step that has not yet been automated is the electrode configuration selection step. Thus far, electrode configurations have been selected manually based on the electrode distance-vs.-frequency curves (DVFs). The DVF is a technique developed by our group to facilitate the visualization of electrode interaction in individual patients [11]. It is a 2D plot that captures important information about the patient-specific spatial relationship between the electrodes and the spiral ganglion (SG) nerves such as is shown in 3D in Figure 5.1b. Figure 5.2a is an example of DVFs for a 7 electrode array. The horizontal axis represents position along the length of the modiolus in terms of the characteristic frequencies of adjacent SG nerves. Each DVF is labeled with a number representing its electrode number. The height of each DVF on the vertical axis represents the distance from the corresponding electrode to the frequency mapped modiolar surface. Thus, a DVF is constructed for a given electrode by finding the distance to that electrode from nearby, frequency-mapped sites on the modiolus. From this visualization technique, we can see that electrode 3 is approximately 1 mm from the modiolus, and the characteristic frequencies of the SG nerves closest to electrode 3 are around 1 kHz. Our current electrode configuration selection method is based on the assumption that if an electrode's DVF is not the closest DVF in the region around its minimum, it is likely that its stimulation region overlaps with other electrodes and thus it is negatively affecting hearing performance. As shown in Figure 5.2, we can see that since the minimum of the DVF for electrode 4 is entirely above the DVF for electrode 3, it is likely that electrode 4 is stimulating the same neural region as electrode 3. Also, while the minimum of the DVF for electrode 6 falls below the other curves, its depth of concavity relative to the minimum envelope of the other neighboring DVFs is small, so it is likely that electrode 6 has an overlapping stimulation region with electrodes 5 and 7. Our active electrode set selection approach is to keep active the largest subset of electrodes that are not likely to cause stimulation overlap. Thus, in the example, we would remove electrodes 4 and 6 from the active electrode set. The DVFs of the updated electrode configuration are shown in Figure 5.2b.

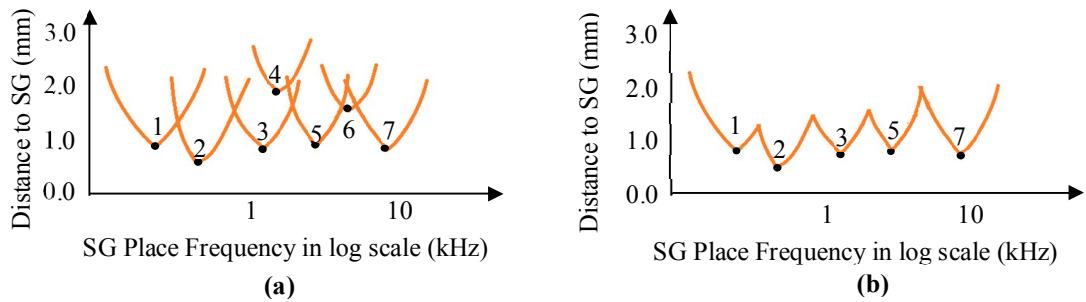

**Figure 5.2**. Visualization of DVFs. (a) shows an example of a combination of the DVFs formed by 7 electrodes. Each single curve represents the distance from the corresponding electrode to the frequency mapped sites along the length of the modiolus. (b) shows the DVFs after electrode configuration adjustment.

As discussed above, we have shown in clinical studies that our manual approach for selecting active electrode set results in significant improvement in hearing performance. While selecting the electrode set manually can usually be done relatively quickly (0.5-2 minutes), it requires specialized expertise, and training new individuals to become experts is time consuming. In order to develop an automated system that implements our approach and can be widely deployed for clinical use, we need an automated method that performs as well as an expert on average for selecting the electrode configuration. To solve this problem, we have developed a DVF-based cost function and propose to optimize it using an exhaustive search method. The rest of this paper presents our approach.

## 5.2  Methods

Our dataset consists of DVFs and expert-defined optimal and acceptable electrode configurations for 96 cases. We divided the dataset into a training and a testing dataset. The training dataset contains 12 subjects implanted with arrays manufactured by Med-El (MD) (Innsbruck, Austria), 10 subjects implanted with arrays manufactured by Advanced Bionics (AB) (Valencia, California, USA), and 14 subjects implanted with arrays manufactured by Cochlear (CO) (New South Wales, Australia). Our testing dataset contains 20 subjects of arrays manufactured by MD, 20 subjects of arrays manufactured by AB, and 20 subjects of arrays manufactured by CO. In our training dataset, we have 18 male and 18 female subjects. Subject age ranges from 18 to 84 with a mean age of 57.9 and standard deviation of 14.69 years. In our testing dataset, we have 28 male and 32 female subjects. The age range is 21 to 84 with a mean age of 58.1 and a standard deviation of 14.6 years.

Our approach is to develop a cost function that assigns a cost for a given electrode configuration, i.e., a particular set of "on" and "off" electrodes, for a subject based on the electrode DVFs. We then can use an exhaustive search method in which all possible configurations are generated, compute the cost for each configuration, and select the one with the minimum cost. In this work, we have chosen to design the cost function to be a linear combination of a set of DVF-based features that capture the heuristics we use for manually producing electrode configuration plans. The features aim at reducing the cross-electrode neural stimulation overlap as described in Section 1. We have defined a total of $N = 10$ feature cost terms. The weighted sum of the $N$ feature cost terms is determined as the final cost value. The weights $\{w_i\}_{i=1}^N$ for the $N$ feature cost terms are determined through a training process using the subjects in the training dataset. Each of the three electrode arrays types has a different number of electrodes (Med-El has 12, Advanced Bionics has 16, and Contour Advance has 22 electrodes) and a different geometry. Thus, they create different DVF patterns, which leads us to estimate the set of weights separately for each electrode type. After generating the estimates of the weights $\{w_i\}_{i=1}^N$, we apply the weights to the testing dataset for validation.

The feature cost terms $\{f_i\}_{i=1}^N$ are defined as follows. First,

$$f_1 = \begin{cases} 0 & \text{If the most apical electrode} \in \text{active set} \\ 1 & \text{If the most apical electrode} \notin \text{active set} \end{cases}, \quad (5.1)$$

which assigns a zero cost to configurations whose most apical electrode, i.e., the deepest electrode in the cochlea (see Figure 5.1b), is activated and a non-zero cost otherwise. $f_1$ is included because deactivating the most apical electrode, which stimulates nerves with lower characteristic frequencies, can result in an up-shift in perceived sound frequency. This affects hearing quality and is usually not preferred. Next,

$$f_2 = \frac{1}{K_a}, \qquad (5.2)$$

where $K_a$ is the number of electrodes that are active in the configuration. While other terms below are designed to deactivate electrodes to increase channel independence, $f_2$ captures the heuristic that keeping more electrodes active is desirable because it results in less frequency compression and better outcomes if those electrodes provide independent stimulation. Next,

$$f_3 = (\sum_{i=1}^{K} e^{-Area\_Term_i})/K, \qquad (5.3)$$

where K is the total number of electrodes, and

$$Area\_Term_i = \begin{cases} \dfrac{T(D_i)}{Area_i} & \text{if electrode } i \in \text{active set} \\ \dfrac{Area_i}{T(D_i)} & \text{if electrode } i \notin \text{active set} \end{cases}, \quad (5.4)$$

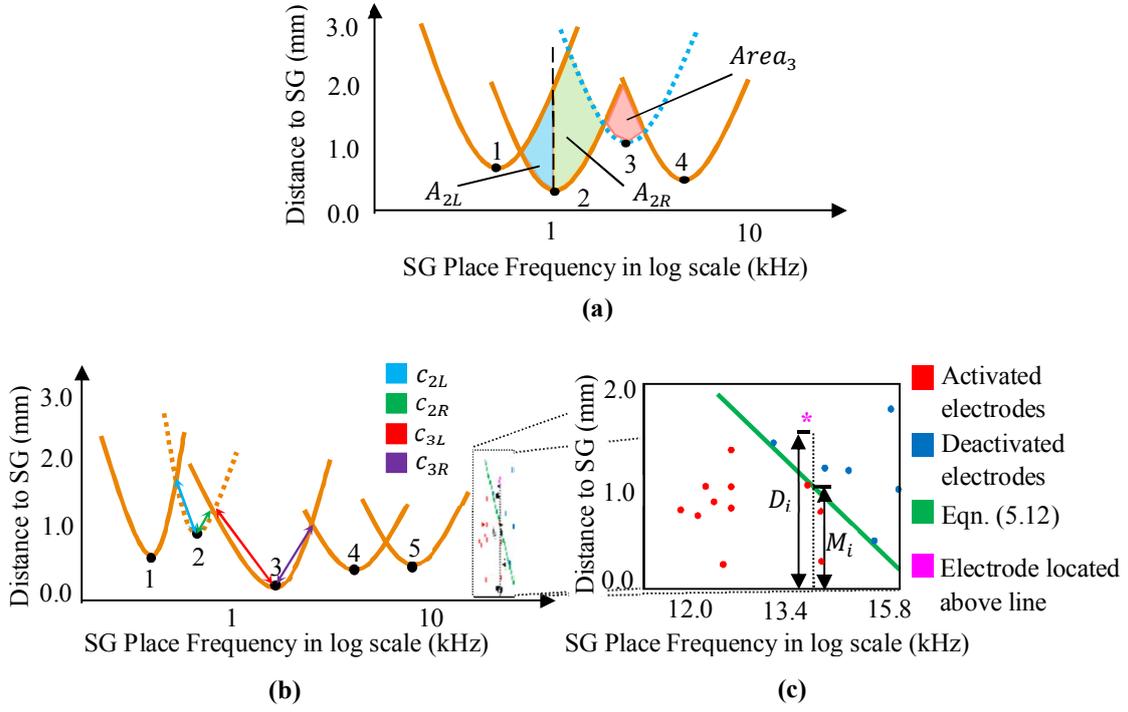

**Figure 5.3.** Visualization of three DVF-based features.

$Area_i$ is a term that captures the channel independence of electrode i by measuring the area above the DVF for electrode i and below the envelope of the other DVF curves, and $T(D_i)$ is a term that defines the value of $Area_i$ at which activating or deactivating electrode $i$ is equally desirable as a function of the distance $D_i$ between electrode $i$ and modiolus. Eqn. (5.3) is designed to assign a lower cost for activating (deactivating) electrodes with DVFs whose $Area_i$ is larger (smaller) than the threshold value $T(D_i)$. Figure 5.3a shows qualitatively the term $Area_i$ for several DVF curves. In this example, $Area_2 > Area_3$, $Area_2 > T(D_2)$, and $Area_3 < T(D_3)$, which leads to electrode 2 having a small cost for being active and 3 having a large cost for being active. This will favor configurations with electrode 2 being activated and electrode 3 being deactivated. Optimal electrode configurations will thus tend to consist of electrodes with DVF curves that have larger $Area_i$ values. To compute $Area_i$, we sum the squared distances measured between the DVF for the ith electrode and the envelope of the other DVFs at discrete positions sampled along the frequency axis. We found empirically that defining $Area_i$ as the sum of the squared distances between the curves is better than a sum of direct distances for describing expert-perceived channel overlap because the sum of squared distances is larger for DVFs that have at least some regions that lie relatively far below the envelope of the other DVFs. $T(\cdot)$ is a function that is defined using a subset of electrodes in our training dataset as follows. Figure 5.4a shows a scatter plot of Electrodes-Of-Interest (EOIs), which are a subset of electrodes from our training dataset for which the expert identified that the decision to keep them active or not was driven by the DVF area. $Area_i$ is shown on the y-axis and the electrode distance to the modiolus, $D_i$, is shown on the x-axis. As observed in the plot, the activation decision is a function of both $D_i$ and $Area_i$. This is because when $D_i$ is larger, the electrode is farther from the modiolus, and we expect wider spread of excitation. Thus we would require a greater $Area_i$ to obtain adequate channel independence and to want to keep the electrode active. Thus, we define $T(\cdot)$ as a polynomial function of modiolar distance that best separates the active and inactive EOIs from the training dataset in this plot in a least squares sense:

$$T(D_i) = -0.2660 + 1.4125 D_i + 0.5398 D_i^2, \tag{5.5}$$

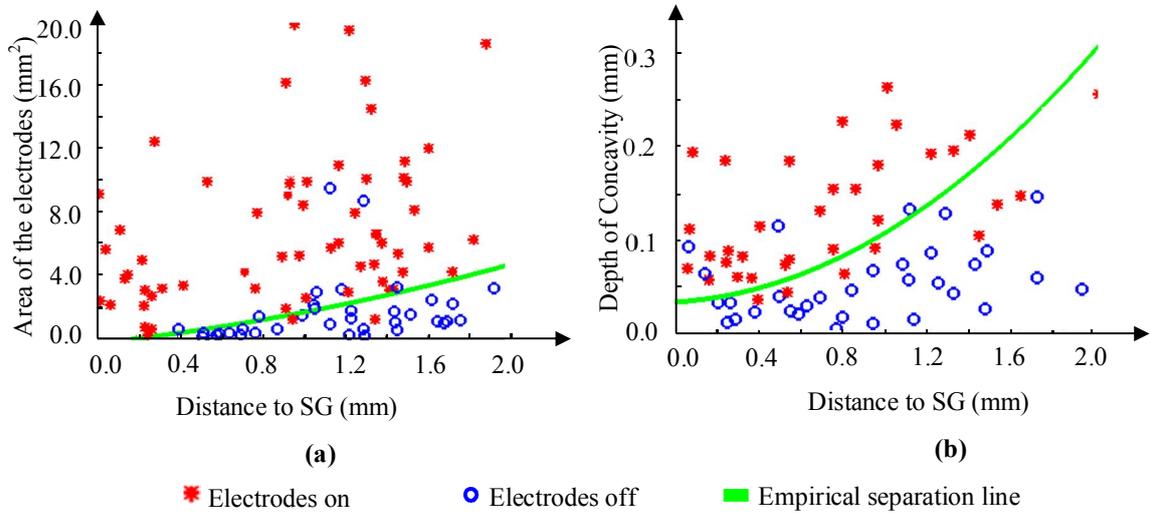

**Figure 5.4.** The visualization of (a) Area-distance and (b) Depth of concavity-distance relationship and the empirical separation line for electrodes in the training dataset.

$T_i$ is shown as the green curve in Figure 5.4a. The coefficients and the order of the polynomial function are determined with our training dataset. First, we randomly separate the EOIs into 90 training EOIs and 20 validation EOIs. Next, we investigated first order, second order and third order polynomials as candidate functions. The coefficients of each polynomial are chosen so that the polynomial best separates the active and inactive training EOIs in a least squares sense. Next, we evaluated each of the three candidate polynomials with the validation EOIs. We found that the second order polynomial correctly classified the largest number of testing EOIs. Thus, we chose to use the second order polynomial as $T(\cdot)$ and this as shown as the green curve in Figure 5.4a. Next,

$$f_4 = \sum_{i=1}^{K} e^{-Depth\_Term_i}/K, \tag{5.6}$$

where $K$ is the total number of electrodes,

$$Depth\_Term_i = \begin{cases} \dfrac{Depth_i}{R(D_i)} & \text{if electrode } i \in \text{active set} \\ \dfrac{R(D_i)}{Depth_i} & \text{if electrode } i \notin \text{active set} \end{cases}, \tag{5.7}$$

$$Depth_i = \min(C_{iL}, C_{iR}), \tag{5.8}$$

$C_{iL}$ and $C_{iR}$ are the depth of concavity of the $i^{th}$ electrode DVF relative to its left and right neighbors, $Depth_i$ is the overall depth of concavity for the curve, and $R(D_i)$ is the value of $Depth_i$ for which activating and deactivating the electrode are equally desirable as a function of the distance $D_i$ between electrode $i$ and modiolus. Eqn. (5.7) is designed to assign a lower cost for activating (deactivating) electrodes with DVFs whose depth of concavity $Depth_i$ is larger (smaller) than the threshold value $R(D_i)$. This term captures the property that optimal configurations consist of electrodes whose DVFs have large depth of concavity. Figure 5.3b shows an example of the depth of concavity measurement. In this example, $Depth_2 = c_{2R} < R(D_2)$, $Depth_3 = c_{3L} > R(D_3)$, which leads to a large cost for activating electrode 2 and a small cost for activating electrode 3. This will favor solutions in which electrode 3 is activated and electrode 2 is deactivated. $R(\cdot)$ is a polynomial function that is defined using a subset of electrodes selected from our training dataset in a manner identical to $T$ as:

$$R(D_i) = 0.0328 + 0.005D_i + 0.0351D_i^2, \tag{5.9}$$

Figure 5.4b shows a scatter plot of EOIs in our training dataset for which the expert

decision to keep them active or not was driven by the depth of concavity. $Depth_i$ is shown on the y-axis, the electrode distance to the modiolus $D_i$ is shown on the x-axis, and $R$ is shown in green. As observed in the plot, the activation decision is a function of both electrode distance $D_i$ and $Depth_i$. This is because, similarly to Eqn. (5.4) above, when $D_i$ is larger and we expect wider spread of excitation, we would require larger $Depth_i$ to indicate adequate channel independence and to keep the electrode active. Next,

$$f_5 = \sum_{i=1}^{K}(D_i - M_i)\, u(D_i - M_i), \qquad (5.10)$$

where $D_i$ is the distance from the electrode $i$ to the modiolus, $M_i$ is a linear function defined as:

$$M_i = -17.29 \log_{10}(Freq_i) + 72.81, \qquad (5.11)$$

in which $Freq_i$ is the place frequency of the nerves closest to electrode i, and $u(\cdot)$ is the unit step function. $f_5$ is designed to assign a cost to electrodes that fall above the line defined by Eqn. (5.11). This line is shown in Figure 5.3c, which shows a small, zoomed in portion of the plot shown in Figure 5.3b. Since the line is steep, electrodes located above it are located in the very high frequency region (>13 kHz) near the entrance of the cochlea. These electrodes are often deactivated clinically because they are outside or nearly outside the cochlea or provide abnormal perception due to loss of neural survival that is common in this region. Thus, $f_5$ is used to indicate that electrodes in this region are less desirable. As shown in Figure 5.3c, Eqn. (5.12) was designed by finding the least squares fit line that separates the groups of electrodes in the training electrode configurations that were set as activated (red) and deactivated (blue). Also shown are distances $D_i$ and $M_i$ for one electrode (magenta). Next,

$$f_6 = K_I/K_a, \tag{5.12}$$

where $K_a$ is the number of active electrodes in the configuration, and $K_I$ is the number of DVFs that have a minimum that falls above the envelope of other electrodes' DVFs (see electrode 4 in Figure 5.2a). When this term is larger than 0, it is a strong indicator that one or more electrodes is stimulating the same frequency range as other electrodes but less effectively since it is located further away from the modiolar surface. Next,

$$f_7 = \left(\sum_{i=1}^{K_s} e^{-ratio_{S(i)}}\right)/K_s, \tag{5.13}$$

where $S$ is the set of $K_s$ active electrodes that have active neighbors on both the left and right side,

$$ratio_i = \min(A_{iL},\ A_{iR})/\max(A_{iL},\ A_{iR}), \tag{5.14}$$

and $A_{iL}$ and $A_{iR}$ indicate the left and right half area terms of the DVF curve of electrode $i$ (see Figure 5.4a). $A_{iL}$ and $A_{iR}$ are defined as the sum of the distances measured between the DVF curve for the $i^{\text{th}}$ electrode and the envelope of the other DVFs at the discrete positions sampled along the frequency axis to the left and right of the minimum, respectively. Eqn. (5.14) assigns a low cost to the configurations with symmetric DVFs, and a high cost to the configurations with one or more highly asymmetric DVFs. Finally, $f_8 = \sqrt{f_3}$, $f_9 = \sqrt{f_4}$, and $f_{10} = \sqrt{f_7}$. These terms were included after testing all combinations of squares and square roots of $f_{1-7}$ and finding that including these terms led to better results.

A linear combination of the cost terms is used to define an overall cost function for a given configuration, i.e.,

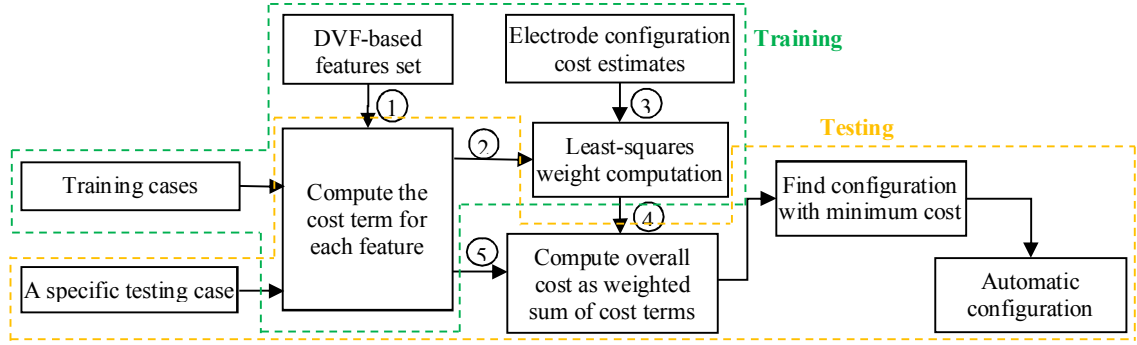

**Figure 5.5** The workflow of the automatic electrode configuration selection method

$$C = \sum_{i=1}^{N} w_i f_i, \qquad (5.15)$$

Because current electrode arrays have ~22 electrodes, it is practical to find the globally optimal configuration through an exhaustive search that evaluates all possible configurations. The values for the set of scalar coefficients $\{w_i\}_{i=1}^{N}$ used to weigh each of the cost terms in Eqn. (5.16) are estimated using a training set of existing manually selected electrode configurations and a least-squares technique.

Our methods are summarized in Figure 5.5. As can be seen in the figure, there is a training stage and a testing stage. The training stage is used to determine the parameter (weight) values $\{w_i\}_{i=1}^{N}$ that control the contribution of each feature term in the overall cost function. Input 1 is the DVF-based feature set. Using this feature set, a cost term is computed for each feature for all the possible electrode configurations in the set of training cases. The resulting cost terms (Output 2) are passed to the least-squares solver, which solves equations of the form:

$$\left\{ \sum_{i=1}^{N} f_i^{m,o} w_i + \delta = C^{m,o} \right\}_{m=1, o=1}^{M,O}, \qquad (5.16)$$

where $\{f_i^{m,o}\}$ is the set of $N$ cost terms for each of the $M$ electrode configurations for the $O$ subjects in our training dataset, $\{C^{m,o}\}$ is the set of cost estimates for each configuration, and $\delta$ is a constant. We compute $\{C^{m,o}\}$ using a piecewise function defined as

$$C^{m,o} = \begin{cases} 0 & e_{m,o} = e_{opt,o} \\ \frac{1}{2} & e_{m,o} \in \{e_{acc,o}\}, \\ \text{dist}(e_{m,o}, e_{opt,o}) & \text{otherwise} \end{cases} \quad (5.17)$$

where $e_{opt,o}$ is the electrode configuration chosen manually by an expert for the $o^{th}$ subject, $\{e_{acc,o}\}$ is a set of other electrode configurations that were identified by the expert as being acceptable for the $o^{th}$ subject, and $\text{dist}(e_{m,o}, e_{opt,o})$ is an electrode configuration distance metric we have defined on all other electrode configurations. $\text{dist}(\cdot, \cdot)$ needs to capture the difference in quality between configurations and is thus a critical element of our method. A straightforward approach would be to use the hamming distance between the electrode configurations. However, we found this to be sub-optimal as certain configuration patterns, such as on-off-on-off vs off-on-off-on would be assigned the highest possible distance value even though this often does not lead to very different stimulation patterns. To address this issue $\text{dist}(e_{m,o}, e_{opt,o})$ is computed in this work in two steps as shown in Figure 5.6: (1) The activation status of each electrode in $e_{m,o}$ is compared with the corresponding electrode in $e_{opt,o}$. For each $j$th electrode $e_{m,o,j}$ in $e_{m,o}$ that does not match $e_{opt,o,j}$, we compute the distance, in terms of the number of electrodes, to the nearest electrode in $e_{opt,o}$ that does match $e_{m,o,j}$. This results in an array of distances, $\vec{d} = \{d_j\}$, where $d_j = |j - k|$ is the distance from $e_{m,o,j}$ to $e_{opt,o,k}$, the closest electrode in $e_{opt,o}$ that matches $e_{m,o,j}$. (2) We then compute $\text{dist}(e_{m,o}, e_{opt,o})$ as the sum of the local maxima in $\vec{d}$. This metric is designed to assign a higher cost to configurations that have more distant mismatches, which indicates greater disagreement with the optimal configuration. In summary, our approach assigns higher values to $C^{m,o}$ for less desirable electrode configurations and lower values to $C^{m,o}$ for more desirable electrode configurations.

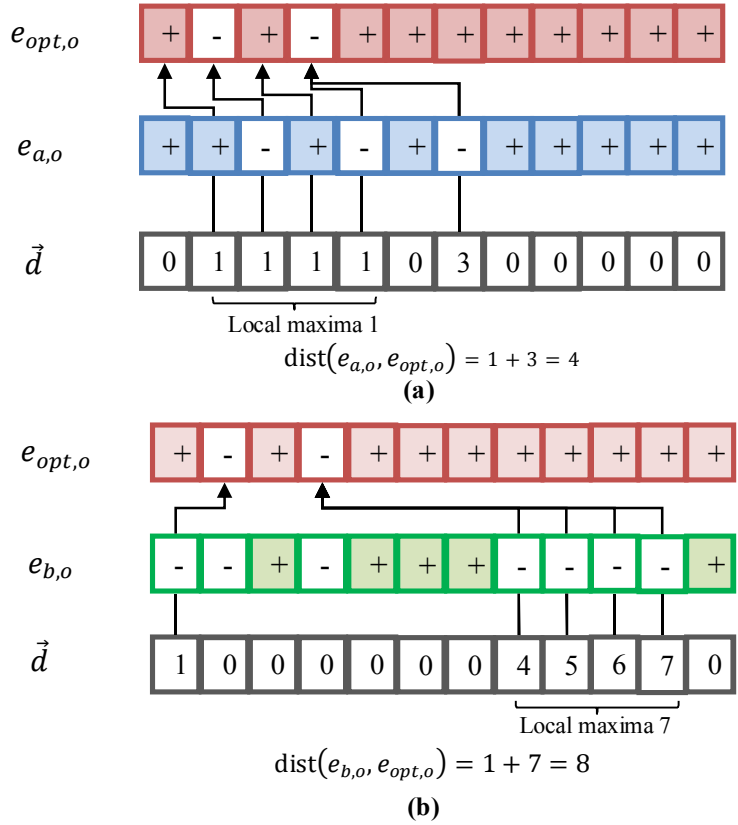

**Figure 5.6.** The distance metric between electrode configuration patterns (Marker $+$: Electrodes activated; Marker $-$: Electrodes deactivated). Both configuration have 5 differences in the electrode activation patterns. With the optimal distance metric, configuration $e_{b,o}$ is assigned with larger distance compared to configuration $e_{a,o}$ to the optimal configuration $e_{opt,o}$.

The set of weights $\{w_i\}_{i=1}^{N}$ can be determined by solving Eqn. (5.16) once offline using a constrained least-squares linear system solver in MATLAB 2014b (Mathworks, Inc. Natick, MA), with the constraint $w_i \geq 0 \ \forall \ i = [1, N]$. This constraint represents an additional piece of a priori knowledge that captures the fact that the cost function should increase when feature terms increase since, as designed, the value of the features increases for less desirable electrode configurations. We have found that this constraint leads to better results. Once the weights are defined by using the training dataset, the optimal electrode configuration for a new subject is determined automatically by finding the global minimum of the cost function through an exhaustive search.

We performed a validation study to show the robustness of our method. To evaluate our method on our testing dataset, we asked two electrode configuration selection experts (JHN and YZ) who currently verify all the configurations used in our clinical studies to perform a blinded and randomized evaluation of the automatic configurations against control configurations. To do this, for the 60 subjects in our testing dataset, we generated three sets of electrode configurations: Manual, automatic, and control electrode configurations. The manual electrode configurations were manually selected by JHN and have been implemented in patients in our previous clinical research studies. The automatic electrode configurations were generated by running our proposed method on the subjects in testing dataset. Control electrode configurations were constructed for each subject in the testing set by the experts by manually selecting a configuration that is not "acceptable" but "close" to acceptable for all testing subjects. An electrode configuration is judged as "acceptable" when the expert believes it can be used for CI programming and is likely to lead to hearing outcomes that are nearly as good as those that would be achieved using the best possible configuration. For each test subject, two tests were done in which each expert was presented with a pair of electrode configurations and asked to rank them in terms of quality and rate whether each configuration was acceptable. In one test, the pair of configurations consists of the automatic and manual plan. In the other test, the control and the manual plan are ranked and rated. The ordering of all tests across all test subjects was randomized and the expert was masked to the identity of each configuration. The control configurations used for tests with one expert were generated by the other expert. Masking the identity of all the configurations, including control configurations, and randomizing the order of tests were steps done to minimize the potential for the experts to be biased towards evaluating all configurations as acceptable and so that the presence of such a bias could be detected in the results. Rating a significant portion of the control plans as acceptable would be indicative of such a bias. Two experts were included so that inter-rater variability could be characterized.

Table 5.1. The feature cost terms generated for Med-El, Advanced Bionics, and Cochlear arrays

| Cost Terms | Expert Heuristics | MedEl | Advanced Bionics | Cochlear |
|---|---|---|---|---|
| 1 | Activate the most apical electrode | 0.68 | 0.28 | 1.09 |
| 2 | Active as many electrodes | 0 | $1.55 \times 10^{-9}$ | $1.62 \times 10^{-4}$ |
| 3 | Activate electrodes with large area terms in DVFs | $3.72 \times 10^{-3}$ | $8.89 \times 10^{-5}$ | $8.95 \times 10^{-4}$ |
| 4 | Active electrodes with large depth of concavities in DVFs | 0 | $6.02 \times 10^{-9}$ | $5.09 \times 10^{-5}$ |
| 5 | Deactivate electrodes outside of cochlea | $3.23 \times 10^{-4}$ | $8.32 \times 10^{-2}$ | $4.89 \times 10^{-4}$ |
| 6 | Deactivate electrodes with minimum of DVFs above others | 2.28 | 1.37 | 5.43 |
| 7 | Tends to activate electrodes with symmetric DVFs | 0 | $6.75 \times 10^{-10}$ | $4.75 \times 10^{-5}$ |
| 8 | Square root of term 3 | 0 | $3.58 \times 10^{-9}$ | $5.55 \times 10^{-5}$ |
| 9 | Square root of term 4 | 0 | $5.07 \times 10^{-10}$ | $4.98 \times 10^{-5}$ |
| 10 | Square root of term 7 | 0 | $1.18 \times 10^{-9}$ | $4.80 \times 10^{-5}$ |

## 5.3   Results

The parameter training process was implemented in MATLAB (Mathworks Inc., Natick, MA) and the electrode configuration selection algorithm was implemented in C++. The training process is an offline process, which generates the feature cost term weights in 1 minute, 4 minutes, and 7 hours 40 minutes on a standard Windows Server PC (Intel (R), Xeon (R) CPU X5570, 2.93GHz, 48GB RAM) for MD, AB, and CO arrays, respectively. The electrode configuration selection algorithm required 15 seconds, 30 seconds, and 2 minutes for MD, AB, and CO arrays, respectively. Compared to the manual selection done by expert (requires 0.5-2 minutes), our automatic electrode configuration selection algorithm is comparable but does not require any specialized training. The feature weight values computed for the MD, AB, and CO arrays from the training dataset are shown in Table 5.1. As can be seen from the table, the feature that prevents deactivating the most apical electrode ($f_1$) and one of the channel interaction features ($f_6$) were assigned the highest weight values for all three types of implants. For the AB and the CO arrays, the other feature that was assigned a high weight value is the term punishing electrodes falling around the entrance of the cochlea ($f_5$). The other features were assigned relatively low weight values. For Med-El, the term punishing activating electrodes that fall around the entrance of the cochlea ($f_5$) and the term favoring a large area for each DVF curve ($f_3$) were assigned moderately high weight values. The remaining features were assigned weights with very low magnitude ($\leq 10^{-13}$). In experiments on the MD training set we found that removing the features that were assigned the very low weights produced identical electrode configurations. This confirms that the features with low magnitude weights ($\leq 10^{-13}$) do not play a significant role in achieving the best results and can be ignored. Thus, for MD, we only kept $f_1$, $f_3$, $f_5$ and $f_6$ and remove the other features by setting their weights as 0.

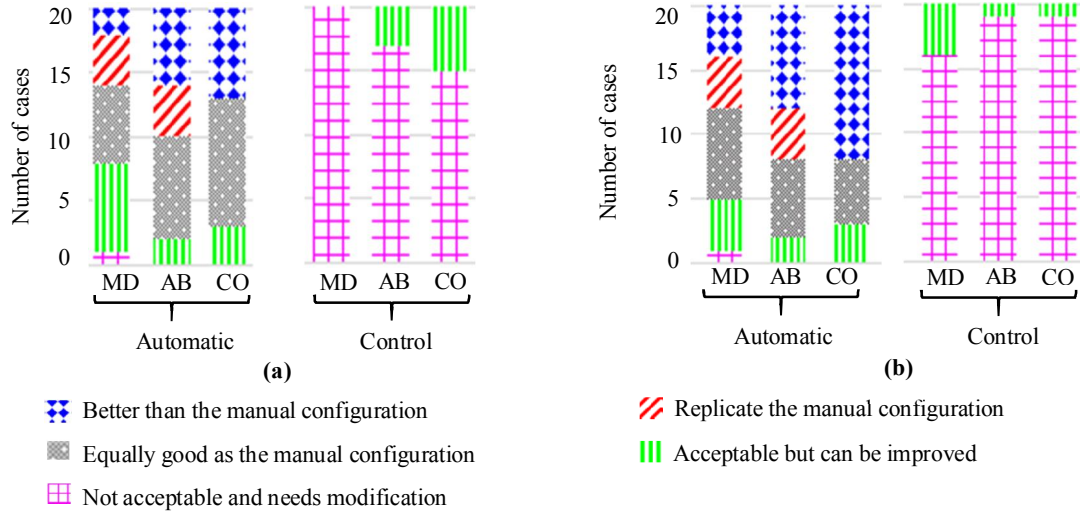

**Figure 5.7.** Validation study results. Panel (a) and (b) visualize the results of validation studies performed by expert 1(JHN) and expert 2 (YZ) on automatic and control electrode configurations, respectively.

The results of our validation study on our testing set are shown in Figure 5.7. As can be seen from Figure 5.7a, according to expert 1 (JHN), across the 60 subjects in our testing dataset, 14 of the automatically generated electrode configurations were found to be better than the manually selected configuration. In these tests, the manual configuration was found to be acceptable, but not as good as the automatic configuration. In the remaining tests, 33 automatic configurations were found to be equivalent to or exactly the same as the manual configurations, 12 were found to be not as good as the manual configuration but still acceptable, and only 1 was found to be not acceptable. None of the control configurations was evaluated as equivalent to or better than the manual configuration, 8 were evaluated as acceptable, and 52 were evaluated as not acceptable. For expert 2 (YZ), 24 automatic configurations were found to be better than the manual configurations, 26 were found to be equivalent to or exactly the same as the manual configurations, 9 were found to be acceptable, and only 1 was evaluated as not acceptable. The same automatic case was rated as not acceptable by both experts. None of the control configurations was rated as equivalent to or better than manual configurations, only 6 were evaluated as acceptable, and the remaining 54 were rated as not acceptable. These results show that, with the exception of one unacceptable result, our method performs similarly to an expert. On average, the automatic method slightly outperforms an expert since more automatic plans are ranked better than manual plans than vice versa. Two-tailed paired-sign tests were used to compare the acceptance rate for control vs. automatic plans and showed that the rate at which the automatic plans are judged to be acceptable was significantly better for both expert 1 ($p = 1 \times 10^{-15}$) and expert 2 ($p = 1 \times 10^{-16}$). No statistically significant differences were found when comparing ratings of the automatic electrode configurations across the two raters ($p = 1$).

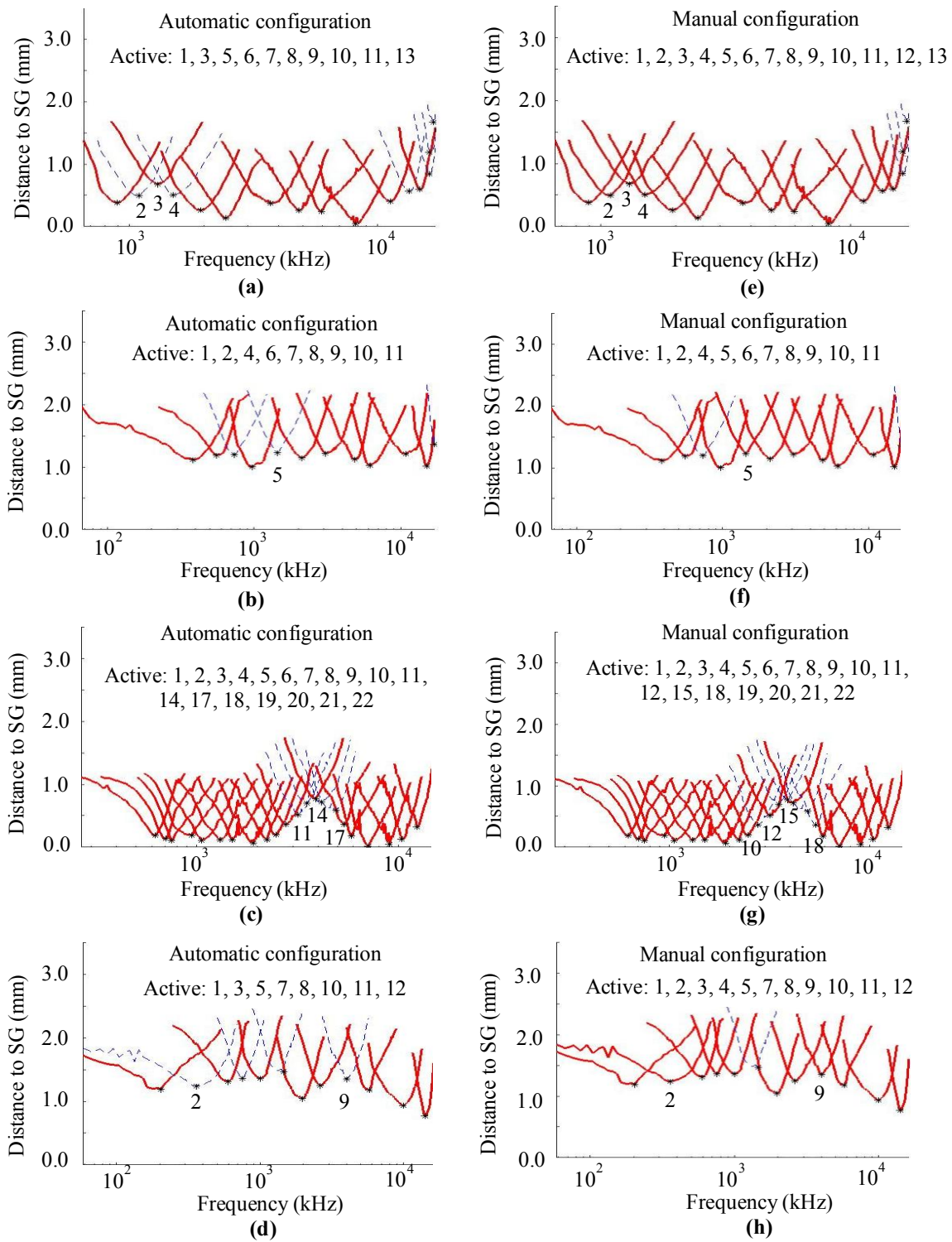

**Figure 5.8.** Visualization of automatically selected (a-d) and corresponding manual (e-h) electrode configurations for several cases. An automatic AB plan that was judged as better than the manual plan is shown in (a). An automatic MD plans judged to be equivalently good are shown in (b). An automatic CO plans judged as acceptable is shown in (c). An automatic MD result that was judged as not acceptable is shown in (d).

In Figure 5.8, we show the DVFs for automatically determined electrode configurations for several cases. The blue dotted curves represent DVFs for electrodes that are removed from the active electrode configuration, and the red solid curves represent DVFs for electrodes that are active. The electrode numbers are in increasing order from the left to the right. To facilitate interpretation, we label the electrodes of interest in the figure. In Figure 5.8a, a result for an AB case is shown that is identified as better than the manual configuration because the 2nd and the 4th electrodes are deactivated in the automatically generated configuration. Deactivating those electrodes is good because they are likely to interfere with electrode 3. Figure 5.8b presents a result for a MD case that is identified as equivalent to the manual configuration. The automatic plan deactivates electrode 5 while the manual plan keeps it. The plans are judged to be equivalent because it is hypothesized that reducing channel interaction artifacts by turning off electrodes will be offset by an increase in frequency compression artifacts resulting in equivalent outcomes. Figure 5.8c presents a result for a CO case that is judged to be not optimal compared with the manual configuration but still acceptable. The 11-14-17 configuration in the automatic plan is not as good as the 10-12-15-18 because the minimum of electrode 11 and 17 in the automatic plan are very close to the curves of the neighbor electrodes 10 and 18. Thus, the 11-14-17 configuration in the automatic plan does not adequately address the channel interaction problem between electrodes 10 and 11 and electrodes 17 and 18. Figure 5.8d presents the only automatic configuration for a MD case that is judged to be not acceptable. In Figure 5.8d, the automatic configuration deactivates electrode 2 and 9. This is not desirable because of the relatively large distances between electrodes 1 and 3 and 8 and 10. This plan is likely to cause frequency compression artifacts.

## 5.4 Conclusions

In this study, we propose the first approach for automatic selection of electrode configurations for image-guided cochlear implant programming. This is a crucial step towards clinical translation of our image-guided cochlear implant programming system that has been shown in clinical studies to lead to significant improvement in outcomes. Our approach is to design a DVF-feature-based cost function and to train its parameters using existing electrode configuration plans that we have accumulated in our database. Our validation study has shown that our method generalizes well on a large-scale testing dataset and can produce acceptable electrode configurations in the vast majority of cases. In the validation tests with implant models from the 3 major CI manufacturers, our automatic method produces acceptable configurations for 98.3% of the arrays tested. According to the evaluation results from two experts in our group, around 83% of the configurations produced by our automatic method were ranked as at least equivalent to the manual configurations. Around 33% of the configurations produced by our automatic method were ranked as better than the manual configurations, wheras only 17% of the manual configurations were ranked as better than the automatic. These results suggest that our method is a viable approach for automatically selecting electrode configurations for image-guided cochlear implant programming with similar performance to a trained expert. While the best approach to assess our IGCIP system would be to analyze a collection of hearing outcomes data from CI recipients before and after using IGCIP with the automatic and the manual electrode configuration selection methods, such data is difficult to obtain. This is so because it would require subjects to come back once for re-programming and again to re-evaluate outcomes 3-6 weeks after re-programing. In the future, we plan to perform such a study with a limited number of recipients who live in close proximity to our institution. While our method generates acceptable configurations for the vast majority of cases tested, it is still capable of producing unacceptable configurations. Thus, in future work we will investigate developing an automatic method to evaluate the quality of the electrode configuration generated by our method. This would enable our IGCIP system to notify the user that expert intervention might be needed to select the electrode configuration when our automatic method fails.

Chapter VI

# VALIDATION OF IMAGE-GUIDED COCHLEAR IMPLANT PROGRAMMING TECHNIQUES

Yiyuan Zhao[1], Jianing Wang[1], Rui Li[1], Robert F. Labadie[2], Benoit M. Dawant[1], and Jack H. Noble[1]

[1]Department of Electrical Engineering and Computer Science, Vanderbilt University, Nashville, TN, 37232, USA

[2]Department of Otolaryngology – Head & Neck Surgery, Vanderbilt University, Nashville, TN, 37232, USA

Abstract

Cochlear implants (CIs) are a standard treatment for patients who experience severe to profound hearing loss. Recent studies have shown that hearing outcome is correlated with the intra-cochlear locations of CI electrodes. Our group has developed image-guided CI programming (IGCIP) techniques that use image analysis techniques to analyze the patient-specific intra-cochlear locations of the implanted CI electrodes to assist audiologist with CI programming by selecting a subset of active electrodes. The image analysis techniques in IGCIP include the identification electrode locations in post-implantation CTs, and the segmentation of intra-cochlear anatomy in pre- and post-implantation CTs. Clinical studies have shown that IGCIP can improve hearing outcomes for CI recipients. However, the sensitivity of IGCIP with respect to the accuracy of the two major steps, electrode localization and intra-cochlear anatomy segmentation, is unknown. In this article, we create a ground truth dataset by using conventional and μCT pairs of 35 temporal bone specimens to rigorously characterize the accuracy of these two steps and then use those dataset for IGCIP sensitivity analyses. The validation study results show that with pre- and post-implantation CTs available, IGCIP can generate acceptable active electrode sets in 86.7% of the subjects tested. With only post-implantation CTs available, IGCIP can generate acceptable active electrode sets in 83.3% of the subjects tested.

6.1. Introduction

Cochlear implants (CIs) are neural prosthetic devices that are the standard of care treatment for patients experiencing severe to profound hearing loss [1]. The external components of a CI device include a microphone, a signal processor, and a signal transmitter, which are used to receive and process sounds, and send signals to implanted CI electrodes. The major internal component is the implanted CI electrode array. The implanted CI electrodes bypass the damaged cochlea and directly stimulate the auditory nerves to induce a sense of hearing for the recipient. During CI surgery, a surgeon threads a CI electrode array into a recipient's cochlea. After the surgery, an audiologist needs to program the CI device which includes determining a series of CI instructions. The programming procedure involves specifying the stimulation levels for each electrode based on the recipient's perceived loudness, and the selection of a frequency allocation table, which determines which electrode is to be activated when a specific frequency is detected in the incoming sound [2]. CIs lead to remarkable success in hearing restoration among the majority of recipients [3-4]. However, there are still a significant number of CI recipients experiencing only marginal benefit.

Recent studies have indicated that hearing outcomes with CI devices are correlated with the intra-cochlear locations of CI electrodes [5-10]. As the electrode array is blindly inserted by a surgeon, the intra-cochlear locations of CI electrodes are generally unknown. Thus, audiologists do not have information about locations of CI electrodes with respect to the auditory nerves. In the traditional CI programming procedure, the audiologist assumes the electrodes are optimally situated and selects a default frequency allocation table. This leads to an artifact named "electrode interaction" [11-12], as shown in Figure 6.1 as overlapping stimulation of electrodes. Electrode interaction occurs when multiple CI electrodes are stimulating the same group of auditory nerves. In natural hearing, a specific group of nerves are activated in response to a specific frequency band. In a CI-assisted hearing process with electrode interaction, the same nerve group is activated in response to multiple frequency bands, which is thought to create spectral smearing and negatively affect hearing outcomes. It is possible to alleviate the negative effect of electrode interaction, by selecting a subset of the available electrodes to keep active, aka the "electrode configuration", that do not have overlapping stimulation patterns. However, without the benefit of knowing the spatial relationship between the electrodes and the auditory neural sites, selecting such an electrode configuration is not possible and audiologists typically leave active all available electrodes.

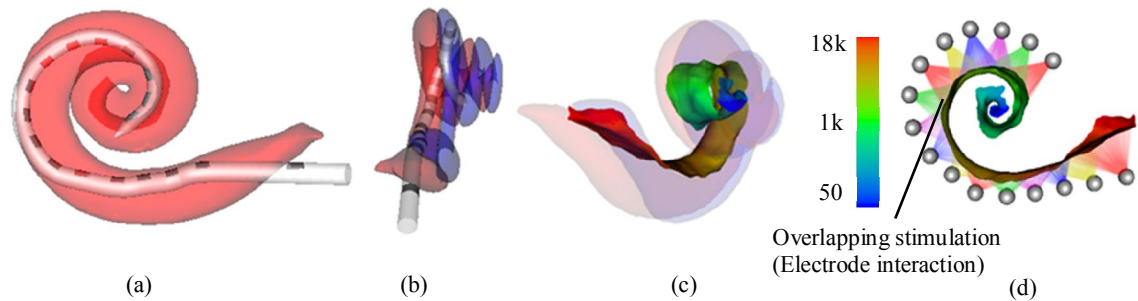

(a)          (b)          (c)          (d)

**Figure 6.1.** Panels (a) and (b) show a CI electrode array superimposed on the scala tympani (red) and scala vestibuli (blue) cavities of the cochlea in posterior-to-anterior and lateral-to-medial views, respectively. Panel (c) shows the scalae and neural activation region color-coded by place frequency in Hz. Panel (d) illustrates overlapping stimulation patterns (electrode interaction) from the implanted electrodes as they stimulate neural regions.

Our group has been developing an image-guided cochlear implant programming (IGCIP) system [2], which uses image analysis techniques to assist audiologists with electrode interaction analysis and electrode configuration selection [18, 24] during the CI programming procedure. Figure 6.2 shows the workflow of IGCIP. We use whole head computed tomography (CT) images of CI recipients as input for IGCIP. For recipients having both pre- and post-implantation CTs, we firstly use a mutual information-based method to register the pre-implantation CT with a reference CT, where the intra-cochlear anatomy could be segmented by using [13]. In the post-implantation CT, the locations of electrodes can be identified by using [14] or [15]. Then, we register the pre- and post-implantation CTs together to analyze the possibility for electrode interactions. For recipients that do not have pre-implantation CTs, we developed two methods [16] and [17] that can segment the intra-cochlear anatomy directly from post-implantation CTs. After segmenting the intra-cochlear anatomy using one of these techniques, we localize the electrodes in the same post-implantation CTs by using [14] or [15] and then proceed to electrode interaction analysis process. To analyze the electrode interactions, our group has develop a technique named distance-vs.-frequency curves (DVFs). The DVF is a 2D plot for facilitating the visualization of electrode interaction in individuals. It captures the patient-specific spatial relationship between the electrodes and the auditory nerves [2], as shown in Figure 6.2. The DVFs show the distance from each electrodes to neural stimulation sites along the length of the cochlea. Based on the DVFs, our group has developed an automatic electrode configuration selection method [18] to select a subset of active electrodes that have reduced electrode interaction. Recent clinical studies we performed indicated that by using our IGCIP-generated electrode configuration, hearing outcomes can be significantly improved [19-21]. The electrode configuration generated by IGCIP is affected by the accuracy of the anatomy and electrode segmentation techniques. To better understand the limitations of IGCIP, in this work, we rigorously characterize the accuracy of the electrode localization and intra-cochlear anatomy segmentation procedures. These results enable determining which automatic processes are the most accurate, and thus the most preferable, and enable the evaluation of the sensitivity of IGCIP with respect to the automatic image processing techniques.

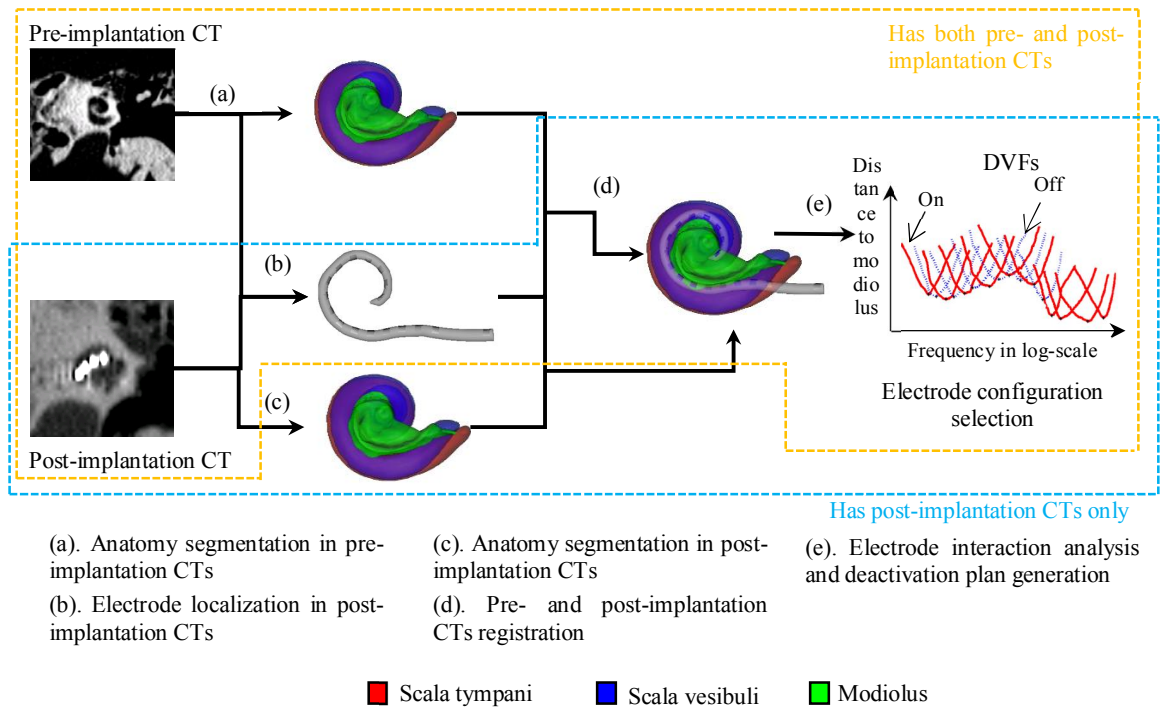

**Figure 6.2.** Workflow of Image-guided cochlear implant programming (IGCIP) techniques.

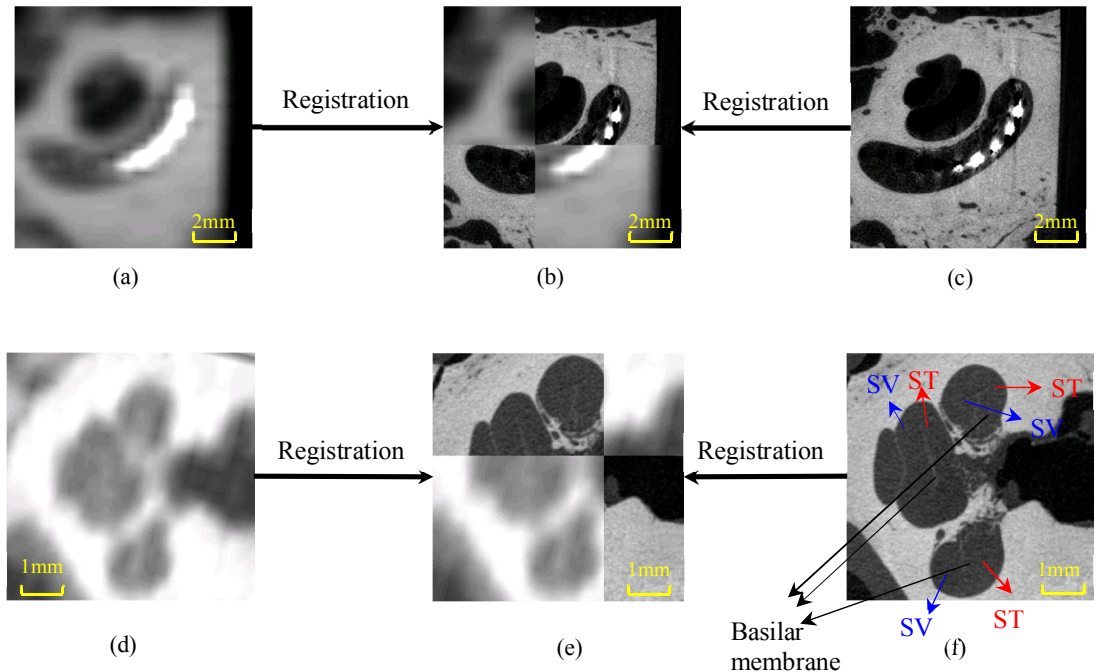

**Figure 6.3.** Panels a-c show three post-implantation CTs: a conventional CT (a), the registered µCT (c), and a checkerboard combination of the two (b). As can be seen, electrodes are more separable in the µCT because of the higher resolution and less partial volume artifacts. Panels d-f show three pre-implantation CTs: a conventional CT (d), the registered µCT (f), and a checkerboard combination of the two (e). As can be seen in panel (f) and (d), the basilar membrane is visible in µCTs but not visible in clinical CTs. This makes it possible for generating ground truth anatomy segmentation results for ST and SV, and then MOD.

The electrode localization method being evaluated in this study is a graph-based path-finding algorithm [14]. We refer to this method as $M_E$ in the remainder of this article. In post-implantation CTs, the CI electrodes appear as high intensity voxel groups, as shown in Figure 6.3. $M_E$ firstly extracts the volume of interest (VOI) that contains the cochlea by using a reference image. Next, it generates candidates of interest (COIs) that represent the potential locations of electrodes. The COIs are used as nodes in a graph for the following path-finding algorithms. Then, it uses path-finding algorithms to find a path constructed by a subset of COIs representing the centroids of CI electrodes on the array. The intra-cochlear anatomy segmentation step in IGCIP focuses on the segmentation of three anatomical structures in cochlea: scala tympani (ST), scala vestibuli (SV), and the active region (AR) of the modiolus (MOD). ST and SV are the two principal cavities of the cochlea. The MOD is the anatomical region housing the auditory nerves. AR is the interface between the MOD and the union of the ST and SV. The auditory nerves stimulated by the electrodes are located in immediate proximity to AR within MOD. In conventional clinical pre-implantation CTs, the basilar membrane that separates ST and SV is not visible, as shown in Figure 6.3d, which makes the segmentation of the intra-cochlear anatomy difficult. When pre-implantation CTs are not available, the segmentation of intra-cochlear anatomy becomes more difficult. This is because in post-implantation CTs, the artifacts caused by metallic electrodes obscure the anatomy structures. Thus, for intra-cochlear anatomy segmentations in both pre- and post-implantation CTs, our group had proposed three automatic methods: (1) a statistical shape model-based method [13], (2) a library-based method [16], and (3) a method [17] based on the Conditional Generative Adversarial Network (cGAN) [18]. We refer to them as $M_{A1}$, $M_{A2}$, and $M_{A3}$, respectively.

$M_{A1}$ is used on pre-implantation CTs if available. In $M_{A1}$, we create an active shape model for ST, SV, and MOD by using the manually delineated anatomical surfaces from 9 high resolution μCTs [13]. Then, the model is fit to the partial structures that are available in conventional CTs, and used to estimate the position of structures not visible in these CTs. When pre-implantation CTs are not available, we apply $M_{A2}$ or $M_{A3}$ directly to post-implantation CTs for intra-cochlear anatomy segmentation. $M_{A2}$ leverages a library of shapes of cochlear labyrinth and intra-cochlear anatomy. Given a target post-implantation CT, first, $M_{A2}$ segments the portions of the cochlear labyrinth that are not typically affected by image artifacts. Then, it selects a subset of labyrinth shapes from the library based on the similarity of the regions not affected by the artifacts. Using this subset of shapes, the method builds a weighted active shape model (wASM) of the cochlear labyrinth to localize the labyrinth in the target image. Then weights of the vertices that are close to (or distant to) the image artifacts are assigned 0 (or 1), respectively. Last, it uses another pre-defined active shape model of ST, SV, and MOD to segment the intra-cochlear anatomy based on the localized labyrinth. $M_{A3}$ uses a cGAN [18] to translate the given post-implantation CT, in which the intra-cochlear anatomy is corrupted by artifacts, to a synthesized pre-implantation CT in which the artifacts are removed. Then on the recovered pre-implantation CT image, we apply $M_{A1}$ to generate the ST, SV and MOD surfaces.

As has been discussed above, to analyze the accuracy of IGCIP, we need to rigorously characterize the accuracy of the automatic image processing techniques. In previous studies, $M_E$, $M_{A2}$, and $M_{A3}$ have only been validated by using reference segmentation results on conventional CTs that have limited accuracy. In [14], to evaluate the accuracy of $M_E$, we used a set of manual localization results generated by an expert on post-implantation clinical CTs. The clinical CTs have a limited resolution (the typical voxel size is 0.2×0.2×0.3mm$^3$). When localizing small-sized objects such as CI electrodes (typical size is 0.3×0.3×0.1mm$^3$), partial volume artifacts (see Figure 6.3a) in clinical CTs limit the accuracy of the localization, even with care and expertise. Other image quality issues, such as the beam hardening artifacts, also complicate localizing CI electrodes. In previous studies for intra-cochlear anatomy segmentation, $M_{A2}$ and $M_{A3}$ were only validated by using reference anatomical structures generated by $M_{A1}$ on corresponding pre-implantation CTs. These limited reference segmentations used in prior studies could only be as accurate as the conventional CT images on which they were defined.

In this article, we create a high accuracy ground truth dataset using μCT imaging to rigorously evaluate the accuracy of our automatic techniques in IGCIP and the sensitivity of IGCIP with respect to them. In Section 2, we describe the creation of the ground truth dataset and the design of the validation approaches. In Section 3, we present and analyze the validation results. In Section 4, we summarize the contribution of this work and discuss potential improvements for the IGCIP process.

## 6.2. Methods

### 6.2.1. Image data

Our image data consists of CTs and μCTs of 35 temporal bone specimens implanted with 4 different types of CI electrode arrays by an experienced otologist. The detailed specifications of the 35 specimens are shown in Table 6.1. Among the 35 specimens, 20 (Specimen 16 to 35 in Table 6.1) were implanted with an array type that our electrode localization method had been trained to localize, and the remaining 15 were implanted with three other array types (5 specimens each, Specimen 1 to 15 in Table 6.1) on which our method was not trained. Every specimen underwent pre- and post-implantation CT imaging and post-implantation μCT imaging. Six specimens underwent pre-implantation μCT imaging (Specimen 30 to 35). The typical voxel size for CT images and μCT images are $0.30 \times 0.30 \times 0.30$mm³ and $0.02 \times 0.02 \times 0.02$mm³, respectively.

**Table 6.1.** The specifications of the CT images of the 35 temporal bone specimens

| # | Conventional CT voxel size (mm$^2$) | | μCT voxel size (mm$^2$) | | Migration | Data Group # |
|---|---|---|---|---|---|---|
| | Pre-op CT | Post-op CT | Pre-op CT | Post-op CT | | |
| 1 | 0.26 × 0.26 × 0.30 | 0.26 × 0.26 × 0.30 | | 0.02 × 0.02 × 0.02 | | 1,3 |
| 2 | 0.28 × 0.28 × 0.30 | 0.24 × 0.24 × 0.30 | | 0.02 × 0.02 × 0.02 | | 1,3 |
| 3 | 0.30 × 0.30 × 0.30 | 0.34 × 0.34 × 0.30 | | 0.02 × 0.02 × 0.02 | Yes | 3 |
| 4 | 0.27 × 0.27 × 0.30 | 0.34 × 0.34 × 0.30 | | 0.02 × 0.02 × 0.02 | | 1,3 |
| 5 | 0.26 × 0.26 × 0.30 | 0.21 × 0.21 × 0.30 | | 0.02 × 0.02 × 0.02 | | 1,3 |
| 6 | 0.27 × 0.27 × 0.30 | 0.31 × 0.31 × 0.30 | | 0.02 × 0.02 × 0.02 | | 1,3 |
| 7 | 0.25 × 0.25 × 0.30 | 0.34 × 0.34 × 0.30 | | 0.02 × 0.02 × 0.02 | | 1,3 |
| 8 | 0.32 × 0.32 × 0.30 | 0.30 × 0.30 × 0.30 | | 0.02 × 0.02 × 0.02 | | 1,3 |
| 9 | 0.24 × 0.24 × 0.30 | 0.32 × 0.32 × 0.30 | | 0.02 × 0.02 × 0.02 | | 1,3 |
| 10 | 0.21 × 0.21 × 0.30 | 0.30 × 0.30 × 0.30 | | 0.02 × 0.02 × 0.02 | | 1,3 |
| 11 | 0.35 × 0.35 × 0.30 | 0.28 × 0.28 × 0.30 | | 0.02 × 0.02 × 0.02 | | 1,3 |
| 12 | 0.35 × 0.35 × 0.30 | 0.21 × 0.21 × 0.30 | | 0.02 × 0.02 × 0.02 | | 1,3 |
| 13 | 0.38 × 0.38 × 0.30 | 0.23 × 0.23 × 0.30 | | 0.02 × 0.02 × 0.02 | | 1,3 |
| 14 | 0.40 × 0.40 × 0.30 | 0.27 × 0.27 × 0.30 | | 0.02 × 0.02 × 0.02 | | 1,3 |
| 15 | 0.25 × 0.25 × 0.30 | 0.26 × 0.26 × 0.30 | | 0.02 × 0.02 × 0.02 | | 1,3 |
| 16 | 0.40 × 0.40 × 0.40 | 0.40 × 0.40 × 0.40 | | 0.03 × 0.03 × 0.03 | | 1,3 |
| 17 | 0.40 × 0.40 × 0.40 | 0.40 × 0.40 × 0.40 | | 0.03 × 0.03 × 0.03 | | 1,3 |
| 18 | 0.40 × 0.40 × 0.40 | 0.40 × 0.40 × 0.40 | | 0.03 × 0.03 × 0.03 | | 1,3 |
| 19 | 0.40 × 0.40 × 0.40 | 0.40 × 0.40 × 0.40 | | 0.03 × 0.03 × 0.03 | | 1,3 |
| 20 | 0.40 × 0.40 × 0.40 | 0.40 × 0.40 × 0.40 | | 0.03 × 0.03 × 0.03 | | 1,3 |
| 21 | 0.40 × 0.40 × 0.40 | 0.40 × 0.40 × 0.40 | | 0.03 × 0.03 × 0.03 | | 1,3 |
| 22 | 0.40 × 0.40 × 0.40 | 0.40 × 0.40 × 0.40 | | 0.03 × 0.03 × 0.03 | | 1,3 |
| 23 | 0.40 × 0.40 × 0.40 | 0.40 × 0.40 × 0.40 | | 0.03 × 0.03 × 0.03 | | 1,3 |
| 24 | 0.40 × 0.40 × 0.40 | 0.40 × 0.40 × 0.40 | | 0.03 × 0.03 × 0.03 | | 1,3 |
| 25 | 0.40 × 0.40 × 0.40 | 0.40 × 0.40 × 0.40 | | 0.03 × 0.03 × 0.03 | | 1,3 |
| 26 | 0.34 × 0.34 × 0.29 | 0.15 × 0.15 × 0.30 | | 0.02 × 0.02 × 0.02 | | 1,3 |
| 27 | 0.31 × 0.31 × 0.30 | 0.25 × 0.25 × 0.30 | | 0.02 × 0.02 × 0.02 | | 1,3 |
| 28 | 0.32 × 0.32 × 0.30 | 0.16 × 0.16 × 0.30 | | 0.02 × 0.02 × 0.02 | Yes | 3 |
| 29 | 0.30 × 0.30 × 0.30 | 0.20 × 0.20 × 0.30 | | 0.02 × 0.02 × 0.02 | Yes | 3 |
| 30 | 0.38 × 0.38 × 0.30 | 0.19 × 0.19 × 0.30 | 0.02 × 0.02 × 0.02 | 0.02 × 0.02 × 0.02 | Yes | 2,3 |
| 31 | 0.39 × 0.39 × 0.30 | 0.14 × 0.14 × 0.30 | 0.02 × 0.02 × 0.02 | 0.02 × 0.02 × 0.02 | | 1,2,3,4 |
| 32 | 0.33 × 0.33 × 0.30 | 0.20 × 0.20 × 0.30 | 0.02 × 0.02 × 0.02 | 0.02 × 0.02 × 0.02 | | 1,2,3,4 |
| 33 | 0.29 × 0.29 × 0.40 | 0.32 × 0.32 × 0.30 | 0.02 × 0.02 × 0.02 | 0.02 × 0.02 × 0.02 | | 1,2,3,4 |
| 34 | 0.32 × 0.32 × 0.30 | 0.23 × 0.23 × 0.30 | 0.02 × 0.02 × 0.02 | 0.02 × 0.02 × 0.02 | | 1,2,3,4 |
| 35 | 0.29 × 0.29 × 0.30 | 0.17 × 0.17 × 0.30 | 0.02 × 0.02 × 0.02 | 0.02 × 0.02 × 0.02 | Yes | 2,3 |

6.2.2. Ground truth dataset creation

Figure 6.3 show examples of pre- and post-implantation CTs and μCTs. As can be seen, the individual electrodes in a post-implantation μCT are more separable than in a conventional post-implantation CT because the μCT has 3 orders of magnitude better resolution and little partial volume artifact. It is also easier to segment the intra-cochlear anatomy in a pre-implantation μCT because the image quality of μCTs is higher and the basilar membrane is visible in a μCT. Thus, our ground truths are manually generated on pre- and post-implantation μCTs.

We use the dataset for four validation purposes: (1) Characterize the accuracy of the electrode localization method $M_E$. (2) Characterize the accuracy of the three existing intra-cochlear anatomy segmentation methods $M_{A1}$, $M_{A2}$, and $M_{A3}$. (3) Analyze the sensitivity of IGCIP with respect to the accuracy of the methods in (1) and (2). (4) Assess the quality of the IGCIP-generated electrode configurations generated by using the complete automatic process, including both the electrode localization and anatomy segmentation. Using the image of the 35 specimens, we create 4 dataset groups and one "electrode configuration dataset". The 4 groups of validation datasets are shown in Table 6.1. The details of each group and the electrode configuration dataset are explained in Section 6.2.3.

6.2.3 Validation approaches

*6.2.3.1 Error analysis for electrode localization method*

We use Group 1 (see Table 6.1) to characterize the accuracy of $M_E$. It consists of 30 out of 35 specimens with pre-, post-implantation CTs and post-implantation μCTs. An expert manually delineated the ground truth locations (GL) of electrodes on the post-implantation μCTs of these 30 specimens. Then, we apply $M_E$ to the corresponding 30 conventional post-implantation CTs of specimens in Group 1 to generate the automatic localization (AL) of electrodes. Post-implantation conventional and μCTs were registered to facilitate comparison between automatic and gold-standard ground truth localizations using mutual information-based registration techniques. The registrations were visually inspected and confirmed to be accurate, as shown in Figure 6.3b. We do not include specimens 3, 28, 29, 30, and 35 in Group 1 because we observed that the CI electrode arrays had clearly moved between the conventional and the μCTs during visual inspection, which makes those 5 subjects not available for the evaluating the accuracy of $M_E$. One example of specimen that has electrode migration between post-implantation μCT and CT is shown in Figure 6.4a. We hypothesize that this motion occurred due to the fact that the specimen cochlea do not have fluid that could typically stabilize the array. Thus, when the specimens being transferred between different imaging sites, the electrode arrays were not internally fixed and may have moved. In addition to GL and AL, we also created an image-based localization (IL) as the average of multiple expert localizations in the CT images. To create IL, an expert manually generated electrode localization results for each case repeatedly until adding a new instance changes the position of each electrode in the average localization by no more than 0.05mm (approximately ¼ the width of a CT voxel). This indicated that the expert's localizations converge to the best localization manually achievable when using the conventional CTs. To compare two electrode localizations, we measured Euclidean distances between the centroids of the corresponding electrode points and compared AL and GL to evaluate the overall accuracy achieved when using our automatic approach. However, the overall localization error is a function of algorithmic errors and errors due to image-based errors. The algorithmic errors exist due to the limitation of the automatic techniques. The image-based errors exist due to the limitation in the quality of the conventional CTs. Thus, we compared IL and AL to estimate algorithmic errors. We also compared AL and GL to measure image-based errors. In Section 6.3.1 we present the validation results of $M_E$.

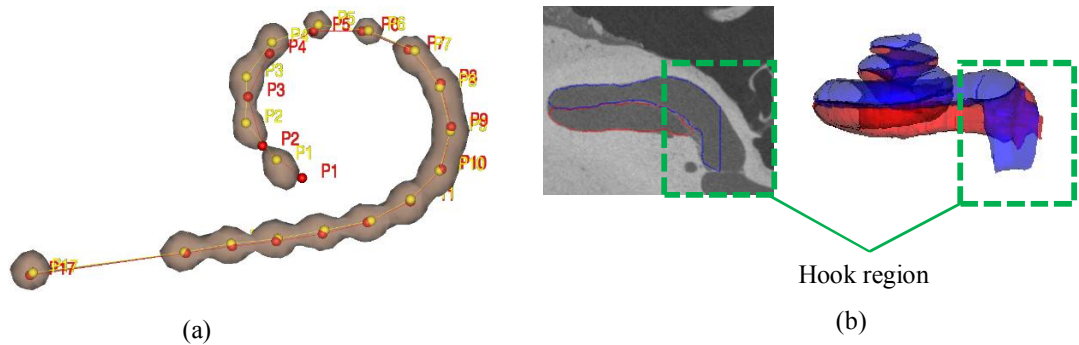

**Figure 6.4.** Panels (a) shows electrode migration in Specimen 3. The CT iso-surface of the highest intensity voxels is shown in orange. The automatically (yellow) and manually (red) localized electrodes from the CT and μCT are different from electrode P1 to P6. Panel (b) shows an axial slice of a μCT around the "hook region" of SV. The blue and red contours in the CT are the manual delineations of SV and ST generated by an expert. The corresponding 3D meshes are shown on the right side. As can be seen, the "hook region" of SV is guessed by the expert.

*6.2.3.2 Validation for intra-cochlear anatomy segmentation methods*

We use Group 2 (see Table 6.1) to evaluate the accuracy of the three intra-cochlear anatomy segmentation methods. Group 2 consists of 6 specimens with post-implantation CTs, pre-implantation CTs, and pre-implantation μCTs available. We apply $M_{A1}$ to the pre-implantation CTs, and $M_{A2}$ and $M_{A3}$ to the post-implantation CTs of the 6 specimens in Group 2, respectively. On the pre-implantation μCTs, an expert manually delineated the ST, SV, and MOD to serve as gold-standard ground truth for intra-cochlear anatomy. We registered pre-implantation and post-implantation CTs, and the pre-implantation μCTs together to facilitate comparing gold-standard segmentation results and automatic segmentation results. The automatic intra-cochlear anatomy segmentation methods generate surface meshes for ST, SV, and MOD that have pre-defined numbers of vertices.

Those pre-defined numbers are different from the number of vertices in the manually generated surface meshes. To enable a point-to-point error estimation for manually and automatically generated meshes, we used an ICP-based [26] iterative non-rigid surface registration method developed in house to register the active shape model used to localize the ST, SV, and MOD to the manually delineated ST, SV, and MOD surfaces in the μCTs. This process results in a set of ground truth ST, SV, and MOD surfaces that have a one-to-one point correspondence with the surfaces generated by our automatic methods. For each intra-cochlear anatomy segmentation method, we then measured the Euclidean distance from each vertex on the automatically localized surfaces to the corresponding point on the gold-standard surfaces. The SV in the cochlea is a cavity with an open region on the side that is close to the round window membrane of the cochlea. In both CT and μCT, the border of the SV in the "hook region" (see Figure 6.4b) that is close to the round window membrane of cochlea cannot be delineated consistently because the SV is an open cavity without an anatomical boundary at the hook region. Thus, the border must be estimated somewhat arbitrarily by the expert when generating the ground truth. Since the accuracy of the segmentation in this region is not important for intra-cochlear electrode localization or IGCIP, we exclude approximately 1.5cm$^3$ around the SV hook region when estimating the SV segmentation error. In the remainder of this article, we denote the gold-standard intra-cochlear anatomy surfaces as $S_0$, and the surfaces generated by using $M_{A1}$, $M_{A2}$, and $M_{A3}$ as $S_1$, $S_2$, and $S_3$. In Section 6.3.2 we analyze the results for the validation studies of the accuracy of $M_{A1}$, $M_{A2}$, and $M_{A3}$.

*6.2.3.3 Sensitivity of intra-cochlear electrode position estimation to processing errors*

We conduct three studies to analyze the sensitivity of IGCIP by using different groups of specimens, as shown in Table 6.2. As is shown in Figure 6.2, one electrode localization and one intra-cochlear anatomy segmentation define one estimation of the spatial relationship between the electrodes and auditory nerves. This relationship can be described by measuring locations of electrodes relative to intra-cochlear structures using an electrode coordinate system proposed by Verbist et al. [25]. As is discussed in Section 1, the intra-cochlear location of electrodes and their relationship to hearing outcomes has been a subject of intense study in recent years [5-10]. Thus, independently of IGCIP, it is of interest to quantify the accuracy of the processing methods for estimating intra-cochlear position to understand the limitations of these techniques for use in such large scale analyses of how electrode position affects accuracy. Thus, in this study, we quantify errors in estimating intra-cochlear electrode position when using $M_E$, $M_{A1}$, $M_{A2}$, and $M_{A3}$. Electrode position is measured in terms of angular depth-of-insertion (DOI), the distance to modiolar surface (DtoM), and the distance to the basilar membrane (DtoBM). As the cochlea has a spiral shape with 2.5 turns in human, the depth of any position within it can be quantified in the terms of a DOI value from 0 to 900 degrees. The DtoM values are directly computed as the Euclidean distances between the centroids of electrodes and the vertices on the modiolar surface. The DtoBM value is computed as the signed Euclidean distance between the centroids of electrodes and basilar membrane, which lies between ST and SV. Figure 6.5 shows the measurements of the three values. Among the three values, DOI and DtoM values are directly related with the construction of DVFs as they correspond to the horizontal and vertical axes of DVFs. DtoM values are not directly related but are still have important information of the intra-cochlear locations of the implanted electrodes.

**Table 6.2.** Electrode configuration names in sensitivity analysis studies

| Study | Data group # | Intra-cochlear anatomy | Electrode locations | Configuration name |
|---|---|---|---|---|
| (a). Electrode localization sensitivity | 1 | $S_1$ | GL | $C_{G1}$ (Reference) |
| | | | AL | $C_{A1}$ |
| (b). Anatomy segmentation sensitivity | 2 | $S_0$ | GL | $C_{G0}$ (Reference) |
| | | $S_1$ | | $C_{G1}$ |
| | | $S_2$ | | $C_{G2}$ |
| | | $S_3$ | | $C_{G3}$ |
| | 3 | $S_1$ | GL | $C_{G1}$ (Reference) |
| | | $S_1'$ | | $C_{G1}'$ |
| | | $S_2'$ | | $C_{G2}'$ |
| | | $S_3'$ | | $C_{G3}'$ |
| (c). Overall sensitivity | 4 | $S_0$ | GL | $C_{G0}$ (Reference) |
| | | $S_1$ | AL | $C_{A1}$ |
| | | $S_2$ | AL | $C_{A2}$ |
| | | $S_3$ | AL | $C_{A3}$ |
| | 1 | $S_1$ | GL | $C_{G1}$ (Reference) |
| | | $S_1'$ | AL | $C_{A1}'$ |
| | | $S_2'$ | AL | $C_{A2}'$ |
| | | $S_3'$ | AL | $C_{A3}'$ |

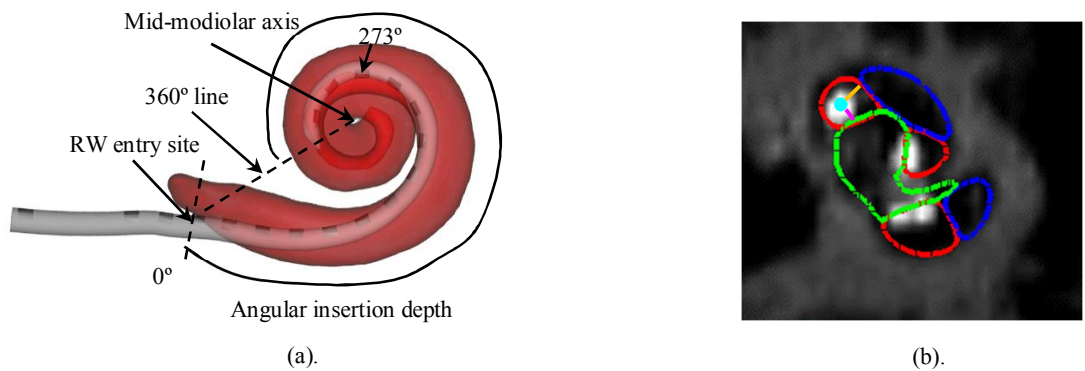

**Figure 6.5.** Panel (a) shows the measurement of the DOI value for the 3$^{rd}$ most apical electrode in the coordinate system proposed by Verbist et al. [25]. The ST is shown in the red. The electrode array carrier is shown in light grey and the contacts are shown in dark grey. Panel (b) shows the measurements of DtoM (magenta line) and DtoBM (orange line) values for a given electrode (cyan point) in a CT slice in coronal view. The ST, SV and MOD are shown in red, blue, and green, respectively.

*6.2.3.4 Sensitivity of IGCIP to processing errors*

The spatial relationship between the electrodes and the intra-cochlear anatomy defines a set of DVFs. Based on the DVFs, an electrode deactivation plan, the "electrode configuration" is generated by using our automatic electrode configuration selection method [18]. In each study shown in Table 6.2, the sensitivity of IGCIP is defined as the difference between the electrode configurations generated by using "automatic" and "reference" intra-cochlear electrode position estimation. Table 6.2 defines the automatic and reference electrode position estimation techniques for each study and denotes the name for each resulting electrode configuration.

In study (a), we evaluate the sensitivity of IGCIP with respect to the electrode localization method by using the specimens in Group 1. The reference configurations in study (a) are defined as $C_{G1}$, which are generated by using $S_1$, together with GL. The automatic configurations are defined as $C_{A1}$, which are generated by using $S_1$ together with AL. In study (b), we evaluate the sensitivity of IGCIP with respect to the intra-cochlear anatomy segmentation methods by using specimens in Groups 2 and 3. In Group 2, which consists of the 6 subjects with pre-implantation CTs, the reference configurations $C_{G0}$ are generated by $S_0$ together with the GL. The three sets of automatic configurations $C_{G1}$, $C_{G2}$, $C_{G3}$ are generated by using $S_1, S_2, S_3$ together with GL, respectively. Due to the limited number of pre-implantation µCTs acquired for subjects in our dataset, we use Group 3 to generate synthesized surfaces for $M_{A1}$, $M_{A2}$, and $M_{A3}$ so that we can analyze the sensitivity of IGCIP with respect to the errors of the three intra-cochlear anatomy segmentation methods on a larger dataset. For the specimens in Group 3, we select $S_1$ of all the 35 specimens as our reference intra-cochlear anatomical surfaces. Then, for each subject, we deform $S_1$ to generate the synthesized surfaces $S_1', S_2', S_3'$ that simulate the segmentation errors of method $M_{A1}$, $M_{A2}$, and $M_{A3}$. To build synthesized surfaces $S_1'$ for $M_{A1}$, we firstly build a gamma distribution by using the mean and the standard deviation of the segmentation error of $M_{A1}$, which is estimated by using specimens in Group 2 and the error measurement approach described in sub-section 6.2.3.2. Then, for each specimen in Group

3, we draw a random number from the defined gamma distribution and set this number as the "desired mean segmentation error" between the synthesized surfaces and the reference surfaces of the selected subject. We randomly adjust the shape control parameters in the active shape model [22] so that we deform the reference surfaces to the synthesized surfaces with a mean point-to-point difference equal to the desired mean segmentation error. The same process is used to generate $S_2'$ and $S_3'$. We use an active shape model to perform this deformation, instead of directly adding errors to each vertices on the reference surface $S_1$, so that the changes in the deformed surfaces have realistic anatomical constraints. In Group 3, the reference configurations $C_{G1}$ are generated by using $S_1$ and GL. The three sets of automatic configurations $C_{G1}'$, $C_{G2}'$, $C_{G3}'$ are generated by using $S_1'$, $S_2'$, $S_3'$, together with GL, respectively. In study (c), we evaluate the sensitivity of IGCIP with respect to both the electrode and anatomy segmentation methods by using specimens in Group 4 and 1. Group 4 consists of the 4 specimens that have pre-implantation μCTs and do not have electrode migration. The reference configurations $C_{G0}$ in Group 4 in study (c) are generated by using the anatomy $S_G$, together with the GL. The three sets of automatic configurations $C_{A1}$, $C_{A2}$, and $C_{A3}$ are generated by using $S_1$, $S_2$, $S_3$, together with AL, respectively. Due to the same issue with the limited pre-implantation μCTs in study (b), for study (c), we use Group 1, which consists of the 30 specimens that do not have electrode migration to expand the size of our dataset for overall sensitivity analysis. The reference configurations $C_{G1}$ in Group 1 are generated by using $S_1$ and GL. The three sets of automatic configurations $C_{A1}'$, $C_{A2}'$, $C_{A3}'$ are generated by using $S_1'$, $S_2'$, $S_3'$, together with AL, respectively.

The most direct way to show the difference of two electrode configurations is to use a binary code (use "1" to indicate an electrode being "activated" and "0" to indicate an electrode being "deactivated") to represent the two configurations and then compute the hamming distance between them. This directly shows the differences between two given configurations. However, sometimes a configuration of "on-off-on-off-on" has an equal quality stimulation pattern with a configuration of "off-on-off-on-off", even though they result in large hamming distance. Thus, we use two other metrics to compare the automatic and reference configurations to evaluate the sensitivity of IGCIP. The first metric we use is the difference between "cost values" of the two configurations. In our automatic electrode deactivation strategy [18], we have developed a cost function which assigns a cost value to a specific electrode configuration. In our design, a lower cost value indicates a configuration that is less likely to cause electrode interaction and more likely to stimulate a broad frequency range. Thus, the difference between the cost values of two configurations is an indicator for the difference between the automatic and the reference electrode configurations. The second metric is the difference between the quality of the automatic and reference electrode configurations. The quality of the electrode configurations is evaluated by an expert (JHN) through an electrode configuration quality assessment study, which is discussed in details in the next subsection.

### 6.3. Results

#### 6.3.1 Accuracy of the electrode localization technique

Validation of the electrode localization technique was presented in [23], and the results are summarized here. Figure 6.6a shows boxplots of the mean, median, maximum, and the standard deviation of localization errors between AL and GL across the 30 specimens in

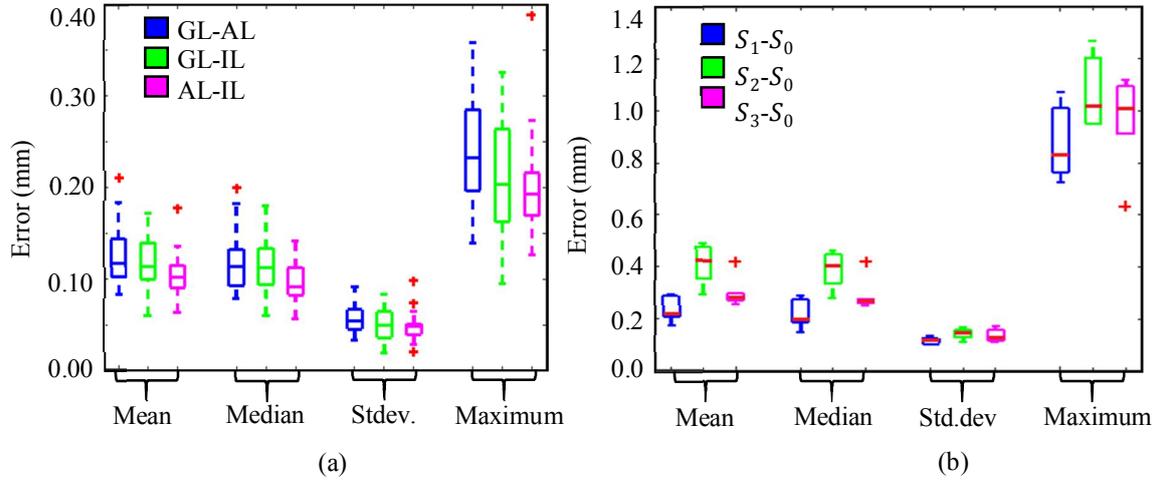

**Figure 6.6.** Panel (a) shows the boxplots for localization errors between AL-GL, IL-GL, and AL-IL. Panels (b) shows the segmentation errors between $S_1$-$S_0$, $S_2$-$S_0$, and $S_3$-$S_0$.

Group 1. In each boxplot, the median value is given as a red line, 25th and 75th percentiles are indicated by the blue box, whiskers show the range of data points that fall within 1.5x the interquartile range from the 25th or 75th percentiles but are not considered outliers, and red crosses indicate outlier data points. Comparing AL and GL, we found mean electrode localization errors of 0.13mm and a maximum localization error of 0.36mm. Comparing IL and GT, we found the mean electrode localization error was 0.12mm and the maximum localization error was 0.32mm. Comparing AL and IL, we found the mean and maximum localization errors are 0.10mm and 0.39mm, respectively. This shows that our automatic method generated localization results close to the optimal localization results that can be generated by an expert from clinical post-implantation CTs. All localization errors were smaller than the length of one voxel diagonal of the conventional post-implantation CTs in our dataset. We performed a paired t-test between the mean localization errors between AL-GL and AL-IL and found the $p$ value was $4.96 \times 10^{-5}$. This shows that the algorithmic errors that would be estimated if using the CT image to create a ground truth would be significantly different from the errors measured when using the μCT to serve as ground truth. However, the errors between AL-GL are still small. Thus, even by using imperfect CT images with limited resolution, our electrode localization method in IGCIP can still generate accurate localization results.

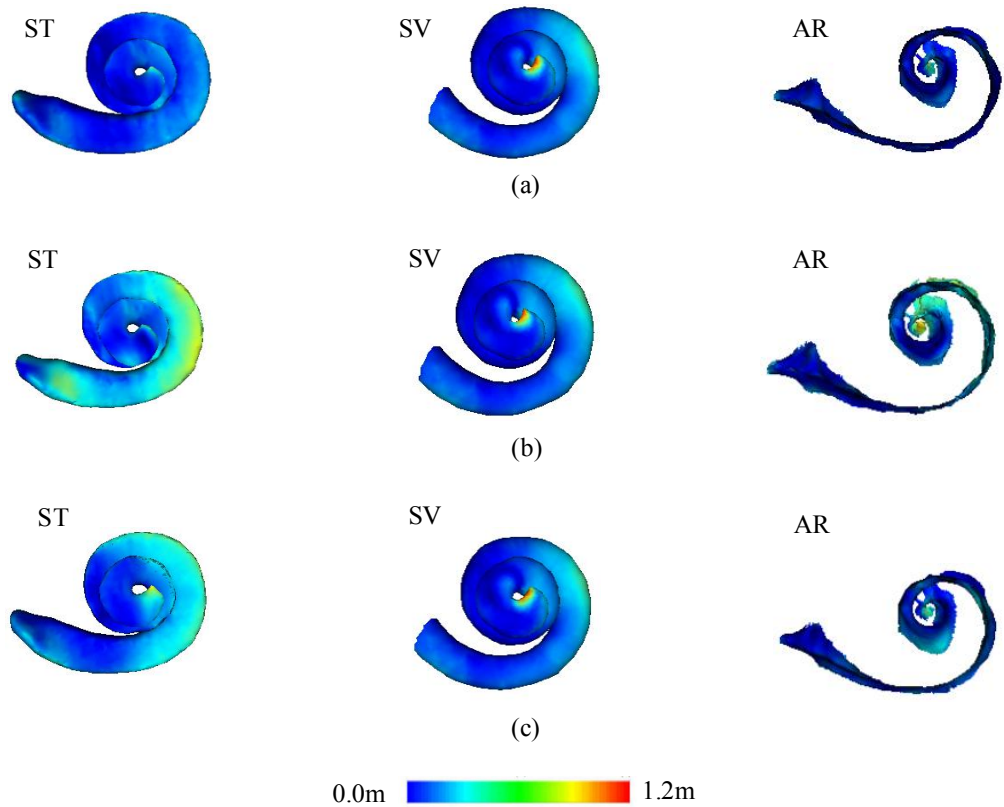

**Figure 6.7.** Panels (a), (b), (c) show qualitative segmentation results ($S_1$, $S_2$, and $S_3$) generated by IGCIP automatic methods $M_{A1}$, $M_{A2}$, and $M_{A3}$ for a representative subject in Group 2. The three surfaces of intra-cochlear anatomical structures are color-coded by the segmentation errors computed by using $S_0$.

6.3.2 Accuracy of intra-cochlear anatomy segmentation methods among the three existing automatic methods in IGCIP and our gold-standard ground truth, we found the most accurate method was $M_{A1}$. This is because $M_{A1}$ is implemented on pre-implantation CTs in which the metallic artifacts caused by electrodes do not exist. Among the rest two methods $M_{A2}$ and $M_{A3}$ implemented on post-implantation CTs, $M_{A3}$ results in better mean segmentation errors than $M_{A2}$ on post-implantation CTs. $M_{A2}$ is less accurate on post-implantation CTs because it relies on using the shape of the cochlear labyrinth to localize the intra-cochlear anatomy and the shape of the cochlear labyrinth may not be a good predictor for the positions of the intra-cochlear anatomy. Overall, all three methods had <0.5mm mean segmentation errors. Figure 6.7 shows the segmentations of ST, SV, and AR from one case generated by all the methods. The surfaces are color-coded by using the segmentation errors computed by using $S_0$.

Figure 6.6b show the boxplots of the mean, the maximum, the median, and the standard deviation of anatomy segmentation errors between automatic methods and the ground truth across the 6 specimens in Group 2. Comparing $S_0$ and $S_1$, the mean and standard deviation of the segmentation errors was 0.23±0.12mm. Comparing $S_0$ and $S_2$, the mean and the standard deviation of the segmentation errors was 0.41±0.15mm. Comparing $S_0$ and $S_3$, the mean and the standard deviation of the segmentation errors was 0.30±0.14mm. Finally,

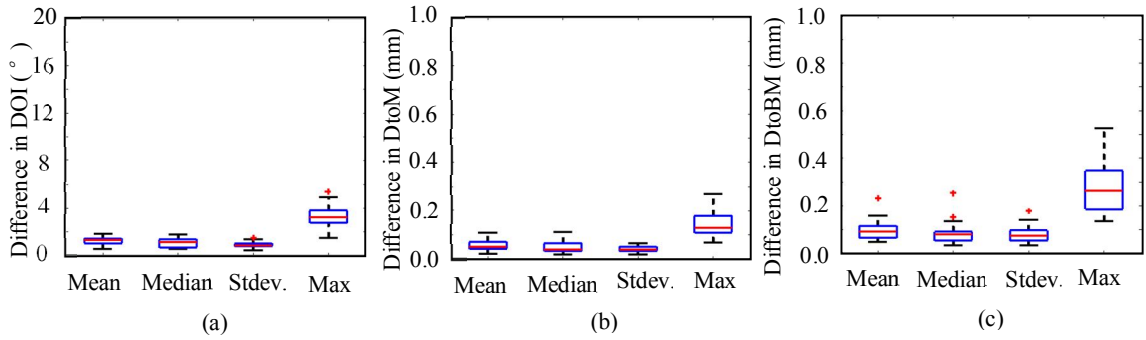

**Figure 6.8.** Panels (a), (b), and (c) show the boxplots for the differences in the DOIs, the DtoM, and the DtoBM of the automatic ($C_{A1}$) and the reference ($C_{G1}$) configurations generated by IGCIP for sensitivity analysis with respect to the electrode localization method (study (a) in Table 6.2).

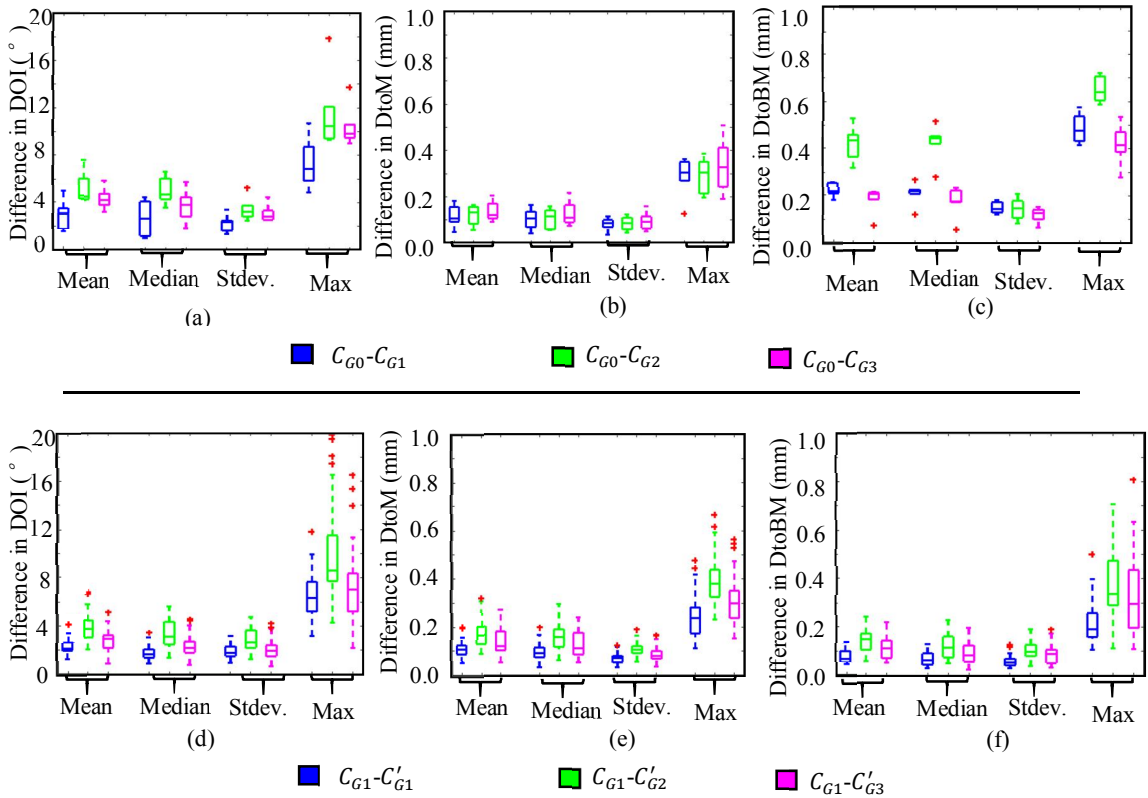

**Figure 6.9.** Panels (a-c) show the boxplots for the differences in the DOIs, the DtoM, and the DtoBM of the electrodes generated by using automatic ($C_{G1}$, $C_{G2}$, $C_{G3}$) and the reference ($C_{G0}$) processing methods on the 6 specimens in Group 2. Panels (d-f) show the boxplots for the differences in the DOIs, the DtoM, and the DtoBM of the electrodes generated by using automatic ($C'_{G1}$, $C'_{G2}$, $C'_{G3}$) and the reference ($C_{G1}$) processing methods on the 35 specimens in Group 3 with the synthesized anatomy surfaces. These results are the IGCIP sensitivity analysis study with respect to the intra-anatomy segmentation method (study (b) in Table 6.2).

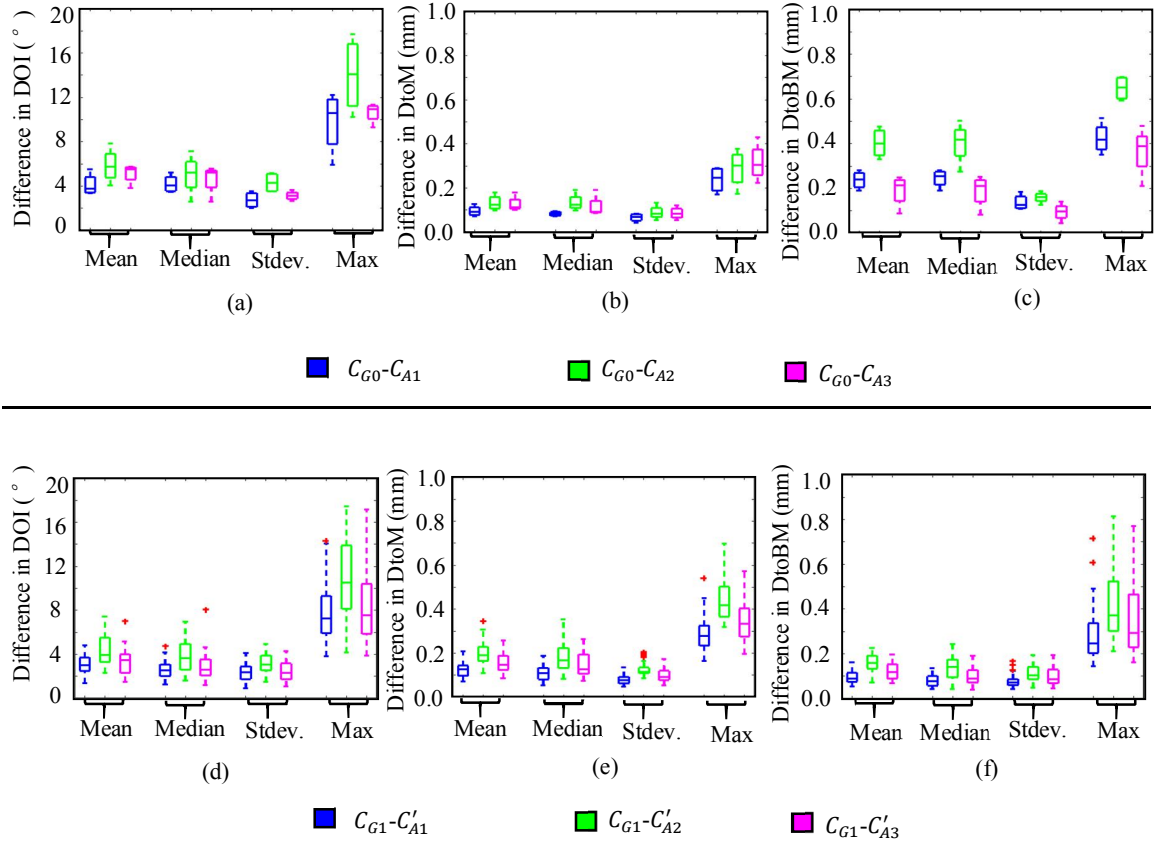

**Figure 6.10.** Panels (a-c) show the boxplots for the differences in the DOIs, the DtoM, and the DtoBM of the electrodes generated by using automatic ($C_{A1}$, $C_{A2}$, $C_{A3}$) and the reference ($C_{G0}$) processing methods on the 4 specimens in Group 4. Panels (d-f) show the boxplots for the differences in the DOIs, the DtoM, and the DtoBM of the electrodes generated by using automatic ($C'_{A1}$, $C'_{A2}$, $C'_{A3}$) and the reference ($C_{G1}$) processing methods on the 30 specimens in Group 1 with the synthesized anatomy surfaces. These results are the IGCIP sensitivity analysis study with respect to the intra-anatomy segmentation method (study (c) in Table 6.2).

6.3.3 Sensitivity of intra-cochlear electrode position estimation to processing errors

Figure 6.8, Figure 6.9, and Figure 6.10 show boxplots for the difference between the intra-cochlear locations of the electrodes identified by using the automatic and the reference processing methods defined in study (a), (b), and (c) in Table 6.2. Comparing the results presented in Figure 6.8 and Figure 6.9, we find that the intra-cochlear locations of the electrodes are less sensitive to the electrode localization method than to the intra-cochlear anatomy segmentation methods. Among the three intra-cochlear anatomy segmentation methods, $M_{A1}$ is the most reliable method for generating accurate intra-cochlear locations, then $M_{A3}$, followed by $M_{A2}$. Comparing the results presented in Figure 6.8-6.10, we find that the overall errors of both the electrode localization and intra-cochlear anatomy segmentation techniques are not substantially larger than the errors due to the intra-cochlear anatomy segmentation alone.

6.3.4 Sensitivity of IGCIP to processing errors

In Figure 6.11, we show the boxplots for the cost values of automatic, reference, and the control configurations defined in sub-section 6.2.3.5. The name of the configurations are indexed in Table 6.2. From Figure 6.11, we can see that besides the outliers, the average cost values for all the automatic configurations are close to the average cost values for the reference configurations. The average cost values for the control configurations are significantly larger than the ones for the reference and the automatic configurations. These results show that the automatic image processing techniques in our IGCIP can generate configurations that have similar quality to the configurations generated by using the reference anatomy and electrode locations. From Figure 6.11b-e, we see that $M_{A1}$ generates the intra-cochlear anatomy that lead to the lower average cost than $M_{A2}$ and $M_{A3}$. This is because $M_{A1}$ is applied on pre-implantation CTs, where the intra-cochlear anatomy are not obscured by the metallic artifacts. For the two methods designed for post-implantation CTs, $M_{A3}$ generates intra-cochlear anatomy that lead to lower average cost than $M_{A2}$. This indicates that $M_{A3}$ is more reliable than $M_{A2}$. This is also shown in the differences in the DOI and the DtoBM values in Figure 6.9 and Figure 6.10.

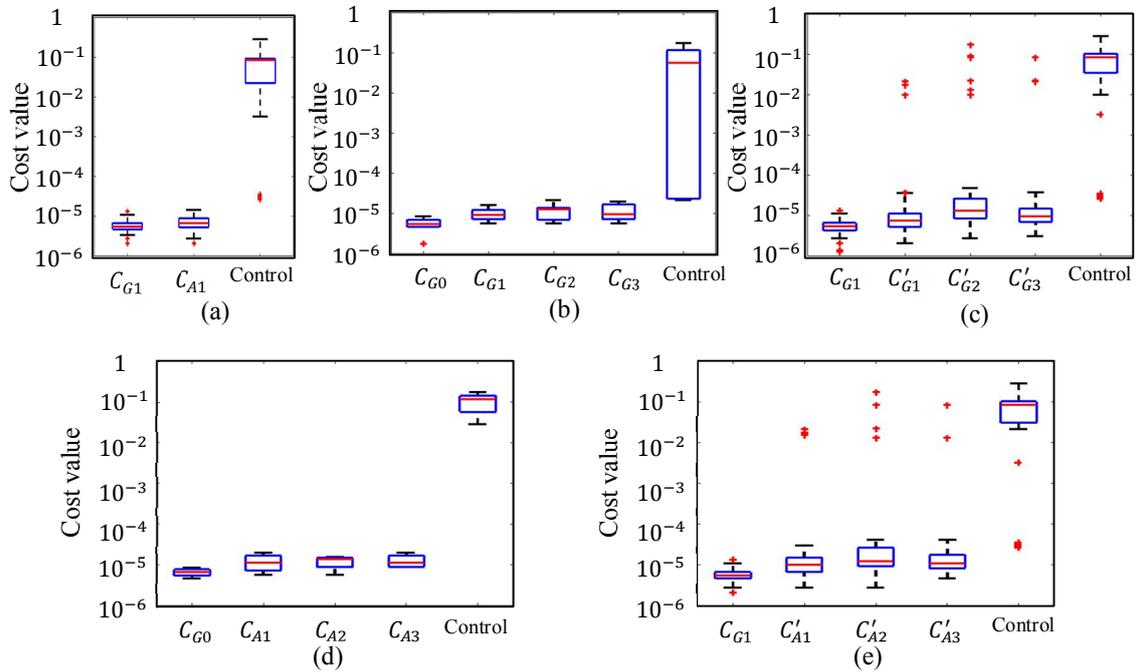

**Figure 6.11.** Panels (a-e) show the boxplots for the cost values (in log-scale) of automatic, reference, and control configurations for subjects in the data being used in the three studies in Table 6.2 for IGCIP sensitivity analysis.

Figure 6.12 shows the evaluation results for the 255 electrode configuration sets inour electrode configuration dataset discussed in sub-section 6.2.3.5. In Figure 6.12, panel (a) shows the evaluation results of the configurations generated for the sensitivity analysis of IGCIP with respect to the electrode localization method. These configurations belong to study (a) in Table 6.2. Panel (b) and (c) show the evaluation results of the configurations generated for the sensitivity analysis of IGCIP with respect to the three intra-cochlear anatomy segmentation methods. These configurations belong to study (b). Panel (d) and (e) show the evaluation results of the configurations generated for the overall sensitivity analysis of IGCIP with respect both the electrode and anatomy segmentation methods for study (c). As can be seen in Figure 6.12a, among the 30 automatic electrode configurations in $C_{A1}$ generated by using AL, none of them in is rated as not acceptable, and 21 out of 30 automatic configurations in $C_{A1}$ are rated as at least equally good as the reference configurations $C_{G1}$. This shows that the errors in the electrode localization method is robust enough to generate localization results that lead to acceptable electrode deactivation configurations.

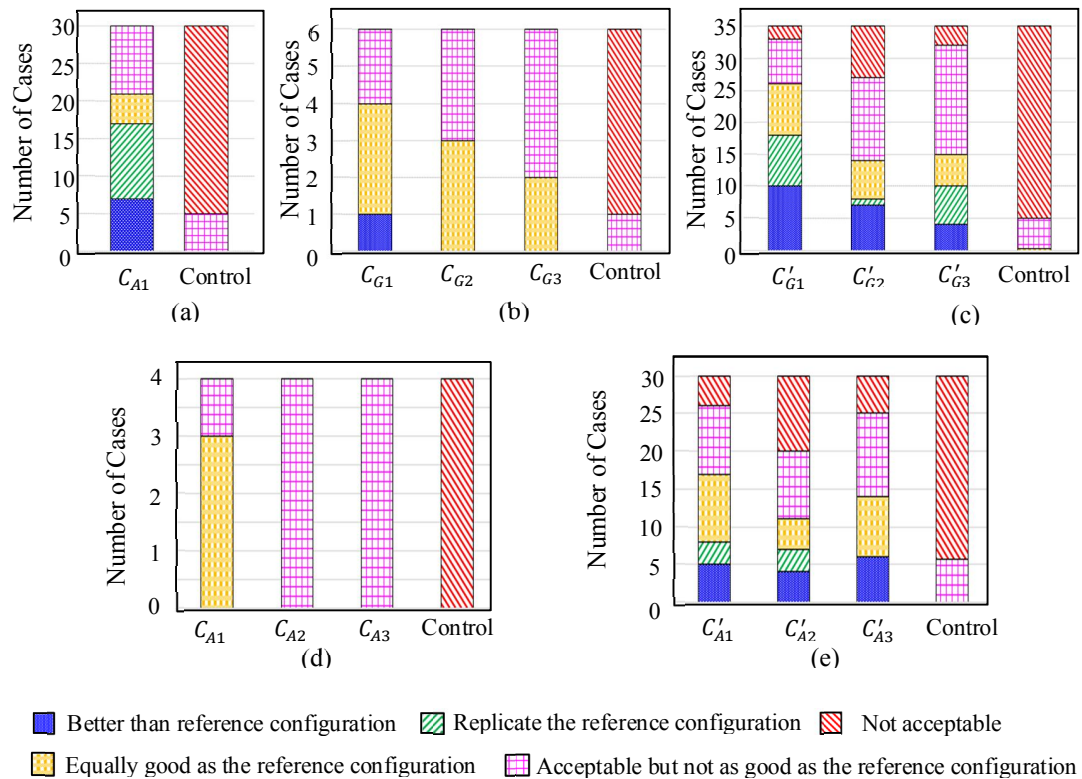

**Figure 6.12.** Evaluation results of the configurations generated for the sensitivity analysis of IGCIP with respect to (a) the electrode localization method, (b-c) the three intra-cochlear anatomy segmentation methods, and (d-e) the overall automatic image processing techniques in IGCIP.

In Figure 6.12b, among the automatic configurations generated by using GL and $S_1$, $S_2$, and $S_3$, none of the automatic configurations in $C_{G1}$, $C_{G2}$, and $C_{G3}$ is rated as not acceptable. Meanwhile, $C_{G1}$, $C_{G2}$, and $C_{G3}$ have generated 4, 3, and 2 configurations that are at least equally as good as the reference configurations $C_{G0}$. In Figure 6.12c, among the automatic configurations generated by using GL and $S_1'$, $S_2'$, $S_3'$, 2, 8, and 3 automatic configurations in $C_{G1}'$, $C_{G2}'$, and $C_{G3}'$ are rated as not acceptable, and 26, 14, and 15 automatic configurations in $C_{G1}'$, $C_{G2}'$, and $C_{G3}'$ are rated as at least equally good as the reference configurations $C_{G1}$. The results shown in Figure 6.12a-c show that the quality of the IGCIP-generated electrode configurations generated are less sensitive to the errors in the electrode localization method than to the intra-cochlear anatomy segmentation methods. In Figure 6.12d, among the automatic configurations generated by using AL and $S_1, S_2, S_3$, none of them in $C_{A1}$, $C_{A2}$, and $C_{A3}$ is rated as unacceptable. Three automatic configurations in $C_{A1}$ are rated as equally good as the reference configurations in $C_{G0}$. In Figure 6.12e, among the automatic configurations generated by using AL and $S_1', S_2', S_3'$, 4, 10, and 5 automatic configurations in $C_{A1}'$, $C_{A2}'$, and $C_{A3}'$ are rated as not acceptable, and 17, 11, and 14 automatic configurations in $C_{A1}'$, $C_{A2}'$, and $C_{A3}'$ are rated at least as good as the reference configurations $C_{G1}$. Altogether, these results suggest that $M_{A1}$ is the most reliable anatomy localization method to generate acceptable electrode configurations. Further, $M_{A3}$ should be used as the secondary choice for anatomy segmentation when pre-implantation CTs are not available and $M_{A1}$ cannot be directly used.

In the results shown in Figure 6.12e, the expert evaluated 26 out of 30 (86.7%)

automatic configurations generated by $M_E + M_{A1}$ as acceptable, and 25 out of 30 (83.3%) automatic configurations generated by $M_E + M_{A3}$ as acceptable. These results, together with the results presented in Section 6.3.3, indicate that our IGCIP can generate reliable electrode configurations by using the automatic image processing techniques. To further improve the reliability of IGCIP, we should increase the accuracy of the intra-cochlear anatomy segmentation methods.

In Figure 6.12a-e, we find that among all the control configurations in all the experiments, 83.3%, 83.3%, 85.7%, 100%, and 81.1% are rated as unacceptable by the expert. This suggests that the evaluation results generated by the expert shown above are not biased towards the tendency for evaluating every configuration as acceptable.

### 6.4. Conclusion

In this article, we create a ground truth dataset with high accuracy and use it for a validation study on an image-guided cochlear implant programming (IGCIP) system developed by our group. The two major image processing techniques in IGCIP are the CI electrode localization and intra-cochlear anatomy segmentation methods. The validation study results show that among 30 cases in our dataset, our localization method can generate results that are highly accurate with mean and maximum electrode localization errors of 0.13mm and 0.36mm. Our three intra-cochlear anatomy localization methods can generate results that have mean errors of 0.23mm, 0.41mm, and 0.30mm. In a sensitivity analysis for IGCIP, we found that our IGCIP is less sensitive to the electrode localization method than to the intra-cochlear anatomy segmentation method. Among the three intra-cochlear anatomy segmentation methods, we found that IGCIP is the least sensitive to method $M_{A1}$, then $M_{A3}$, then $M_{A2}$. In an overall IGCIP-generated automatic electrode configuration quality evaluation study, we found that IGCIP can generate configurations that are 86.7% acceptable when the pre-implantation CTs are available, and 83.3% acceptable when the pre-implantation CTs are not available. One limitation of this study is that while it includes several models of CI electrode arrays, they were produced by only one manufacturer. In the future, we plan to expand the validation dataset by acquiring pre- and post-implantation CTs and μCTs of temporal bone specimens implanted with electrode arrays from different CI manufacturers. We also will study hearing outcomes of CI recipients using IGCIP-generated configurations and the manually selected configurations to show the effectiveness of IGCIP-generated configurations.

# SUMMARY AND FUTURE WORK

This dissertation introduces several innovative image processing and image-based automatic techniques for fully automating our image-guided cochlear implant (CI) programming (IGCIP) system [1]. Prior to this dissertation, the electrode localization and electrode deactivation configuration selection steps in IGCIP were not fully automated. In this dissertation, we have made three major contributions: (1) We propose several automatic methods for localizing different types of CI electrode arrays in post-implantation CTs [2-4], as described in Chapter II, III, and IV. (2) We develop an automatic method for electrode deactivation configuration selection that can generate configurations that are comparable to the ones selected by experts [5], as described in Chapter V. (3) We perform the first thorough validation of IGCIP by using a highly accurate ground truth dataset [6] and analyze the sensitivity of IGCIP to errors introduced by the image processing techniques we have developed, as described in Chapter VI.

In Chapter II, we propose a snake-based method [2] for localizing one type of the closely-spaced CI electrode arrays. First, this method uses a reference image to locate the VOI that contains the cochlea region from a whole head clinical CT image. Then, it uses a Maximum Likelihood Estimation-based (MLE-based) method to estimate a threshold for the VOI. By applying the threshold to the VOI, the method generates ROIs that possibly contain the electrode array. The assumption is that the ROI that contains the largest number of voxels is occupied by the electrode array. Next, we apply a voxel thinning method [7] to the largest ROI to generate the medial axis line, which is treated as the initial centerline of the implanted CI electrode array. The centerline endpoints are first localized within the neighborhood of their initialized positions using an endpoint detection filter we have designed [2]. Then, the endpoints are fixed and the points in the rest of the centerline are optimized by using a snake [8] with its external energy defined using the output of a vesselness filter that is applied to the original VOI to enhance the centerline of the electrode array. The final step is a resampling step on the extracted centerline to determine the position of each electrode using *a-priori* knowledge about the distance between neighboring electrodes. Out of 15 cases, our testing results show that the average electrode localization error with this method is 0.21mm. This method is a preliminary method for localizing closely-spaced CI electrode arrays in CIs. It shows the feasibility of using the centerline of the implanted array to estimate the individual locations of closely-spaced electrodes. In a more comprehensive evaluation of the snake-based method on a large scale dataset, we discovered several limitations of this method and have proposed a more refined method for localizing closely-spaced electrode arrays that is presented in Chapter IV.

In Chapter III, we propose a graph-based method for localizing distantly-spaced CI electrode arrays in clinical CTs with sub-voxel accuracy [3]. This method is extended from a graph-based path finding algorithm [15] developed earlier. The first step of this method [3] is also the localization of the VOI in a whole head clinical CT image using a reference image. We up-sample the VOI and the subsequent procedures are performed on the up-sampled VOI. Next, we identify the ROIs by thresholding a set of feature images, which are created with a weighted sum of the up-sampled VOI and the blob filter response of the up-sampled VOI. The weighting scalars are determined using *a-priori* knowledge of the geometry of the electrode array model. Then, we identify the ROIs by using the feature images. We perform a voxel thinning method [7] on each of the ROIs to generate the medial axis points as COIs. Once the COIs are extracted, we treat them as nodes in a graph.

We use a coarse path-finding algorithm to firstly find a fixed-length candidate path with the $N$ COIs on that path representing the $N$ electrodes on the array. The candidate path selected minimizes a cost function we designed. Finally, we use a second path-finding algorithm to locally refine the location of each coarsely localized electrode. The final path minimizes another cost function designed for this purpose. The validation study performed to validate this method shows that among 125 clinical CTs, this method generate final localization results with a mean error of 0.12mm when comparing them with the average of two manual localization results generated by an expert. The mean localization error of this method outperforms the other existing electrode localization methods and it is close to the mean rater's consistency error. Another validation study performed on 28 CTs of a cadaveric specimen acquired with different acquisition parameters (dose, resolution, extended or limited Hounsfield range, and the types of electrode array) shows that this method is not sensitive to acquisition parameters [9]. This method represents the state-of-the-art for the automatic localization of CI electrodes in distantly-spaced arrays. It is also a crucial step for fully automating IGCIP.

In Chapter IV, we present a generic method for localizing closely-spaced electrode arrays in clinical CTs [4]. This method is a generalization of the preliminary method presented in Chapter II that can be applied to a range of closely-spaced array types and to images acquired with different CT scanners. It firstly generates the VOI using a reference image. Then, a feature image is computed using the weighted sum of the intensity of VOI and the Frangi vesselness filter response [10]. We threshold the feature image to generate the ROIs that contain electrodes and false positive voxels. For each ROI, we perform a voxel thinning step [7] to generate its medial axis line. A particular connection of medial axes is denoted as a "centerline candidate". We propose an approach to find the centerline of the implanted array by exhaustively searching all the centerline candidates for the positions of the most basal and apical electrodes, such that the centerline defined by those two points and the points between them minimizes a cost function we have designed. After finding the centerline of the implanted array, we resample it by using the known electrode spacing distance of the array. The points on the resampled curve correspond to the centers of the electrodes. On a testing dataset consisting of 129 clinical CTs implanted with three types of electrode arrays, our centerline-based method generates localization results with mean localization error of 0.13mm. 98% of our results have a maximum localization error lower than one voxel diagonal. This method can generate localization results for closely-spaced arrays with errors that are close to the rater's consistency errors and are smaller than the snake-based method discussed in Chapter II. This method is the state-of-the-art for the automatic localization of CI electrodes in closely-spaced arrays. With the methods presented in Chapter III and IV, we are now capable of fully automating the electrode localization step in IGCIP.

Chapter V presents an automatic method [5] for automatic electrode configuration selection in IGCIP. The method captures the heuristics used by the expert when selecting electrode configurations with the assistance of a method to visualize the spatial relationship between electrodes and the auditory nerves determined with the image analysis techniques presented in Chapter III and IV. In this method, we design a DVF-feature-based cost function and train its parameters using existing electrode configurations in our database. In the testing stage, given a set of DVF curves, our method computes the cost values for all the possible configurations and selects the configuration with the lowest cost as the automatic electrode configuration. The validation study has shown that our method generalizes well on a large-scale testing dataset and that it can produce acceptable electrode configurations in most cases. 98.3% of the automatic configurations generated by our method in our testing dataset are rated as acceptable by two experts. These results suggest that our method is a viable approach for automatic selection of electrode configuration in IGCIP. This is the first method that is capable of automatically generating electrode configurations that are comparable to those manually selected by human experts. Our fully automated electrode localization methods (presented in Chapter III and IV) and our automated electrode configuration selection method are critical to permit translation of IGCIP from the laboratory to clinical use.

In Chapter VI, we create a highly accurate ground truth dataset to characterize the accuracy of the electrode localization and the intra-cochlear anatomy segmentation methods we have developed for IGCIP [6]. The ground truth dataset is created with 35 temporal bone specimens. All specimens underwent pre- and post-implantation CT imaging and post-implantation μCT imaging. Six of them underwent pre-implantation μCT imaging. We use the post-implantation μCTs to manually localize the electrodes and the pre-implantation μCTs to manually segment the anatomy. Manual localizations and segmentations serve as ground truth. The mean localization error of our electrode localization methods evaluated with the gold-standard ground truth is 0.13mm. The mean segmentation errors of our three intra-cochlear anatomy segmentation methods ([11], [12], and [13]) are 0.24mm, 0.41mm, and 0.31mm, respectively. In our sensitivity analysis for IGCIP, we found that IGCIP is not sensitive to the electrode localization method. For intra-cochlear anatomy segmentation method, we found that IGCIP achieves the best performances when using method [11] and [13], on pre- and post-implantation CTs, respectively. In a qualitative evaluation of the automatic electrode configurations generated by IGCIP using the most advanced automatic image processing techniques, we found that

IGCIP can generate configurations that are 86.7% acceptable when the pre-implantation CTs are available, and 83.3% acceptable when the pre-implantation CTs are not available. This shows that our automatic techniques for IGCIP can, in most cases, lead to reliable electrode deactivation configurations for improving hearing outcomes for CI recipients. This is the first thorough validation study on the sensitivity of IGCIP to the errors introduced by the IGCIP-related automatic image processing techniques we have developed. We have also created a highly accurate ground truth dataset made of 35 temporal bone specimens. The ground truth dataset includes expert localization of electrode positions in post-implantation μCTs and expert segmentation of the intra-cochlear anatomy in pre-implantation μCTs. This dataset and the validation framework we have developed can be used for other validation studies related to other aspects of IGCIP.

Even though we have made substantial progresses in automating IGCIP, further improvements are possible. With regards to electrode configuration selection, our proposed method relies on three sets of parameters, the values of which are separately estimated with three sets of DVF curves corresponding to the three arrays models produced by the three major manufacturers, respectively. This design limits the potential of this method to be used for other arrays with different numbers of electrodes. *Zhang et al.* proposed a generic algorithm for electrode configuration selection [14] that uses a set of DVF curves with known expert-approved configurations to build a DVF patch library. This library is used by a template matching-based method for selecting electrode deactivation configurations for a new set of DVFs. The validation study results presented in [14] show that the template matching-based method generates configurations with quality that are comparable to the ones obtained by our proposed method. In the future, the assessment of the effectiveness of the configurations generated by these two methods should also be done by comparing hearing outcomes in the same group of CI recipients when using the two configurations recommended by the two methods.

The validation study in Chapter VI includes ground truths for both electrode localization and intra-cochlear anatomy segmentation. One limitation is that the electrode arrays we have in this dataset are produced by only one of the major manufacturers (Advanced Bionics®, Valencia, CA, USA) and the electrodes in those arrays are all distantly-spaced. Thus, only the distantly-spaced array localization method has been validated by using the dataset. In the future, a larger study should be done with an expanded dataset that contains both distantly- and closely-spaced CI electrode arrays. Another limitation is that one intra-cochlear anatomy segmentation method [16] is not validated. This method requires a clinical CT containing both ears with only one implanted ear. It would be desirable to acquire more specimens to enable the validation of this specific method. We also note that the best approach to assess the quality of different electrode configurations is to compare hearing outcomes obtained with each of them. This is difficult to do because it require CI recipients to commute between home and the Vanderbilt University Medical Center several times for reprogramming and hearing outcomes evaluation. In the future, such study could be done with a limited number of recipients who live close by Vanderbilt University and are willing to participate in our research.

The current assumption on which IGCIP is based on is that the electrode interaction is associated with the distance between the electrodes and the modiolar surface. The DVF curves are also designed to visualize a simplified group of stimulation patterns based on this distance information. In the future, a more complicated electrode stimulation model can be created for the electrodes in the different locations within cochlea. Thus, a better method for characterizing electrode interaction can be one direction of future research.

The automatic electrode localization techniques presented in this paper also enables a thorough investigation of the correlation between intra-cochlear locations of CI electrodes and hearing outcomes. These studies can be conducted by using hearing outcome data and the clinical whole head CTs of a large number of CI recipients. The results of these studies could inform the design of future CI arrays and provide valuable information for the implantation phase of the procedure.

This dissertation presents methods that have automated two crucial steps in IGCIP: electrode localization in post-implantation CTs and automatic electrode configuration selection for CI programming. The automatic techniques presented in Chapter III, IV and V have been integrated in the latest version of the IGCIP software. The inclusion of these two procedures is key to make IGCIP a fully automatic end-to-end system. Although the automatic techniques that have been presented herein may not be the final solutions for IGCIP, we believe the work that has been accomplished has made valuable contributions towards improving hearing outcomes for CI recipients and that it provides efficient tools for future research related to image-guided cochlear implant programming.